\documentclass[final,3p,times]{elsearticle}

\usepackage{graphicx}
\usepackage{sidecap}

\usepackage{amssymb}
\usepackage{version}

\usepackage[english,francais]{babel}

\def\og{\leavevmode\raise.3ex\hbox{$\scriptscriptstyle\langle\!\langle$~}}
\def\fg{\leavevmode\raise.3ex\hbox{~$\!\scriptscriptstyle\,\rangle\!\rangle$}}
\newcommand{\g}[1]{\og~#1\fg}

\usepackage[utf8]{inputenc}
\usepackage[T1]{fontenc}
\usepackage{amsmath}
\usepackage{color}
\usepackage{hyperref}
\hypersetup{breaklinks=true}
\hypersetup{colorlinks=true}
\hypersetup{urlcolor=blue}
\hypersetup{citecolor=black}
\hypersetup{linkcolor=black}
\usepackage{cleveref}
\newcommand{\bea}{\begin{eqnarray}}
\newcommand{\eea}{\end{eqnarray}}
\newcommand{\be}{\begin{equation}}
\newcommand{\ee}{\end{equation}}
\newcommand{\rr}{\mathbf{r}}
\newcommand{\RR}{\mathbf{R}}
\newcommand{\kk}{\mathbf{k}}

\newcommand{\uu}{\mathbf{u}}
\newcommand{\vv}{\mathbf{v}}

\newcommand{\mD}{\mathcal{D}}
\newcommand{\mA}{\mathcal{A}}
\newcommand{\mB}{\mathcal{B}}
\newcommand{\mE}{\mathcal{E}}
\newcommand{\mJ}{\mathcal{J}}
\newcommand{\mK}{\mathcal{K}}
\newcommand{\mT}{\mathcal{T}}
\newcommand{\mR}{\mathcal{R}}
\newcommand{\mI}{\mathcal{I}}
\newcommand{\mC}{\mathcal{C}}
\newcommand{\KKK}{\mathbb{K}}
\newcommand{\zero}{\mathbf{0}}

\newcommand{\veps}{\varepsilon}
\newcommand{\omb}{\bar{\omega}}
\newcommand{\mb}{\bar{m}}
\newcommand{\sigg}{\boldsymbol{\sigma}}
\newcommand{\yvan}{\color{black}}

\newcommand{\ii}{\textrm{i}}
\newcommand{\eee}{\textrm{e}}
\newcommand{\dd}{\mathrm{d}}
\def\identit{\mbox{l\hspace{-0.55em}1}}

\DeclareMathOperator\asin{arcsin}
\DeclareMathOperator\acos{arccos}
\DeclareMathOperator\atan{arctan}

\DeclareMathOperator\argth{argth}

\DeclareMathOperator\thf{th}
\DeclareMathOperator\ch{ch}
\DeclareMathOperator\sh{sh}
\DeclareMathOperator\re{Re}
\DeclareMathOperator\im{Im}
\newcommand{\rouge}{\color{black}}
\newcommand{\bleu}{\color{black}}
\usepackage{float}

\begin{document}

\begin{frontmatter}


\selectlanguage{english}
\title{Fourth cluster and virial coefficients of a unitary Fermi gas for an arbitrary mass ratio}


\author[ad1]{Shimpei Endo}
\author[ad2]{Yvan Castin}

\address[ad1]{Physics Department, Tohoku University, Sendai, Japan}
\address[ad2]{Laboratoire Kastler Brossel, ENS-Universit\'e PSL, CNRS, Universit\'e de la Sorbonne and Coll\`ege de France, 24 rue Lhomond, 75231 Paris, France}



\begin{abstract}
We calculate the fourth cluster coefficients of the homogeneous unitary spin 1/2 Fermi gas as functions of the mass ratio {\yvan of spin-up and spin-down states}, over intervals constrained by the 3- or 4-body Efimov effect. For this we use our 2016 conjecture (validated for equal masses by Hou and Drut in 2020) in a numerically efficient formulation making the sum over angular momentum converge faster, which is crucial at large mass ratio. The mean cluster coefficient, relevant for equal chemical potentials, is not of constant sign and increases rapidly close to the Efimovian thresholds. We also get the fourth virial coefficients, which we find to be very poor indicators of interaction-induced 4-body correlations. We obtain analytically for all $n$ the cluster coefficients of order $n+1$ for an {\yvan infinite-mass} impurity fermion, {\yvan and find agreement with} the conjecture for $n=3$. Finally, in a harmonic potential, we predict a non-monotonic behavior of the $3+1$ cluster coefficient with trapping frequency, {\yvan at mass ratios close but not equal to the} mass ratios where this coefficient vanishes in the homogeneous case. 

\noindent 
\vskip 0.5\baselineskip

\end{abstract} 
\end{frontmatter}

\selectlanguage{english}



\section{Introduction to the problem and main results}
\label{sec1}

\paragraph{\rouge The system} Our object of study is a three-dimensional gas of non-relativistic neutral fermions with two {\yvan spin} states $\uparrow$ and $\downarrow$ 
in the regime of maximum interaction allowed in the gas phase, i.e. in the so-called unitary limit \cite{unit1,unit2}: there is no interaction between fermions in the same {\yvan spin state and} a binary zero-range interaction, exclusively in the $s$-wave and of infinite scattering length between fermions of different {\yvan spin} states {\yvan $\uparrow$ and $\downarrow$}.\footnote{\label{note0} This situation corresponds to the replacement of the interaction potential by the Wigner-Bethe-Peierls contact conditions on the wave function $\psi$ of the system in each spin configuration 
$|\uparrow\ldots\uparrow\downarrow\ldots\downarrow\rangle=|\uparrow\rangle^{\otimes n_\uparrow}|\downarrow\rangle^{\otimes n_\downarrow}$
: when the distance $r_{ij}$ between particle $i$ in {\yvan spin} state $\uparrow$ and particle $j$ in {\yvan spin} state $\downarrow$ tends to zero at a fixed position $\RR_{ij}$ of their center of mass, the positions $\rr_k$ of the other particles being fixed at values different from $\RR_{ij}$, there exists a constant $A_{ij}$ (function of $\RR_{ij}$ and of the $\rr_k$'s) such that 
$\psi(\rr_1,\ldots,\rr_{n_\uparrow+n_\downarrow})=A_{ij}(a^{-1}-r_{ij}^{-1})+O(r_{ij})$
where $a$ is the scattering length, and this for all $i$ and for all $j$. Here, we are in the unitary limit $1/a=0$.} We consider the general case where the fermion masses $m_\uparrow$ and $m_\downarrow$ can be different {\yvan for the two spin} states, which is the originality of our work.  This system is realizable in laboratory with trapped cold atomic gases. {\yvan Indeed,} its main properties at equilibrium have been determined experimentally {\yvan for equal masses $m_\uparrow=m_\downarrow$:} its superfluidity at low temperature {\yvan has been observed} in the unpolarized case (with equal numbers of $\uparrow$ and $\downarrow$ fermions) \cite{super1,super2} and its equation of state {\yvan has been measured} at any temperature and polarization \cite{eqet1,eqet2,eqet3}. {\yvan The case $m_\uparrow \neq m_\downarrow$ remains to be explored; a natural way to obtain it is to use a mixture of two fully polarised fermionic atomic species, as has been done in references \cite{m1,m2,m3}.}

\paragraph{\rouge Cluster and virial expansions} Given the strength of the interactions, which provides no obvious small parameter, there are few theoretical tools for quantitatively reliable predictions {\rouge on the unitary gas} that can be compared to measurements.  One of them is the diagrammatic Monte Carlo simulation of the many-body problem on a computer \cite{Svis,MC}. Another is the cluster or virial expansion \cite{Huang} of the pressure $P$ of the spatially homogeneous infinite gas into powers of the fugacities $z_\sigma=\exp(\beta\mu_\sigma)$ or of the phase space densities $\rho_\sigma\lambda_\sigma^3$ {\yvan respectively}, in the strongly non-degenerate regime where they tend to zero, with $\rho_\sigma$ the density of the $\sigma=\uparrow,\downarrow$ component in real space, $\mu_\sigma$ its chemical potential, $\lambda_\sigma=(2\pi\hbar^2/m_\sigma k_B T)^{1/2}$ its de Broglie thermal wavelength at temperature $T$, and $\beta=1/k_B T$ \cite{XLiu} : 
\be
\label{eq000}
\frac{P\lambda^3}{k_B T} = \sum_{(n_\uparrow,n_\downarrow)\in\mathbb{N}^{2*}} b_{n_\uparrow,n_\downarrow} z_\uparrow^{n_\uparrow} z_\downarrow^{n_\downarrow} = \sum_{(n_\uparrow,n_\downarrow)\in\mathbb{N}^{2*}} c_{n_\uparrow,n_\downarrow} (\rho_\uparrow\lambda_\uparrow^3)^{n_\uparrow} (\rho_\downarrow\lambda_\downarrow^3)^{n_\downarrow}
\ee
To scale the pressure, it was necessary to introduce a reference de Broglie thermal wavelength $\lambda=(2\pi\hbar^2/\bar{m}k_B T)^{1/2}$ depending on a mean mass $\bar{m}$ to be specified {\yvan (see equation (\ref{eq012}))}.  One can then try to extrapolate to the non-trivial regime $z_\sigma\approx 1$ with heuristic recipes such as the Pad\'e approximant \cite{Pade} or optimized resummation methods \cite{Werner} which take into account the behavior of the coefficients at large orders.  The cluster or virial expansion {\yvan has the advantage} over the Monte Carlo simulation {\yvan that} it is closer to the analytical calculation, {\yvan since} the coefficients of order $n$ {\yvan can be obtained} from the solution of a problem with at most $n$ interacting fermions, i.e. with few bodies: to obtain $b_{n_\uparrow,n_\downarrow}$, it suffices to determine the canonical partition functions of all systems with $n_\sigma$ or less fermions in each {\yvan spin} state {\yvan $\sigma=\uparrow,\downarrow$}. 

\paragraph{\bleu The unitary limit} The scale invariance of the unitary gas simplifies considerably the calculation {\rouge of cluster coefficients} for $n>2$ (the order two, given by the Beth-Uhlenbeck formula \cite{Beth1,Beth2,Landau}, is not debated). Thus, the third-order coefficients are known analytically, even if the scale invariance at the three-body level is broken by the Efimov effect \cite{b3,EPL}, in sharp contrast to the model of hard sphere interaction of radius $a$ where the coefficients are known analytically (for bosons) only in the limiting cases $\lambda/a\ll 1$ \cite{lpda1,lpda2,lpda3,lpda4} or $\gg 1$ \cite{lgda1,lgda2,lgda3,lgda4}. The harmonic regulator method \cite{reg1,reg2,reg3,reg4}, consisting in trapping each component of the gas in a fictitious isotropic harmonic potential, $U_\sigma(\rr)=m_\sigma\omega^2r^2/2$, whose trapping frequency $\omega$ (common to both {\yvan spin} states) is made to tend to zero at the end of the calculations, allows us to take full advantage of the scale invariance since the $n$-body spectrum in the trap is {\yvan obtained} from the discrete set of scale exponents {\rouge $s_i$} of the zero energy $E=0$ eigenstates in free space {\rouge\cite{unit1,sym1}}. More precisely, we generalize the cluster expansion to the trapped case, replacing the pressure by the grand potential {\rouge $\Omega$} and taking its ratio to the partition function $Z_1=1/[2\sh(\omb/2)]^3$ of a single fermion,\footnote{\label{note2} Indeed, the {\rouge numerator in the} first side of equation (\ref{eq000}) can be seen, in a quantization box of arbitrarily large volume $V$, as the ratio between $PV$, {\yvan i.e.\ minus the gas grand potential,} and the partition function $V/\lambda^3$ of a fictitious single particle of mass $\mb$. } so that 
\be
\label{eq001}
\frac{-\Omega}{k_B T Z_1} = \sum_{(n_\uparrow,n_\downarrow)\in\mathbb{N}^{2*}} B_{n_\uparrow,n_\downarrow}(\omb)\, z_\uparrow^{n_\uparrow} z_\downarrow^{n_\downarrow}
\ee
The coefficients of the trapped case, marked by a capital letter to avoid confusion, depend only on the dimensionless ratio ${\rouge\omb=}\hbar\omega/k_B T$ {\yvan due} to the scale invariance of the unitary gas. They are related to those of the homogeneous case by means of the local density approximation, exact in the limit $\omega\to 0$ {\rouge\cite{reg3,reg4} and giving} \cite{EPL}: 
\be
\label{eq002}
b_{n_\uparrow,n_\downarrow}= \frac{(n_\uparrow m_\uparrow+n_\downarrow m_\downarrow)^{3/2}}{\bar{m}^{3/2}} B_{n_\uparrow,n_\downarrow}(0^+)
\ee
This method leads to an analytic integral expression of the third order coefficients because the transcendental Efimov function {\rouge $\Lambda(s)$}, whose roots are the scale exponents {\rouge $s_i$}, is known explicitly \cite{f1,f2,f3,f4}. {\yvan This} allows to express the coefficients as a contour integral around $\mathbb{R}^+$ by means of the residue theorem and then to unfold the contour on the pure imaginary axis by analyticity on $\mathbb{C}\setminus\mathbb{R}$ \cite{b3}. On the other hand, for non-trivial fourth order coefficients,\footnote{\label{note1} If one of the $n_\sigma$ is zero, the fermions of the $n$-body problem are non-interacting and the corresponding $B$ coefficient reduces to that of the ideal gas. We thus find 
$B_{n,0}(0^+)=B_{0,n}(0^+)=(-1)^{n+1}/n^4$
and, by means of relation (\ref{eq002}), 
$b_{n,0}=(m_\uparrow/\mb)^{3/2} (-1)^{n+1}/n^{5/2}$ and $b_{0,n}=(m_\downarrow/\mb)^{3/2}(-1)^{n+1}/n^{5/2}$.
} the Efimov function is the determinant of operators $M_{3,1}(s)$ \cite{PRL} or $M_{2, 2}(s)$ \cite{PRA} parametrically dependent on $s$; we know how to compute {\yvan it} numerically only on the pure imaginary axis $s=\ii S$ and {\yvan its} imperfectly known analytical properties in the complex plane {\rouge do not {\bleu guarantee} a safe use of the residue theorem}. Therefore reference \cite{JPA} could only produce a conjecture, which we briefly recall. 

\paragraph{\rouge {\bleu Our} 2016 conjecture on the fourth cluster coefficients} {\rouge As in reference \cite{PRA},} we introduce the integral expression modeled on that of the third order {\rouge coefficients} {\rouge ($\Lambda(s)$ is replaced by an operator determinant):} 
\be
\label{eq003}
\boxed{
I_{n_\uparrow,n_\downarrow}(\bar{\omega}) = \sum_{\ell=0}^{+\infty} \sum_{\veps} (\ell+1/2) \int_{-\infty}^{+\infty} \frac{\dd S}{2\pi} \frac{\sin(\bar{\omega}S)}{\sh\bar{\omega}} \frac{\dd}{\dd S} \ln\det [M_{n_\uparrow,n_\downarrow}^{(\ell,\veps)}(\ii S)]
}
\ee
where the sum is taken on the internal angular momentum $\ell$ (i.e. after separation of the center of mass) of the four-body eigenstates and their internal parity $\veps=\pm 1$ (limited to $\veps=1$ for $\ell=0$), and where the operator $M_{n_\uparrow,n_\downarrow}(\ii S)$ is restricted to the corresponding $(\ell,\veps)$ subspace. Then $I_{n_\uparrow,n_\downarrow}(\omb)$ gives exactly $B_{n_\uparrow,n_\downarrow}(\omb)$ (as is the case at order three) when the asymptotic decoupled objects (independent non-monoatomic packets of fermions strongly correlated by interactions, in terms of which the highly excited eigenstates are expressed) are distinguishable; otherwise, the quantum statistical effect of the indistinguishability of these objects, which contributes to the cluster coefficient even if they do not interact with each other, is missing, for the same reason that the cluster coefficients of the quantum ideal gas differ from those of the classical ideal gas. In the $(n_\uparrow,n_\downarrow)=(3,1)$ configuration, called $3+1$ for short {\yvan from here on}, the possible asymptotic objects are a $\uparrow\uparrow\downarrow$ triplon or a $\uparrow\downarrow$ pairon of strongly correlated fermions; there can only be one at a time, which rules out any quantum statistical effect. On the other hand, in the $2+2$ configuration, the fermions can also decouple into two $\uparrow\downarrow$ pairons of correlated fermions, which are indistinguishable bosons; since these bosons do not interact, their partition function is easily calculated \cite{JPA}. The conjecture of reference \cite{JPA} is finally written 
\be
\label{eq004}
\boxed{
B_{3,1}(\bar{\omega})=I_{3,1}(\bar{\omega}) \quad ; \quad B_{2,2}(\bar{\omega})=I_{2,2}(\bar{\omega}) + \frac{1}{32} \frac{1}{\ch\bar{\omega} \ch^3(\bar{\omega}/2)}
}
\ee
the $1+3$ case being deduced from the $3+1$ case by exchanging the two {\yvan spin} states thus changing the mass ratio $\alpha=m_\uparrow/m_\downarrow$ to its inverse $1/\alpha$. For $\alpha=1$, conjecture (\ref{eq004}) is in agreement with the quantum Monte Carlo calculation of reference \cite{Blume} {\yvan down to} the minimal accessible value of $\omb$, $\omb\approx 1$; it is {\yvan also} in agreement with a recent, more powerful numerical calculation {\yvan down to} values of $\omb\ll 1$ \cite{DrutTrap1,DrutTrap2}. After using relation (\ref{eq002}), it is also in agreement with the same powerful numerical calculation performed directly in the spatially homogeneous case of a quantization box \cite{DrutBox}. Conjecture (\ref{eq004}) is thus confirmed for equal masses.

\paragraph{\rouge Content of the study} {\rouge The idea} of the present work is to believe in the validity of conjecture (\ref{eq004}) for any mass ratio $\alpha$ and to access the corresponding fourth cluster and virial coefficients, by numerically calculating the determinant of the operators 
$M^{(\ell,\veps)}_{n_\uparrow,n_\downarrow}(\ii S)$,
and then integrating over $S$ and summing over $\ell$ and $\veps$ {\rouge in expression (\ref{eq003})}. In reality, we still have to put bounds on the mass ratio, because the method of solving the four-body problem in a harmonic trap, {\rouge at the basis of} expression (\ref{eq004}), assumes separability of the internal Schr\"odinger equation in hyperspherical coordinates; this is true only if there is no Efimov scale invariance breaking at the three-body level, which constrains us to the intervals 
\be
\label{eq005}
\mbox{for }B_{3,1}: \alpha < \alpha_c^{\rm 3\, body}\simeq 13.60697 \quad ; \quad \mbox{for }B_{1,3}: \alpha > \frac{1}{\alpha_c^{\rm 3\, body}}\simeq 0.0734917 \quad ; \quad \mbox{for }B_{2,2}: \frac{1}{\alpha_c^{\rm 3\, body}} < \alpha < \alpha_c^{\rm 3\, body}
\ee
where $\alpha_c^{\rm 3\, body}$, also noted $\alpha_c^{2, 1}$, is the threshold of the three-body Efimov effect in the $\uparrow\uparrow\downarrow$ system, which occurs in the three-body internal momentum channel $L=1$ \cite{f1,Petrov}. {\bleu In using form (\ref{eq004}) of the conjecture, it is also important to ensure} that there is no four-body Efimov effect; \footnote{\label{note3} {\yvan Moreover, our assumption of a strictly scale-invariant interaction becomes difficult to satisfy experimentally if $\alpha$, although on the non-Efimovian side, is too close to the tetramer appearance threshold. All this can be improved.} One could complement the $3+1$-body contact condition of the usual zero range interaction model of footnote \ref{note0} by means of a length called \g{$3+1$-body parameter} in the $(\ell,\veps)=(1,+1)$ \cite{PRL} channel where the $3+1$-body Efimov effect occurs. In this case, $B_{3,1}(0^+)$ and $b_{3,1}$ would be smooth functions of $\alpha$ at $\alpha=\alpha_c^{\rm 4\, body}$. {\rouge Reference \cite{EPL} {\bleu proves and implements this} for $B_{2,1}(0^+)$ that is in the $2+1$ body problem.}} {\bleu indeed, it was shown that} such an effect occurs only in the $3+1$ or $1+3$ configuration \cite{PRL,PRA}, at mass ratio $\alpha_c^{\rm 4\, body}=\alpha_c^{3,1}$ or its inverse, hence the additional conditions 
\be
\label{eq006}
\mbox{for }B_{3,1}: \alpha < \alpha_c^{\rm 4\, body}\simeq 13.3842 \quad ; \quad \mbox{for }B_{1,3}: \alpha > \frac{1}{\alpha_c^{\rm 4\, body}}\simeq 0.074715
\ee
Once the cluster {\yvan expansion is} known, the virial coefficients $c_{n_\uparrow,n_\downarrow}$ are easily deduced, as rational functions of {\yvan the cluster coefficients} $b_{n_\uparrow',n_\downarrow'}$ of total order {\rouge $n'_\uparrow+n'_\downarrow$} less than or equal to $n_\uparrow+n_\downarrow$: \footnote{To obtain them, we replace in the expansion of the pressure in the third side of equation (\ref{eq000}) the densities by their virial expansion 
$\rho_\sigma\lambda^3=\sum_{(n_\uparrow,n_\downarrow)\in\mathbb{N}^{2*}} n_\sigma b_{n_\uparrow,n_\downarrow} z_\uparrow^{n_\uparrow} z_\downarrow^{n_\downarrow}$
from the thermodynamic relation {\yvan $\rho_\sigma=\partial P/\partial\mu_\sigma$}, and we adjust the $c_{n_\uparrow,n_\downarrow}$ order by order to find the expansion in the second side of equation (\ref{eq000}).  It is simpler, at first, to write the virial expansion in the form of 
$P\lambda^3/k_B T=\sum_{(n_\uparrow,n_\downarrow)\in\mathbb{N}^{2*}} a_{n_\uparrow,n_\downarrow} (\rho_\uparrow\lambda^3)^{n_\uparrow} (\rho_\downarrow\lambda^3)^{n_\downarrow}$
and then, once the coefficients $a_{n_\uparrow,n_\downarrow}$ have been calculated, to use the relation 
$c_{n_\uparrow,n_\downarrow}=a_{n_\uparrow,n_\downarrow}/(t_\uparrow^{n_\uparrow} t_\downarrow^{n_\downarrow})$
where we have put $t_\sigma=(\mb/m_\sigma)^{3/2}$. Note that {\rouge $b_{1,0}t_\uparrow=b_{0,1}t_\downarrow=1$, and that} $a_{1,0}=a_{0,1}=1$ whatever the reference mass $\mb$ according to the ideal gas law $P=(\rho_\uparrow+\rho_\downarrow) k_B T$. Replacing $b_{n,0}$ and $b_{0,n}$ by their expressions given in footnote \ref{note1}, we finally get 
$c_{3,1}=(-3/4+1/\sqrt{3})b_{1,1}-(3/\sqrt{2})b_{2,1}-3b_{3,1}+3b_{1,1}(b_{2,1}+b_{1,1}/\sqrt{8})t_\downarrow-b_{1,1}^3t_\downarrow^2$, $c_{2,2}=-3[b_{2,2}+(b_{1,2}+b_{2,1})/\sqrt{8}+b_{1,1}/8]+(9\sqrt{2}/8)b_{1,1}^2(t_\uparrow+t_\downarrow)+6b_{1,1}(b_{2,1}t_\uparrow+b_{1,2}t_\downarrow)-3b_{1,1}^3t_\uparrow t_\downarrow$
(the ideal gas coefficients $c_{4,0}$ and $c_{0,4}$ are given in explicit form later in the main text). We easily derive $b_{1,1}$ from equation (\ref{eq002}) knowing that $B_{1,1}(0^+)=1/2$. On the other hand, $b_{2,1}$ and $b_{1,2}$ are computed numerically from explicit integral expressions {\rouge of} reference \cite{EPL}.} 
\be
\label{eq017a}
c_{1,0}=b_{1,0} \quad ; \quad c_{2,0}=-b_{2,0} \quad ; \quad c_{3,0}=\frac{4 b_{2,0}^2}{b_{1,0}}-2b_{3,0} \quad ;\quad c_{4,0}=-3 b_{4,0} -\frac{20 b_{2,0}^3}{b_{1,0}^2}+\frac{18 b_{3,0}b_{2,0}}{b_{1,0}} \quad ; \quad c_{1,1}=-b_{1,1} 
\ee
\be
\label{eq017b}
\quad c_{2,1}=-2b_{2,1}+\frac{4 b_{1,1} b_{2,0}}{b_{1,0}} +\frac{b_{1,1}^2}{b_{0,1}} \quad ; \quad c_{3,1}=-3 b_{3,1}-\frac{b_{1,1}^3}{b_{0,1}^2} - \frac{6 b_{1,1}^2 b_{2,0}}{b_{1,0}b_{0,1}} - \frac{24 b_{1,1} b_{2,0}^2}{b_{1,0}^2} +\frac{3 b_{1,1} b_{2,1}}{b_{0,1}} +\frac{12 b_{2,0} b_{2,1}}{b_{1,0}} + \frac{9 b_{1,1} b_{3,0}}{b_{1,0}}
\ee
\be
\label{eq017c}
c_{2,2}=-3b_{2,2}-\frac{3 b_{1,1}^3}{b_{0,1}b_{1,0}}-9b_{1,1}^2\left(\frac{b_{0,2}}{b_{0,1}^2}+\frac{b_{2,0}}{b_{1,0}^2}\right)+6 b_{1,1}\left(\frac{b_{1,2}}{b_{0,1}}-\frac{2b_{0,2}b_{2,0}}{b_{0,1}b_{1,0}}+\frac{b_{2,1}}{b_{1,0}}\right)+\frac{6b_{0,2}b_{2,1}}{b_{0,1}}+\frac{6b_{1,2}b_{2,0}}{b_{1,0}}
\ee
plus the equations obtained by exchanging the two {\yvan spin} states $\uparrow$ and $\downarrow$. Our expressions of $c_{n,0}$ are in agreement with equation (10.33) of reference \cite{Huang}, except that they extend it to a ratio $m_\uparrow/\mb$ different from unity through the coefficient $b_{1,0}=\lambda^3/\lambda_\uparrow^3$.

Before presenting our results, we need to choose the reference mass $\bar{m}$. To do so, we {\rouge rely on} the particular form of the cluster expansion commonly used in the literature for equal fugacities, which puts the number of {\yvan spin} states of the fermions as a factor of the series, 
\be
\label{eq007}
\frac{P\lambda^3}{k_B T} \stackrel{z_\sigma=z}{\equiv} 2\sum_{n=1}^{+\infty} b_n z^n \quad \mbox{with}\quad b_n=\frac{1}{2}\sum_{n_\uparrow=0}^{n} b_{n_\uparrow,n-n_\uparrow}
\ee
see in particular references \cite{eqet1,eqet3} where $b_4$ is measured for equal masses. \footnote{The values of $b_4$ measured at ENS and MIT are in agreement with each other but within a factor $\simeq 2$ and respectively within $\simeq 2\sigma$ and $\simeq 3.5\sigma$ of the now accepted value \cite{DrutBox}, where $\sigma$ is the experimental uncertainty. In the case of ENS, the discrepancy is due to the fact that all usable pressure data are of fugacity $z>1$ and their naive extrapolation to $z=0$ fails. Indeed, {\yvan series} (\ref{eq007}) converges slowly and cannot be truncated to a good approximation at $n=4$ even for a value as small as $z=0.22$, see the augmented version \cite{aug} of publication \cite{JPA} which draws this conclusion from the high precision equation of state obtained by diagrammatic Monte Carlo \cite{Werner,Svis}.}  {\yvan Pulling out} such a factor {\yvan makes sense} if the coefficient $b_1$ takes the very simple value equal to one. From footnote \ref{note1}, we thus derive the natural choice \footnote{In the context of footnote \ref{note2}, this choice amounts to taking as a reference one-body partition function $\bar{Z}_1$ the arithmetic mean of $Z_{1\uparrow}$ and $Z_{1\downarrow}$. It reduces indeed to $\mb=m$ in the case of equal masses $m_\uparrow=m_\downarrow=m$. It ensures that the factor {\rouge relating} the trapped case to the homogeneous case in equation (\ref{eq002}) is a uniformly bounded function of the masses (bounded from above by $2(n_\uparrow^3+n_\downarrow^3)^{1/2}$ according to H\"older's inequality of parameters $p=3/2$ and $q=3$). These conditions are not satisfied if we take for $\mb$ the reduced mass of two $\uparrow$ and $\downarrow$ fermions as in reference \cite{EPL}.} 
\be
\label{eq012}
b_1=1\quad\Longrightarrow\quad\mb^{3/2}=\frac{1}{2}(m_\uparrow^{3/2}+m_\downarrow^{3/2}) \quad\mbox{hence}\quad \frac{1}{\lambda^3}=\frac{1}{2}\left(\frac{1}{\lambda_\uparrow^3}+\frac{1}{\lambda_\downarrow^3}\right)
\ee
This choice also makes the virial expansion very simple in the case of equal phase space densities in the two {\yvan spin} states: one then has the nice expression $\rho_\sigma\lambda_\sigma^3=\rho\lambda^3/2$ for the joint value, where $\rho=\rho_\uparrow+\rho_\downarrow$ is the total density, and one sets as in equation (\ref{eq007}): 
\be
\label{eq008}
\frac{P\lambda^3}{k_B T} \stackrel{\rho_\sigma\lambda_\sigma^3=\rho\lambda^3/2}{\equiv} 2\sum_{n=1}^{+\infty} c_n (\rho\lambda^3/2)^n \quad \mbox{with}\quad c_n=\frac{1}{2}\sum_{n_\uparrow=0}^{n} c_{n_\uparrow,n-n_\uparrow}
\ee

\begin{figure}[t]
\begin{center}
\includegraphics[width=7.5cm,clip=]{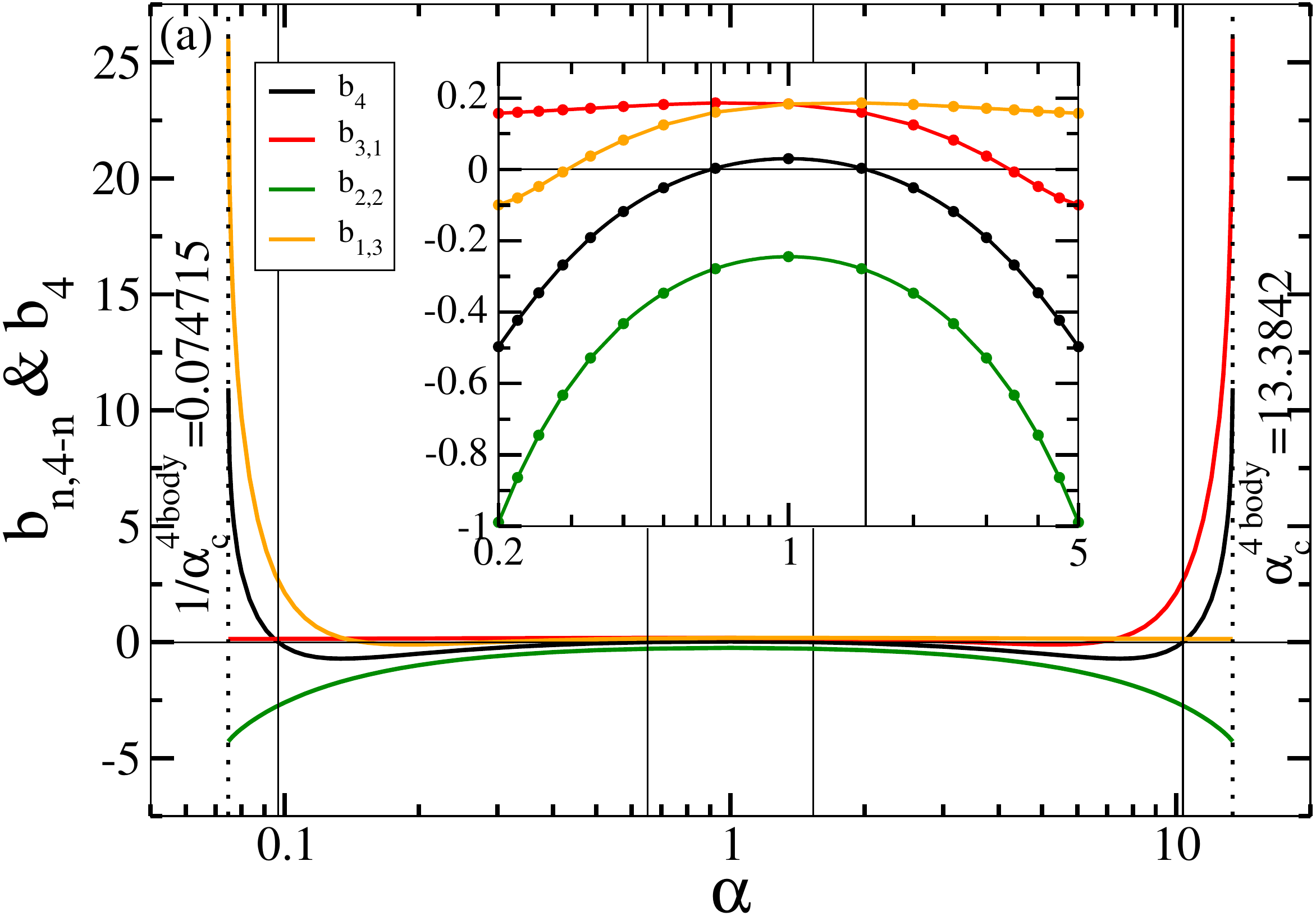}\hspace{5mm}\includegraphics[width=7.5cm,clip=]{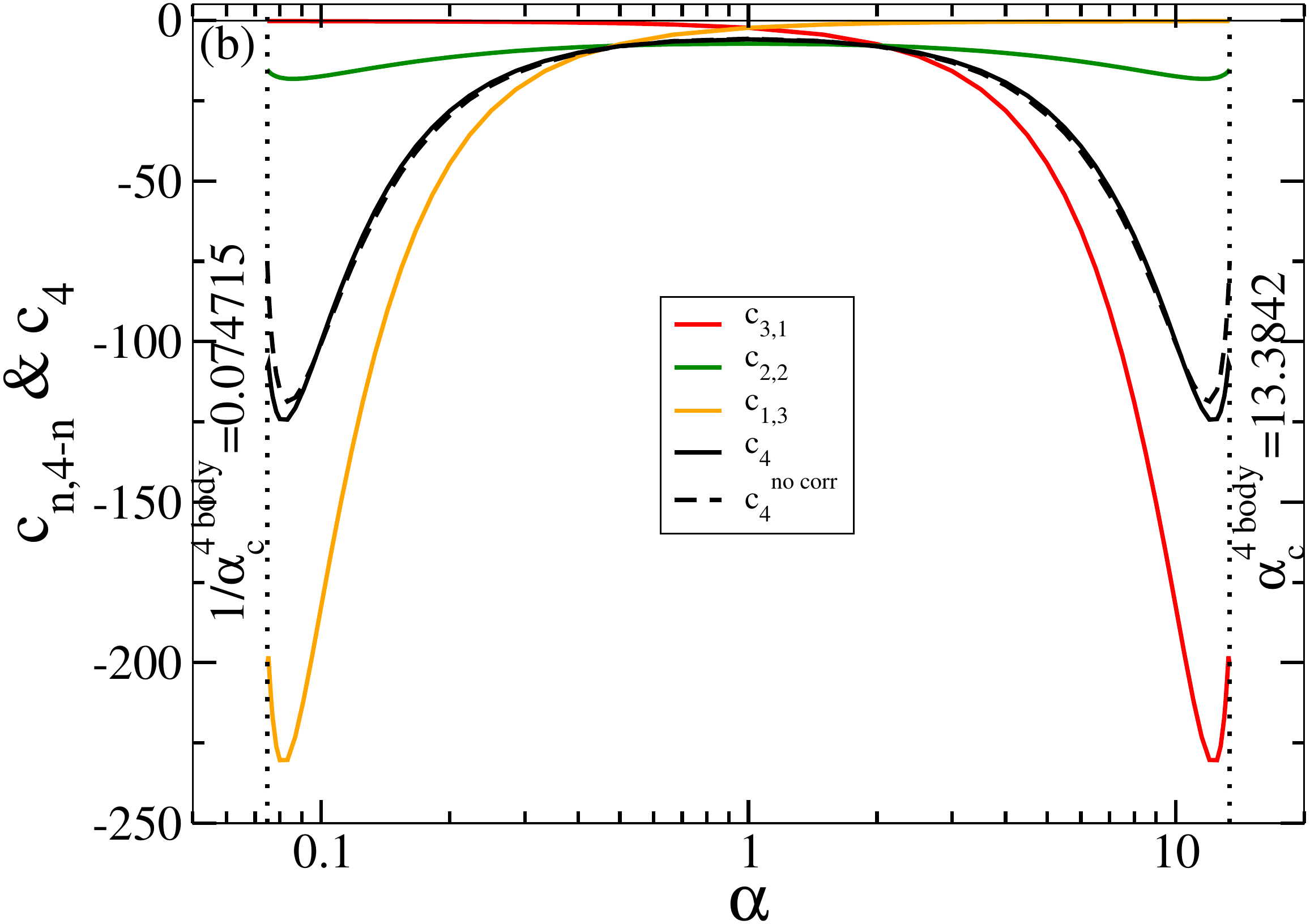}
\end{center}
\caption{For a spatially homogeneous unitary Fermi gas with two {\yvan spin} states $\uparrow$ and $\downarrow$, fourth cluster (a) and virial (b) coefficients defined by equations (\ref{eq000},\ref{eq007},\ref{eq008}) and predicted by conjecture (\ref{eq004}) of reference \cite{JPA}, as functions of the mass ratio $\alpha=m_\uparrow/m_\downarrow$. The $\uparrow\downarrow$ zero-range interaction is assumed to be scale-invariant in the four-body problem, which forces one to restrict oneself to the mass ratios between the critical values of the $3+1$- and $1+3$-body Efimov effect, indicated by the vertical dotted lines. In (a), the thin vertical lines mark the points where $b_4$ vanishes (always with sign change), namely $\alpha\simeq 1.535$ and $\alpha\simeq 10.355$ on the $\alpha>1$ side, and the inset is an enlargement {\rouge (the calculated points are represented by disks on interpolation lines)}. In (b), the dashed curve neglects in $c_4$ the true interaction-induced four-body component, the one \protect
$c_4^{\rm corr}=-3(b_{1,3}+b_{2,2}+b_{3,1})/2$
that depends on $b_{3,1}$, $b_{2,2}$ or $b_{1,3}$, to represent $c_4^{\rm no\, corr}=c_4-c_4^{\rm corr}$.}
\label{fig1}
\end{figure}

\paragraph{\rouge The results} We plot the {\rouge unitary gas} non-trivial fourth cluster coefficients ({\yvan we do not plot here} those of the ideal gas) as functions of the mass ratio $\alpha$ in figure \ref{fig1}a (the values $b_{4, 0}=-(m_\uparrow/\mb)^{3/2}/32$ and $b_{0,4}=-(m_\downarrow/\mb)^{3/2}/32$ taken from footnote \ref{note1} are thus not {\yvan plotted} but of course contribute to $b_4$). The logarithmic scale used on the x-axis highlights the $\alpha\leftrightarrow 1/\alpha$ symmetry. For a mass ratio $\alpha=1$, we find the value $b_4=0.030(1)$ conjectured by reference \cite{JPA} and confirmed numerically by reference \cite{DrutBox}. As we move away from this point, the behavior of $b_4$ is first imposed by $b_{2,2}$ which causes it to change sign (the black curve and the green curve are almost parallel), before $b_{3,1}$ or $b_{1,3}$ prevails at high mass ratios $\alpha$ or $1/\alpha$ and causes $b_4$ to rise to large and positive values. At the $\alpha_c^{\rm 4\, body}$ or $1/\alpha_c^{\rm 4\, body}$ thresholds of the $3+1$- or $1+3$-body Efimov effect, marked by dotted vertical lines, $b_4$ has a finite limit but an infinite derivative like $b_{3,1}$ or $b_{1,3}$, see section \ref{sec2} below and footnote \ref{note3}. On the other hand, $b_{2,2}$ remains a smooth function, and would only exhibit an infinite derivative at the thresholds of the three-body Efimov effect, see section \ref{sec3}. In figure \ref{fig1}b, we see that the non-trivial fourth virial coefficients ({\yvan we do not plot here the trivial ones} $c_{4,0}=A_4(m_\uparrow/\mb)^{3/2}$ and $c_{0, 4}=A_4(m_\downarrow/\mb)^{3/2}$, with $A_4=(18+15\sqrt{2}-16\sqrt{6})/192\simeq 1.11\times 10^{-4}$) have a similar structure even though they are of constant sign. We also notice that coefficient $c_4$ depends little on the four-body correlations induced by the interactions: an approximation of $c_4$ neglecting the contributions of the cluster coefficients $b_{3,1}$, $b_{2,2}$ and $b_{1,3}$, {\yvan plotted} in dashed line in figure \ref{fig1}b, is everywhere close or very close to $c_4$. {\yvan The} fairly recent possibility to prepare spatially homogeneous cold atomic gases in a flat-bottom potential box \cite{box1,box2} makes the measurement of the virial coefficients {\bleu $c_n$} very natural, since the densities $\rho_\sigma$ are directly accessible there. {\yvan This} last prediction makes it {\yvan however} less motivating than the measurement of the cluster coefficients {\bleu $b_n$} in an inhomogeneous gas according to the specific technique for the harmonically trapped case \cite{eqet1}, in which the chemical potentials $\mu_\sigma$ are the {\bleu relevant variables} to be used. \footnote{In a trapping potential $U(\rr)$ common to both {\yvan spin} states and very elongated along the eigenaxis $Oz$, the cold-atom gases are well described by the local density approximation, and thus have well-defined local chemical potentials on the axis $\mu_\sigma^{\rm loc}(z)=\mu_\sigma-U(0, 0,z)$, which makes the measurement of the grand-canonical equation of state, and thus of the cluster coefficients, very straightforward, following the clever proposal of reference \cite{Ho}, which is usable however only {\rouge if $U(\rr)$ is} harmonic.} {\bleu For all practical purposes, we also give the fourth cluster and virial coefficients in numerical form in table \ref{tab:bc}.}

\paragraph{\rouge Outline of the rest of the article} In the following, we explain how we were able to obtain accurate results in a reasonable computation time, in particular by means of a convergence acceleration of the sum on $\ell$ in expression (\ref{eq003}) relying on an asymptotic expansion of the summand and playing an essential role near the Efimovian thresholds. To do so, we had to generalize the analytical method of reference {\rouge\cite{PRA}}, implemented for $3+1$ bodies, to the more difficult case of $2+2$ bodies. In the $3+1$-body case, we also highlight an unexpected non-monotonic dependence of $B_{3,1}(\omb)$ {\yvan on} $\omb$, with change of sign, for mass ratios $\alpha$ close to $3.5$ or $6.6$. Section \ref{sec2} deals with the $3+1$-body case and section \ref{sec3} with the $2+2$-body case in a harmonic potential.

\begin{table}[t]
{\footnotesize
\begin{center}
\begin{tabular}{|c|c|c|c|c|c|c|c|c|c|c|}
\hline
$\alpha$ & 1 & 1.5 & 2 & 2.5 & 3 & 3.5 & 4 & 4.5 & 5 & 5.5\\
\hline
\hline
$b_{3,1}$ & 0.1837 & 0.1604 & 0.1247 & 0.08238 & 0.03726 & $-$0.007203 & $-$0.04762 & $-$0.07982 & $-$0.09941 & $-$0.1016 \\
$b_{1,3}$  & 0.18374 & 0.18672 & 0.18199 & 0.17641 & 0.17133 & 0.16696 & 0.16325 & 0.16010 & 0.15740 & 0.15508 \\
$b_{2,2}$  & $-$0.2445 & $-$0.2781 & $-$0.3469 & $-$0.4321 & $-$0.5283 & $-$0.6328 & $-$0.7448 & $-$0.8637 & $-$0.9890 & $-$1.1209 \\
$b_4$ & 0.03026 & 0.003256 & $-$0.05135 & $-$0.1179 & $-$0.19110 & $-$0.2678 & $-$0.3458 & $-$0.4229 & $-$0.4968 & $-$0.5649 \\
\hline
$c_{3,1}$ & $-$2.2558 & $-$4.4056 & $-$7.3318 & $-$11.100 & $-$15.773 & $-$21.407 & $-$28.050 & $-$35.737 & $-$44.495 & $-$54.332 \\
$c_{1,3}$  & $-$2.2558 & $-$1.2265 & $-$0.8296 & $-$0.6292 & $-$0.5114 & $-$0.4350 & $-$0.3820 & $-$0.3433 & $-$0.3139 & $-$0.2910 \\
$c_{2,2}$ & $-$7.1913 & $-$7.3897 & $-$7.7991 & $-$8.3081 & $-$8.8748 & $-$9.4800 & $-$10.111 & $-$10.758 & $-$11.415 & $-$12.078 \\
$c_4$ & $-$5.8513 & $-$6.5108 & $-$7.9801 & $-$10.018 & $-$12.580 & $-$15.661 & $-$19.271 & $-$23.419 & $-$28.112 & $-$33.350 \\
\hline
\hline
$\alpha$ & 6 & 6.5 & 7 & 7.5 & 8 & 8.5 & 9 & 9.5 & 10 & 10.5 \\
\hline
\hline
$b_{3,1}$  & $-$0.07925 & $-$0.02715 & 0.06412 & 0.2022 & 0.4010 & 0.6724 & 1.0339 & 1.5059 & 2.1195 & 2.9095 \\
$b_{1,3}$ & 0.15305 & 0.15128 & 0.14972 & 0.14833 & 0.14708 & 0.14597 & 0.14496 & 0.14404 & 0.14321 & 0.14244 \\
$b_{2,2}$ & $-$1.2590 & $-$1.4034 & $-$1.5542 & $-$1.7113 & $-$1.8749 & $-$2.0452 & $-$2.2223 & $-$2.4077 & $-$2.6003 & $-$2.8015 \\
$b_4$ & $-$0.6239 & $-$0.6709 & $-$0.7014 & $-$0.7116 & $-$0.6947 & $-$0.6447 & $-$0.5530 & $-$0.4101 & $-$0.2001 & 0.09397 \\
\hline
$c_{3,1}$ & $-$65.248 & $-$77.214 & $-$90.190 & $-$104.10 & $-$118.85 & $-$134.28 & $-$150.20 & $-$166.33 & $-$182.32 & $-$197.64 \\
$c_{1,3}$  & $-$0.2725 & $-$0.2574 & $-$0.2449 & $-$0.2343 & $-$0.2252 & $-$0.2174 & $-$0.2105 & $-$0.2045 & $-$0.1991 & $-$0.1944 \\
$c_{2,2}$ & $-$12.740 & $-$13.397 & $-$14.044 & $-$14.676 & $-$15.288 & $-$15.873 & $-$16.423 & $-$16.926 & $-$17.374 & $-$17.747 \\
$c_4$ & $-$39.130 & $-$45.434 & $-$52.239 & $-$59.505 & $-$67.180 & $-$75.184 & $-$83.417 & $-$91.733 & $-$99.948 & $-$107.79 \\
\hline
\hline
$\alpha$ & 11 & 11.5 & 12 & 12.5 & 12.75 & 13 & 13.1 & 13.2 & 13.3 & 13.3842 \\
\hline
\hline
$b_{3,1}$  & 3.9362 & 5.2839 & 7.0993 & 9.7004 & 11.522 & 14.042 & 15.402 & 17.152 & 19.728 & 26.101 \\
$b_{1,3}$  & 0.14174 & 0.14109 & 0.14049 & 0.13993 & 0.13967 & 0.13941 & 0.13932 & 0.13922 & 0.13912 & 0.13904 \\
$b_{2,2}$ & $-$3.0120 & $-$3.2360 & $-$3.4725 & $-$3.7278 & $-$3.8654 & $-$4.0133 & $-$4.0767 & $-$4.1430 & $-$4.2140 & $-$4.2784 \\
$b_4$ & 0.5017 & 1.0632 & 1.8524 & 3.0250 & 3.8667 & 5.0528 & 5.7008 & 6.5428 & 7.7951 & 10.950 \\
\hline
$c_{3,1}$ & $-$211.58 & $-$223.04 & $-$230.30 & $-$230.37 & $-$225.99 & $-$217.03 & $-$211.71 & $-$205.24 & $-$198.00 & $-$200.77 \\
$c_{1,3}$  & $-$0.1901 & $-$0.1863 & $-$0.1828 & $-$0.1796 & $-$0.1781 & $-$0.1766 & $-$0.1761 & $-$0.1756 & $-$0.1750 & $-$0.1746 \\
$c_{2,2}$ & $-$18.025 & $-$18.163 & $-$18.114 & $-$17.767 & $-$17.414 & $-$16.864 & $-$16.561 & $-$16.189 & $-$15.720 & $-$15.213 \\
$c_4$ & $-$114.90 & $-$120.69 & $-$124.30 & $-$124.16 & $-$121.79 & $-$117.04 & $-$114.22 & $-$110.80 & $-$106.95 & $-$108.08 \\
\hline
\end{tabular}
\end{center}
}
\caption{\bleu Numerical values of the fourth cluster coefficients $b_{n,4-n}$ and virial coefficients $c_{n,4-n}$ of the spatially homogeneous unitary gas of {\yvan two-spin-state} $\uparrow$ and $\downarrow$ fermions, tabulated as functions of the mass ratio $\alpha=m_\uparrow/m_\downarrow$, for choice (\ref{eq012}) of the reference mass $\bar{m}$. As we go from $b_{n,4-n}$ to $b_{4-n,n}$ and from $c_{n,4-n}$ to $c_{4-n,n}$ by changing $\alpha$ to $1/\alpha$, we restrict to $\alpha\geq 1$. We do not give the coefficients associated with the integers $n=0$ and $n=4$ because they are identical to those of the ideal gas, but we give the mean coefficients $b_4$ and $c_4$ useful in the case of equal fugacities or phase-space densities in the two {\yvan spin} states, see equations (\ref{eq007}) and (\ref{eq008}). The uncertainties on $b_{3,1}$, $b_{1,3}$ and $b_{2,2}$, not specified, are less than one percent. The given values are shown graphically in figure \ref{fig1}.}
\label{tab:bc}
\end{table}

\section{Cluster coefficient for $3+1$ fermions in a trap}
\label{sec2}

We explain in this section how to {\rouge perform an efficient numerical calculation of} the quantity $I_{3,1}(\omb)$ defined by equation (\ref{eq003}), for any value (zero or positive) of the reduced trapping frequency $\omb=\hbar\omega/k_B T$. The desired cluster coefficient $B_{3,1}(\omb)$ follows directly from conjecture (\ref{eq004}).

\paragraph{\rouge Formulation of the problem} Let us first recall the expression of the operator $M^{(\ell,\veps)}_{3,1}(\ii S)$ involved in equation (\ref{eq003}), as {\yvan was obtained} in reference \cite{PRL}. We are dealing with the sum of a diagonal part $\mD_{1,3}$ and a kernel operator $\mK_{3,1}$ acting on functions $f_{m_z}(x,u)$ of two continuous variables, the logarithm $x\in\mathbb{R}^+$ of {\rouge the norm ratio of two wavevectors and the cosine $u=\cos\theta\in[-1, 1]$ of the angle between them}, and a discrete variable, the magnetic quantum number $m_z$ along the quantization axis $Oz$, varying in steps of two between $-\ell$ and $\ell$ for {\yvan parity} $\veps=(-1)^\ell$, and between $-\ell+1$ and $\ell-1$ for {\yvan parity} $\veps=(-1)^{\ell-1}$ and $\ell\neq 0$. It reads in a mixed Dirac and Schr\"odinger notation: 
\be
\label{eq100}
\boxed{
\langle x,u|\langle \ell,m_z| M_{3,1}^{(\ell,\veps)}(\ii S)|f\rangle = \mD_{3,1}(x,u) f_{m_z}(x,u) + \int_{0}^{+\infty}\dd x' \int_{-1}^{1}\dd u' \sum_{m_z' \ |\ (-1)^{m_z'}=\veps}  \mK^{(\ell)}_{3,1} (x,u,m_z;x',u',m_z') f_{m_z'}(x',u') 
}
\ee
The diagonal part is independent of angular momentum and scaling exponent $\ii S$. {\yvan As we took advantage of the fermionic exchange symmetry of the state vector to restrict to $x>0$,} the kernel $\mK_{3,1}$ is obtained by symmetrization of a primitive kernel $K_{3,1}$:\footnote{\label{note4} In Dirac notation, {\yvan $\mK_{3,1}^{(\ell)}=(1+U)K_{3,1}^{(\ell)}(1+U)$} with the Hermitian involution $U=-P_x \eee^{\ii\pi L_x/\hbar}$, the parity operator $P_x$ changing $|x\rangle$ into $|-x\rangle$ and the rotation operator of axis $Ox$ of angle $\pi$ such that 
$\exp(\ii\pi L_x/\hbar)|\ell,m_z\rangle=(-1)^\ell |\ell,-m_z\rangle$
\cite{PRA}. } 
\be
\label{eq101}
\mathcal{D}_{3,1}(x,u)=\frac{(1+2\alpha+\alpha u/\ch x)^{1/2}}{1+\alpha} \quad ; \quad \mK^{(\ell)}_{3,1}(x,u,m_z;x',u',m_z')=\sum_{\eta,\eta'=\pm {\rouge 1}} (\eta\eta')^{\ell+1} K^{(\ell)}_{3,1}(\eta x,u,\eta m_z; \eta' x',u',\eta' m_z')
\ee
The matrix elements of the primitive kernel depend on $\ii S$ and contain an integral over a rotation angle $\phi$ around axis $Ox$ {\yvan as follows}, 
\be
\label{eq103}
K_{3,1}^{(\ell)} (x,u,m_z;x',u',m_z')=\frac{(\lambda\lambda')^{3/2}}{[(1+\lambda^2)(1+\lambda'^2)]^{1/4}} \int_0^{2\pi}\frac{\dd\phi}{2\pi^2} \frac{(1+\lambda^2)^{\ii S/2}\eee^{-\ii m_z\theta/2}\langle l,m_z|\eee^{\ii\phi L_x/\hbar} |l,m_z'\rangle\eee^{\ii m_z'\theta'/2}(1+\lambda'^2)^{-\ii S/2}}{1+\lambda^2+\lambda'^2+\frac{2\alpha}{1+\alpha} [\lambda u +\lambda' u' +\lambda\lambda' (uu'+vv'\cos\phi)]}
\ee
{\yvan where $L_x$ is the angular momentum operator along $Ox$. Here,} as in reference \cite{PRL}, we have put for abbreviation: 
\be
\label{eq104}
\lambda=\eee^x,\quad \lambda'=\eee^{x'},\quad \theta=\acos u\in [0,\pi],\quad \theta'=\acos u'\in [0,\pi],\quad v=\sin\theta,\quad v'=\sin\theta'
\ee

To evaluate $I_{3,1}(\omb)$ {\bleu numerically}, one must first replace the operator by a {\yvan finite size} matrix, by truncating the variable $x$ to $x_{\rm max}$ and discretizing it according to the midpoint integration method, then by discretizing the variable $\theta$ (which we prefer to the variable $u$ because it leads to a smooth integrand) according to the Gauss-Legendre integration method. Then we compute the determinant of the matrix by putting it in the Cholesky form, to take advantage of the fact that the operator $M_{3,1}^{(\ell,\veps)}(\ii S)$ is positive in the absence of $3+1$-body Efimov effect. Finally, we compute the integral over $S$ in the interval $[0,S_{\rm max}]$ by the midpoint method (we need to know the logarithm of the determinant at integer multiples of the integration step $\dd S$ to obtain its derivative at half-integer multiples). {\yvan We take} into account the contribution of the omitted interval $[S_{\rm max}, +\infty[$ by means of an exponential approximation $A\exp(-B S)$ of the logarithmic derivative of the determinant {\rouge justified by reference \cite{PRA}}, {\yvan where we have estimated the coefficients $A$ and $B$ by fitting} on {\bleu a neighborhood of $S_{\rm max}$, in practice the interval} $[S_{\rm max}-5/2,S_{\rm max}]$.

\begin{figure}[t]
\includegraphics[width=7.5cm,clip=]{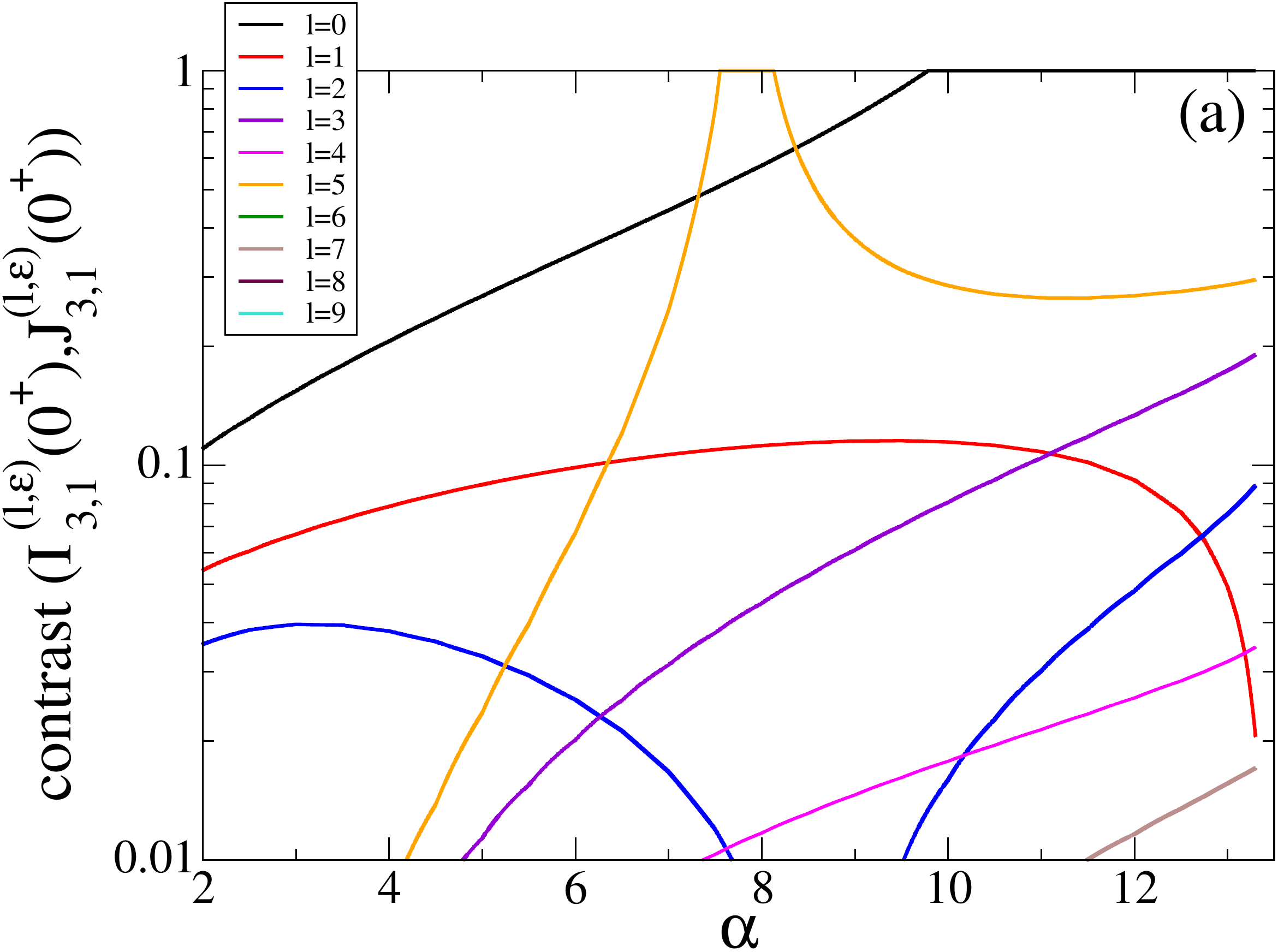}\hspace{5mm}\includegraphics[width=7.5cm,clip=]{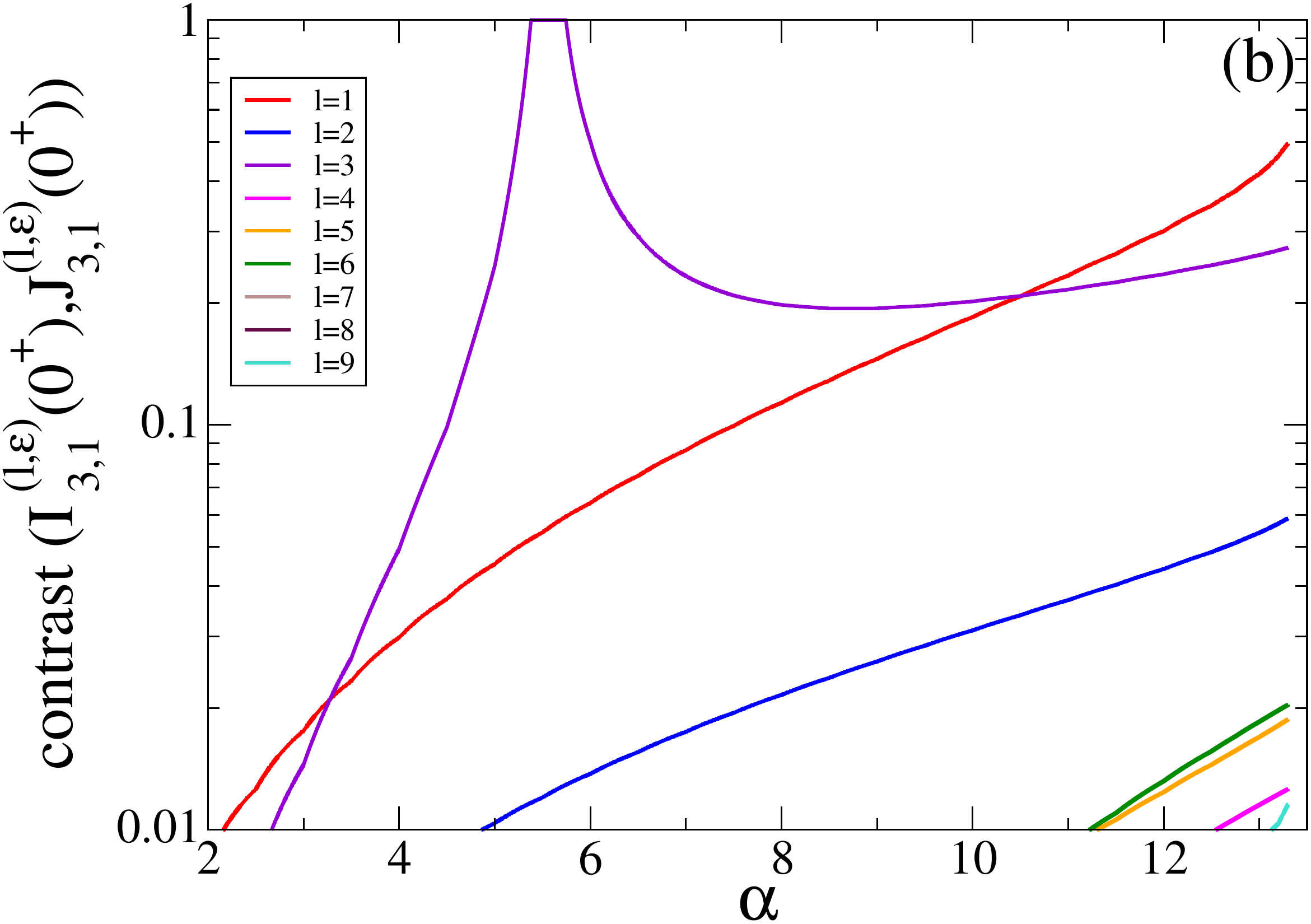}
\caption{Contrast between the contribution $I_{3,1}^{(\ell,\veps)}$ of the angular momentum $\ell$ and parity $\veps$ channel to $I_{3,1}$ and its asymptotic approximant $J_{3, 1}^{(\ell,\veps)}$ given by equation (\ref{eq110}), as a function of the mass ratio $\alpha=m_\uparrow/m_\downarrow$ and $\ell$, in the limit $\omb=0^+$. {\bleu The contrast between two real quantities $a$ and $b$ is here the ratio $|a-b|/(|a|+|b|)$.} (a) Case $\veps=(-1)^\ell$. (b) Case $\veps=(-1)^{(\ell+1)}$ (which imposes $\ell>0$). The numerical truncation and discretization parameters are those of figure \ref{fig3}. {\bleu The contrast} saturates to one (as it happens for $\ell=0$ and $\ell=5$ in (a), for $\ell=3$ in (b)) when {\bleu the two compared quantities} are of opposite signs.}
\label{fig4}
\end{figure}

\paragraph{\rouge Asymptotic approximant and applications} It remains to take into account the truncation on the angular momentum $\ell$ at some $\ell_{\rm max}$, {\yvan which is} unavoidable in a numerical calculation. In practice, it is not reasonable to go beyond $\ell_{\rm max}=15$, {\yvan because high values of $\ell$ lead} to a complexity $O(\ell^3)$ and {\yvan their numerical calculation would be}  very expensive in time. Unfortunately, this cut-off is not {\yvan yet enough} for large mass ratios $\alpha\gg 1$ if one aims at an accuracy on $B_{3,1}(0^+)$ better than one percent. The {\yvan solution to this difficulty} is to determine an asymptotic approximant $J_{3,1}^{(\ell,\veps)}(\omb)$ of the contribution $I_{3,1}^{(\ell,\veps)}(\omb)$ of angular momentum $\ell$ and parity $\veps$ to the desired quantity $I_{3,1}(\omb)$. Then, instead of neglecting completely the terms $I_{3,1}^{(\ell,\veps)}(\omb)$ for $\ell>\ell_{\rm max}$, we replace them by $J_{3,1}^{(\ell,\veps)}(\omb)$ as follows, 
\be
\label{eq105}
I_{3,1}(\omb) \simeq \sum_{\ell=0}^{\ell_{\rm max}}\sum_\veps I_{3,1}^{(\ell,\veps)}(\omb) + \sum_{\ell=\ell_{\rm max}+1}^{+\infty}\sum_\veps J_{3,1}^{(\ell,\veps)}(\omb) = \sum_{\ell=0}^{\ell_{\rm max}}\sum_\veps \left[I_{3,1}^{(\ell,\veps)}(\omb)-J_{3,1}^{(\ell,\veps)}(\omb)\right] + \sum_{\ell=0}^{+\infty}\sum_\veps J_{3,1}^{(\ell,\veps)}(\omb)
\ee
{\yvan By using this method, we accelerate} the convergence of the series: the error tends to zero more quickly with $\ell_{\rm max}$. An exact asymptotic approximant to subleading order in $\ell$ is obtained by generalizing to $\omb\neq 0$ the method of reference \cite{PRA}, which takes the kernel $\mK_{3,1}^{(\ell,\veps)}$ as a small formal parameter and expands the logarithm of the determinant of $M_{3,1}^{(\ell,\veps)}$ to second order: 
\be
\label{eq106}
\ln\det M_{3,1}^{(\ell,\veps)}= \ln\mathrm{det}^\veps[\mD_{3,1}+\mK_{3,1}^{(\ell)}]=\ln\mathrm{det}^\veps[\mathcal{D}_{3,1}]+\ln\mathrm{det}^\veps[\,\identit+\mD_{3,1}^{-1}\mK_{3,1}^{(\ell)}]=\mbox{const}+\mathrm{Tr}^{\veps}\left[\mD_{3,1}^{-1}\mK_{3,1}^{(\ell)}-\frac{1}{2}\mD_{3,1}^{-1}\mK_{3,1}^{(\ell)}\mD_{3,1}^{-1}\mK_{3,1}^{(\ell)}+\ldots\right]
\ee
where the symbol $\veps$ in superscript of the trace {\yvan and} determinant means that we restrict ourselves to the subspace of values of $m_z$ compatible with parity $\veps$. {\bleu The computation is done} in \ref{apA}, and the corresponding expression of $J_{3,1}^{(\ell,\veps)}(\omb)$ is given in equation (\ref{eq110}) {\yvan in the form of} multiple integrals. For this already very elaborate choice, the first values of the approximant ($0\leq\ell\leq\ell_{\rm max}$) are still easy to compute numerically with the same truncation and discretization as for the full determinant, but it would be tedious to go {\bleu to larger values of $\ell$, which is however required by the second sum in the second side of equation (\ref{eq105})}; fortunately, the infinite series in the third side of equation (\ref{eq105}) has a simple integral expression, see equation (\ref{eq125}), which can even be expressed analytically for $\omb=0^+$ in terms of known functions {\bleu such as} the dilogarithm function, see equation (\ref{eq126}). {\yvan The efficiency} of our asymptotic approximant {\yvan can be seen} in figure \ref{fig4}, where it is compared to the numerical result. The figure shows, as a function of the mass ratio $\alpha$, in which angular momentum channels $\ell$ the approximant deviates by more than one percent from the exact value (this is the accuracy on the cluster coefficient {\yvan we aim for in this work}); these channels must be included in the numerical sum from $0$ to $\ell_{\rm max}$.

An interesting {\bleu by-product} of our convergence acceleration method is to obtain an asymptotic equivalent of the angular momentum $\ell$ and parity $\veps$ contribution to the cluster coefficient $B_{3,1}(0^+)$; it suffices to keep the contribution linear in $\mK_{3,1}^{(\ell)}$ in equation (\ref{eq106}) and to determine its dominant behavior at large $\ell$ with Cauchy's integral theorem, as done in \ref{apA}. We find: 
\be
\label{eq333}
\boxed{
I_{3,1}^{(\ell,\veps)}(0^+) \underset{\ell\to+\infty}{\sim} \frac{1}{2} \left\{\frac{(1+\alpha)^2}{\sqrt{\alpha(1+3\alpha)}}\im \left[\left(\frac{\ell}{6\pi C_0}\right)^{1/2} \frac{z_0}{1+z_0}\, z_0^\ell\right]+\veps \frac{(1+\alpha)^2}{\alpha^{3/2}} \re \left[\left(\frac{\ell}{2\pi C_1}\right)^{1/2} \frac{(1-z_1)}{\sqrt{1-\cos\xi_1}}\, z_1^\ell\right]\right\} 
}
\ee
where
\be
\label{eq334}
C_0=\frac{(1-\cos\xi_0)(5/4+\cos\xi_0)}{(1/2+\cos\xi_0)(z_0-1/z_0)}\quad\mbox{and}\quad C_1=4(z_1-1/z_1)(\cos\xi_1-1/2) \left[(1+\alpha^{-1})\cos\xi_1 +\frac{1+3\alpha^{-1}}{4}\right]
\ee
with the shorthand notation $\cos\xi_n\equiv (z_n+1/z_n)/2$, $n=0$ or $1$, the complex number $z_0$ given by equation (\ref{eq133}) and the complex number $z_1=-z_0^*$. This is the generalization to $3+1$ fermions of a result obtained for three bosons in reference {\yvan\cite{b3}}, see equation (42) of this reference. Thus, to within a power law factor, $I_{3,1}^{(\ell,\veps)}(0^+)$ tends exponentially to zero with $\ell$, with irregular oscillations due to the fact that the argument of $z_0$ is in general not commensurable to $\pi$.

\begin{figure}[t]
\includegraphics[width=7.5cm,clip=]{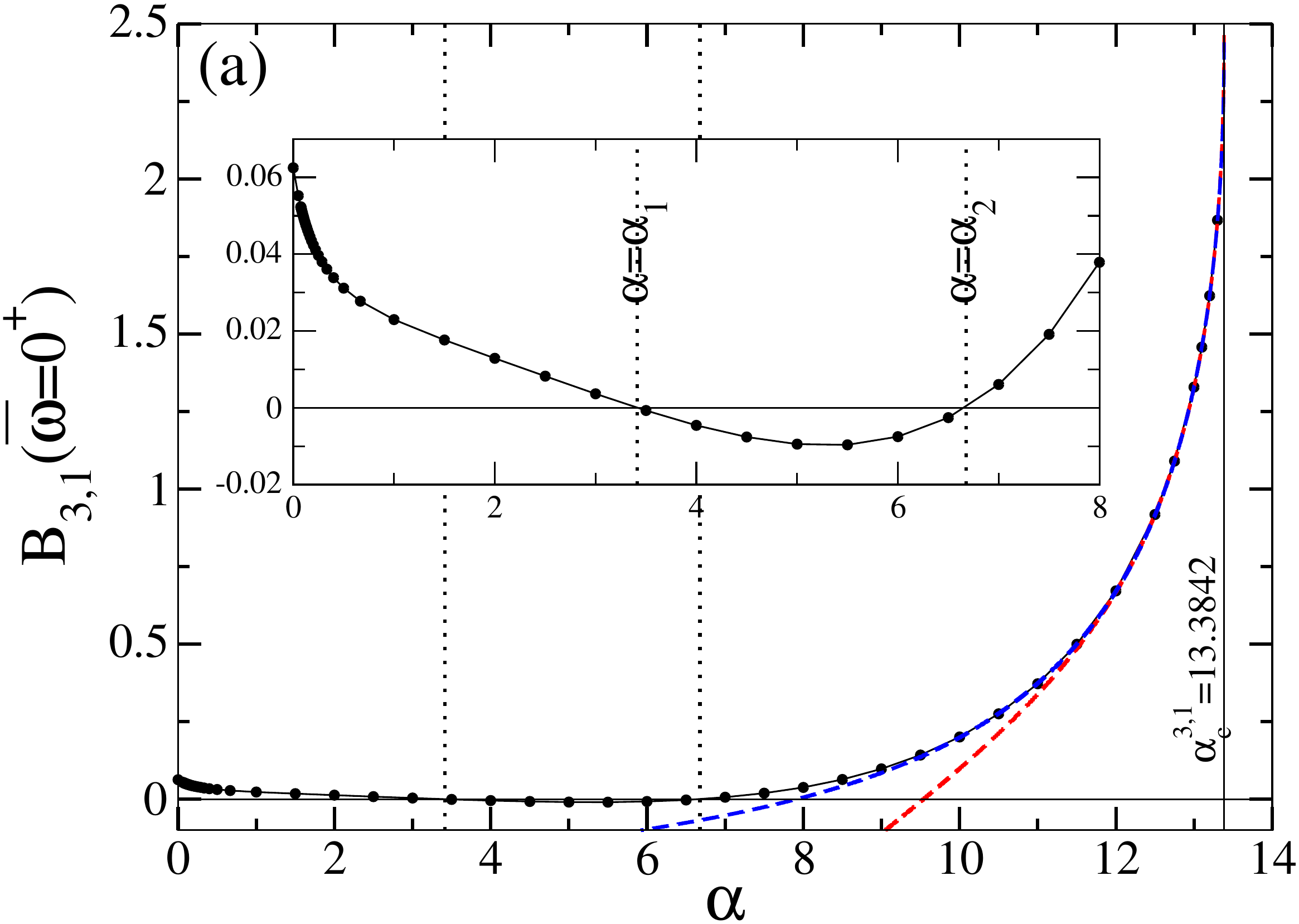}\hspace{5mm}\includegraphics[width=7.5cm,clip=]{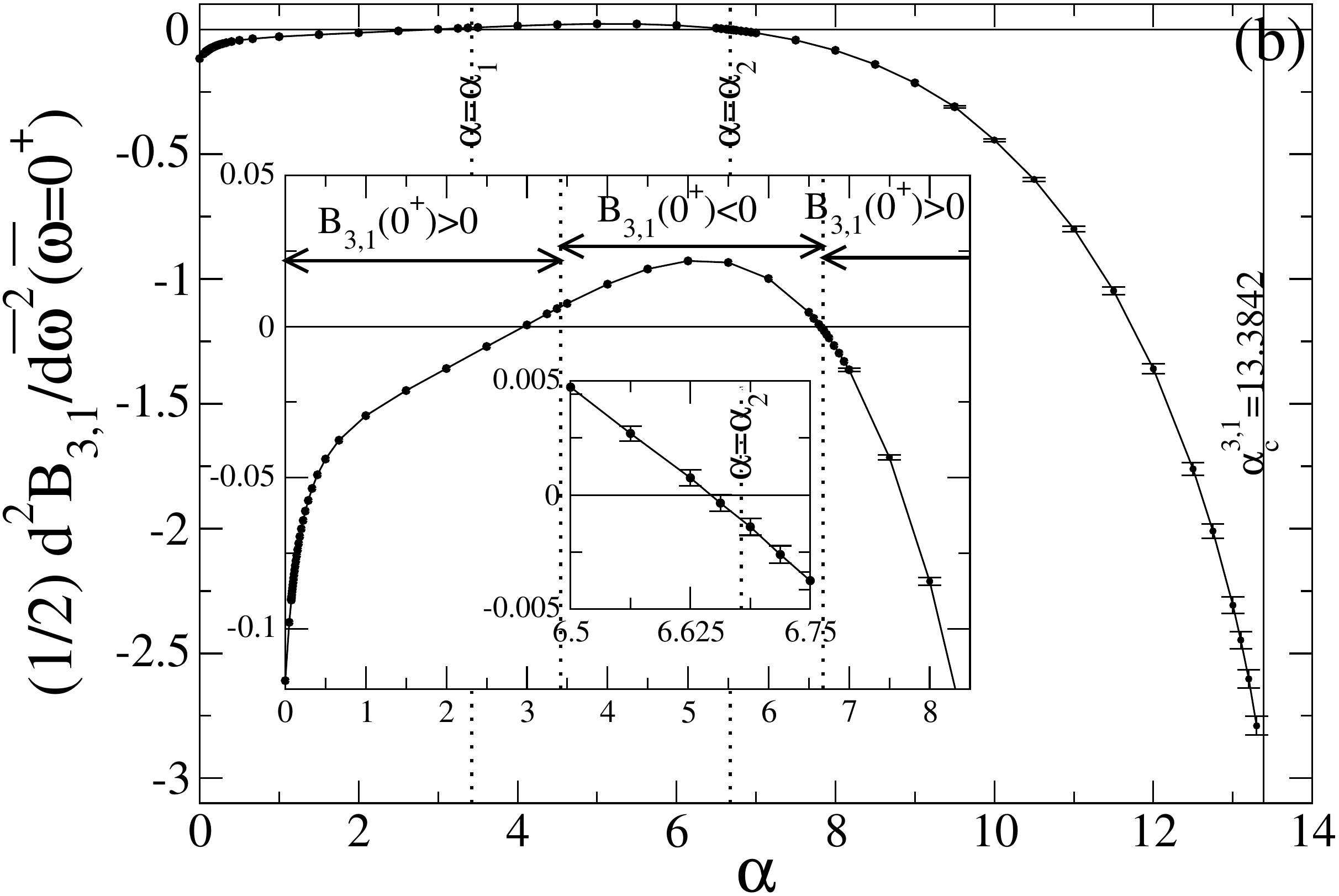}
\caption{(a) Fourth cluster coefficient $B_{3,1}(0^+)$ of the unitary Fermi gas in an infinitesimally stiff harmonic trap, as a function of the mass ratio $\alpha=m_{\uparrow}/m_{\downarrow}$. Disks (connected by a thin line): Our numerical calculation of expression (\ref{eq003}) for $(n_\uparrow,n_\downarrow)=(3,1)$ and application of conjecture (\ref{eq004}), with truncation parameters $x_{\rm max}=5$, $S_{\rm max}=25$, $\ell_{\rm max}=15$ and discretization $\dd x=1/10$, $\dd S=1/8$, $n_\theta=37$ (this is the number of values of $\theta$ in the Gauss-Legendre integration method); for $\alpha>10$ and $0\leq\ell\leq 5$, we take instead $x_{\rm max}=20$ and $n_\theta=25$ to keep the error below one percent. We use the convergence acceleration method (\ref{eq105}) and extrapolate to $S_{\rm max}=+\infty$ as explained in the text. Blue dashed line: cubic fit {\yvan in} the variable $X=(\alpha_c^{\rm 4\, body}-\alpha)^{1/2}$ on the interval $10\leq\alpha\leq 13.3$, i.e. \protect
$B_{3,1}(0^+)=2.4401-2.1627X+0.64137X^2-0.069596X^3$,
where $\alpha_c^{\rm 4\, body}\simeq 13.3842$ is the $3+1$-body Efimovian threshold (vertical solid line). Red dashed line: same on the interval $12.5\leq\alpha\leq 13.3$, i.e. \protect
$B_{3,1}(0^+)=2.4637-2.2834X+0.82318X^2-0.15269X^3$.
The inset is an enlargement showing better $B_{3,1}(0^+)$ vanishing with sign change at $\alpha_1\simeq 3.412$ and $\alpha_2\simeq 6.678$ (vertical black dotted lines). (b) Same for the half-second derivative $(1/2)B_{3,1}''(0^+)$ of the cluster coefficient with respect to the reduced trapping frequency $\omb=\hbar\omega/k_B T$, with the differences that (i) we do not give a fit in the variable $X$, (ii) we put error bars because the relative uncertainty may now exceed one percent, (iii) we include a magnification in the enlargement to make it clear that the second zero (with sign change) of $(1/2)B_{3,1}''(0^+)$ is very close to but distinct from that of $B_{3,1}(0^+)$ (always indicated by a vertical dotted line).}
\label{fig3}
\end{figure}

\paragraph{\rouge Born-Oppenheimer regime} In the limit $\alpha\to 0$, we notice that $z_0$ tends to zero in equation (\ref{eq133}), $z_0\sim \ii\sqrt{\alpha/3}$, so we expect the exponential suppression of the summand in (\ref{eq003}) to become very fast for $3+1$ fermions. This expectation is confirmed numerically and, for $\alpha=0$, only the $\ell=0$ channel contributes. In this case, the single {\bleu spin-$\downarrow$} fermion is infinitely massive and behaves for the {\bleu spin-$\uparrow$} fermions as a fixed pointlike scatterer, of infinite $s$-wave scattering length and placed at the center of the trap. The Born-Oppenheimer approximation becomes exact and gives the {\rouge time-independent} Schr\"odinger equation on the wave function $\Psi(\rr_\downarrow)$ of the heavy particle:\footnote{The particles are at fixed temperature $T$ so, in the $m_\downarrow\to+\infty$ limit, the {\bleu spin-$\downarrow$} fermion occupies a region around the center of the trap of radius $R=O((k_B T/m_\downarrow\omega^2)^{1/2})$. On the other hand, the Born-Oppenheimer potential $W(\rr_\downarrow)$ has an energy scale $k_B T$ and varies with a length scale $\lambda_\uparrow$, the de Broglie thermal wavelength of {\yvan the} light fermions, if $k_B T>\hbar\omega$, and {\yvan it} is of the order of $\hbar\omega$ and varies with a length scale $(\hbar/m_\uparrow\omega)^{1/2}$, the size of the ground vibrational state of a light fermion, otherwise. The variation of $W(\rr_\downarrow)$ becomes negligible in all cases and we can replace it by $W(\mathbf{0})$. For the same reason, the so-called scalar or topological potential, which is added to the potential $W$ in the full adiabatic approximation \cite{Dum}, can be omitted.} 
\be
\label{eq140}
E_{\rm BO} \Psi(\rr_\downarrow) = \left[-\frac{\hbar^2}{2 m_\downarrow} \Delta_{\rr_\downarrow} +\frac{1}{2} m_\downarrow\omega^2 r_\downarrow^2 + W(\mathbf{0)}\right]\Psi(\rr_\downarrow) \quad\quad (m_\downarrow\to+\infty)
\ee
In the $n_\uparrow+1$ fermion trapped problem, $W(\zero)$ is the energy of an eigenstate of $n_\uparrow$ non-interacting fermions in the presence of the scattering center. A {\bleu spin-$\uparrow$} fermion of orbital quantum numbers $(n,\ell,m_z)$ sees the scattering center only if $\ell=0$, in which case its spectrum is lowered by $\hbar\omega$, thus having energy levels 
\be
\label{eq141}
\veps_{n,\ell,m_z} =\left\{
\begin{array}{lcl}
(2n+\ell+3/2)\hbar\omega &\mbox{if}& \ell>0 \\
(2n+1/2)\hbar\omega & \mbox{if}& \ell=0
\end{array}
\right.\quad\quad {\rouge (n\in\mathbb{N}, -\ell\leq m_z \leq\ell)}
\ee
Since the Born-Oppenheimer energy $E_{\rm BO}$ is the sum of $W(\zero)$ and a vibrational energy level of a {\bleu spin-$\downarrow$} particle in the {\rouge trap}, we conclude that 
\be
\label{eq142}
{\rouge \lim_{\alpha\to 0^+} Z_{n_\uparrow,1} = Z_1} Z_{n_\uparrow}^{\rm scat}
\ee
where $Z_{n_\uparrow}^{\rm scat}$ is the canonical partition function of a ideal gas of $n_\uparrow$ fermions with spectrum (\ref{eq141}), {\rouge i.e.\ }in the presence of the fixed scatterer, {\rouge $Z_1$ is as in equation (\ref{eq001}) the partition function of a single fermion} and {\rouge $Z_{n_\uparrow,1}$} is that of {\rouge the unitary gas} of $n_\uparrow+1$ trapped fermions. It is convenient to calculate its deviation {\rouge $\Delta Z_{n_\uparrow,1}$} from that of {\rouge the} ideal gas of $n_\uparrow+1$ trapped fermions, since (\ref{eq141}) differs from the ordinary spectrum only in the zero angular momentum channel. Taking into account Fermi statistics via the Pauli exclusion principle, and equations (80) to (83) of reference \cite{JPA} which relate the cluster coefficients of the trapped system to the few-body partition functions, we find that 
\bea
\label{eq143a}
B_{1,1}(\omb)&\!\!\!\!=\!\!\!\!& {\rouge Z_1^{-1}\Delta Z_{1,1}}\underset{\alpha\to 0^+}{\to} \frac{1}{2\ch(\omb/2)} \quad ; \quad B_{2,1}(\omb)={\rouge Z_1^{-1}\Delta Z_{2,1}}-Z_1 B_{1,1}\underset{\alpha\to 0^+}{\to} -\frac{1}{8\ch^2(\omb/2)\ch\omb} \\
\label{eq143b}
B_{3,1}(\omb)&\!\!\!\!=\!\!\!\!& {\rouge Z_1^{-1}\Delta Z_{3,1}}-{\rouge Z_{2,0}} B_{1,1}-Z_1 B_{2,1} \underset{\alpha\to 0^+}{\to} \frac{1}{16\ch^3(\bar{\omega}/2)(\ch\bar{\omega})(2\ch\bar{\omega}-1)}
\eea
The value (\ref{eq143b}) of the limit is in perfect agreement with our numerical calculation of expression (\ref{eq003}) {\bleu not only for $\omb=0^+$, where $B_{3,1}(0^+)\to 1/16=0.0625$, as seen in figure \ref{fig3}a, but, as we have verified, for all $\omb$}, which is an additional test of conjecture (\ref{eq004}) in the $3+1$-body case. \footnote{In the $2+1$-body case, starting from the analytic integral expression of $B_{2,1}(\omb)$ of reference \cite{EPL} in terms of the transcendental Efimov function $\Lambda_{2,1}^{(\ell)}$ on the pure imaginary axis, and by showing in the limit $\alpha\to 0^+$ that $\Lambda_{2,1}^{(\ell)}(\ii S)$ tends to $1$ for $\ell>0$ and tends to $1+1/\ch(\pi S/2)$ for $\ell=0$ (for example by means of equation (46) of reference \cite{f4}), we find the second result in equation (\ref{eq143a}). } {\yvan As a side remark,} let us point out that it is possible to go to the next orders by using the grand-canonical version of equation (\ref{eq142}), 
\be
\label{eq160}
\Xi \underset{z_\downarrow\to 0\ \mbox{\scriptsize then}\ \alpha\to 0^+}{=} \Xi_\uparrow + z_\downarrow Z_1 \Xi_\uparrow^{\rm scat}+O(z_\downarrow^2)
\ee
where $\Xi_\uparrow$ ($\Xi_\uparrow^{\rm scat}$) is the grand partition function of the ideal gas of {\bleu spin-$\uparrow$} fermions {\rouge of fugacity $z_\uparrow$} in the absence (presence) of the scattering center and $\Xi$ is that of the two-component unitary gas. Expanding the grand potential $\Omega=-k_B T \ln \Xi$ as in equation (\ref{eq001}), we obtain \footnote{A sum over an integer $s$ in the fourth side was introduced by taking the logarithm of the infinite product in the third side, expanding around $1$ the function $\ln$ into an integer series of index $s$ and then summing over $n$ the resulting geometric series.} 
\be
\label{eq161}
\sum_{n_\uparrow=0}^{+\infty} z_\uparrow^{n_\uparrow} \lim_{\alpha\to 0^+} B_{n_\uparrow,1}(\omb) = \frac{\Xi_\uparrow^{\rm scat}}{\Xi_\uparrow} = \prod_{n=0}^{+\infty} \frac{1+z_\uparrow\eee^{-(2n+1/2)\omb}}{1+z_\uparrow\eee^{-(2n+3/2)\omb}} =\exp\left[\sum_{s=1}^{+\infty} \frac{(-1)^{s+1} z_\uparrow^s}{2s\ch(s\omb/2)}\right]
\ee
In the special case $\omb=0^+$, the exponential in the fourth side of (\ref{eq161}) reduces to $\sqrt{1+z_\uparrow}$, easy to expand into powers of $z_\uparrow$, hence the result at all orders in terms of Euler's $\Gamma$ function: 
\be
\label{eq162}
\boxed{
\lim_{\alpha\to 0^+} B_{n_\uparrow,1}(0^+) = \frac{(-1)^{n_\uparrow}\Gamma(n_\uparrow-1/2)}{n_\uparrow!\,\Gamma(-1/2)}  \quad \forall n_\uparrow\in\mathbb{N}
}
\ee

\paragraph{\rouge The results} Our numerical results for $I_{3,1}(0^+)$, hence for the cluster coefficient $B_{3,1}(0^+)$ from equation (\ref{eq004}), {\rouge are given explicitly in table \ref{tab:3p1} and} are plotted as {\rouge functions} of the mass ratio in figure \ref{fig3}a, with no error bars but with an uncertainty of less than one percent. We find that $B_{3,1}(0^+)$ is positive, except over an interval $[\alpha_1,\alpha_2]$, as best seen {\yvan in} the enlargement included in the figure. At large values of $\alpha$, $B_{3,1}(0^+)$ shows a marked growth that we attribute to the three-body Efimov effect; {\yvan it} has a finite limit at the threshold of the $3+1$-body Efimov effect where our computation stops, reached with an infinite slope and that we determine by extrapolation, see the dashed curves in figure \ref{fig3}a and the explanations which follow, 
\be
\label{eq144}
\boxed{
B_{3,1}(0^+) \underset{\alpha\to\left(\alpha_c^{\rm 4\,body}\right)^-}{{\rouge\longrightarrow}} 2.47 \pm 0.03
}
\ee

\begin{table}[t]
{\footnotesize
\begin{tabular}{|c|c|c|c|c|c|c|c|c|c|c|}
\hline
$\alpha$ & 1 & 1.5 & 2 & 2.5 & 3 & 3.5 & 4 & 4.5 & 5 & 5.5\\
\hline
$B_{3,1}$ & 0.02297 & 0.01764 & 0.01289 & 0.008233 & 0.003651 & $-$0.000697 & $-$0.004571 & $-$0.007623 & $-$0.00946 & $-$0.00964 \\
$B_{1,3}$ & 0.02297 & 0.02775 & 0.03116 & 0.03387 & 0.03611 & 0.03802 & 0.03967 & 0.04110 & 0.04236 & 0.04349 \\
\hline
$\alpha$ & 6 & 6.5 & 7 & 7.5 & 8 & 8.5 & 9 & 9.5 & 10 & 10.5 \\
\hline
$B_{3,1}$ & $-$0.00751 & $-$0.00257 & 0.006065 & 0.01912 & 0.0379 & 0.06354 & 0.09769 & 0.1423 & 0.2003 & 0.275 \\
$B_{1,3}$ & 0.04449 & 0.04539 & 0.04621 & 0.04695 & 0.04763 & 0.04825 & 0.04882 & 0.04935 & 0.04983 & 0.05029 \\
\hline
$\alpha$ & 11 & 11.5 & 12 & 12.5 & 12.75 & 13 & 13.1 & 13.2 & 13.3 & 20 \\
\hline
$B_{3,1}$ & 0.3721 & 0.4996 & 0.6714 & 0.9176 & 1.090 & 1.329 & 1.457 & 1.623 & 1.867 & \\
$B_{1,3}$ & 0.05071 & 0.05110 & 0.05147 & 0.05182 & 0.05198 & 0.05214 & 0.05220 & 0.05227 & 0.05233 & 0.05523 \\
\hline
\end{tabular}
}
\caption{\rouge Numerical values of the fourth cluster coefficients $B_{3,1}(0^+)$ and $B_{1,3}(0^+)$ of the unitary Fermi gas in an infinitesimal-stiffness trap, tabulated as functions of the mass ratio $\alpha=m_\uparrow/m_\downarrow$. Since we go from one coefficient to the other by changing $\alpha$ to $1/\alpha$, we limited ourselves to $\alpha\geq 1$. The uncertainties, not specified, are less than one percent. {\bleu These values are shown graphically in figure \ref{fig3}a.}}
\label{tab:3p1}
\end{table}

\paragraph{\rouge Near the $3+1$-body Efimovian threshold} The limit $\alpha\to (\alpha_c^{\rm 4\, body})^{-}$ in equation (\ref{eq144}) is difficult to achieve numerically with precision for several reasons. First, if $\alpha\to+\infty$, the exponential decay rate $\ln(1/|z_0|)$ in equation (\ref{eq333}) tends to zero as $\alpha^{-1/2}$ so the convergence of the series $I_{3,1}(0^+)$ becomes slower and slower with $\ell$; fortunately, as the mass ratio $\alpha$ remains below the four-body Efimovian threshold, $\ln(1/|z_0|)$ does not become so small (it remains above 0.267) and the problem is solved by the convergence acceleration (\ref{eq105}) if one is satisfied with an error on $B_{3,1}(0^+)$ less than one percent (see figure \ref{fig4}). Secondly, the $3+1$-body Efimov effect about to appear in the $\ell=1,\veps=+1$ channel forces to increase $x_{\rm max}$ to the values considered in reference \cite{PRL}, at least in the $\ell\leq 5$ angular momentum channels; {\yvan this} increases the computation time a lot and leads to a singularity of type $(\alpha_c^{\rm 4\, body}-\alpha)^{1/2}$ which has to be taken into account in the extrapolation {\yvan of} $B_{3,1}(0^+)$ to $\alpha=\alpha_c^{\rm 4\, body}$, see dashed lines in figure \ref{fig3}a and the legend of the figure.\footnote{This singularity is present in one of the scaling exponents $s_i$ of the $3+1$-body unitary problem, the one $s_0$ whose square vanishes by changing sign at the Efimovian threshold, in the $\ell_0=1$ channel \cite{PRL}. Now $B_{3,1}(0^+)$ depends linearly on the scaling exponents. We see it well on the universal component of the third cluster coefficient of the unitary Bose gas in equation (36) of reference {\yvan\cite{b3}} by making $\omb$ (denoted $x$ in this reference) tend to zero; we also see it for $2+1$ fermions in reference \cite{EPL}. More precisely, $B_{3,1}(0^+)=-(\ell_0+1/2) s_0+\ldots$ where the ellipse is a smooth function of $\alpha$ in the neighborhood of $\alpha_c^{3,1}$. As 
$s_0^2\simeq c_0(\alpha_c^{3,1}-\alpha)$
with $c_0\simeq 2.2$ near the threshold \cite{PRL}, we find that 
$\dd B_{3,1}(0^+)/\dd X=-(\ell_0+1/2)c_0^{1/2}\simeq -2.2$ at $X\equiv(\alpha_c^{3,1}-\alpha)^{1/2}=0$.
This constraint is fairly well satisfied by the fits in figure \ref{fig3}a.} Third, the threshold for the three-body Efimov effect $\alpha_c^{2,1}\simeq 13.60697$ is close by and affects all angular momentum channels $\ell$ of the four-body problem \cite{PRA}, even though it occurs in the $L=1$ angular momentum channel of the $2+1$ fermion problem. Indeed, the continuum spectrum of the $M_{3,1}^{(\ell,\veps)}(\ii S)$ operator is the union of continua corresponding to $2+1$ fermions strongly correlated by the interactions and a decoupled {\yvan spin-}$\uparrow$ fermion, thus parametrized by a three-body angular momentum $L$ \cite{PRL}. The contribution of the continuum of angular momentum $L$ to the $3+1$ cluster coefficient in the ($\ell,\veps)$ channel is written up to a factor \cite{PRA}: 
\be
\label{eq149}
I_{3,1}^{(\ell,\veps)}(0^+)|_{C^0,L} \propto \int_\mathbb{R} \dd S\int_{\mathbb{R}^+}\dd k\, S \partial_S\theta_L(k,S) \frac{\dd}{\dd k} \ln \Lambda_{2,1}^{(L)}(\ii k)
\ee
The eigenmodes of the continuum have asymptotically a plane wave structure in the space of the variable $x$, i.e. they are superposition, when $x\to+\infty$, of an incident wave $\exp(-\ii kx)$ and a reflected wave $-\exp[\ii{\rouge \theta_L(k,S)}]\exp(\ii kx)$. {\yvan Here} the phase shift {\rouge $\theta_L(k,S)$} is a function of the wavenumber $k>0$ (fictitious because $x$ is not a real position), of the pure imaginary scaling exponent $\ii S$ and of the three-body angular momentum $L$. We see in expression (\ref{eq149}) the transcendental Efimov function $\Lambda_{2,1}^{(L)}$ whose roots are the scaling exponents of the unitary $2+1$-body problem with angular momentum $L$. This is because the continuum modes of wavenumber $k$ correspond to the eigenvalue $\Lambda_{2,1}^{(L)}(\ii k)$ of $M_{3,1}^{(\ell,\veps)}(\ii S)$ \cite{PRL}. \footnote{\label{note11} In the absence of a cutoff in $x$ space ($x_{\rm max}=+\infty$), we fall into a paradox: $k$ spans the continuous set $\mathbb{R}^+$ independently of the variable $S$ and the derivative of the eigenvalue $\Lambda_{2,1}^{(L)}(\ii k)$ with respect to $S$ is zero, so the continuum should not contribute to $I_{3,1}$ in equation (\ref{eq003}). The right way to reason is to put a cutoff $x_{\rm max}$ that we make tend to infinity at the end, with the condition that the eigenmodes vanish at $x=x_{\rm max}$. The resulting equation 
{\rouge $\theta_L(k,S)=-2 k x_{\rm max}\ [2\pi]$}
quantizes $k$, i.e. restricts it to a discrete set, and makes it $S$-dependent as the phase shift {\rouge $\theta_L(k,S)$}.} The key point now is that the lower edge of the $L=1$ continuum, namely the minimum of $\Lambda_{2,1}^{L=1}(\ii k)$ with respect to the variable $k$, tends to zero when $\alpha\to\alpha_c^{2,1}$ because of the $2+1$-body Efimov effect. This has two consequences. First, a practical consequence in the calculation of $I_{3,1}$ on a computer: as the continuum is made discrete by truncating $x$ at $x_{\rm max}$, its lower edge deviates from the true edge by about $1/x_{\rm max}^2$; \footnote{In fact, the wavenumber $k$ varies in steps of the order of $1/x_{\rm max}$, as shown in footnote \ref{note11}, and $\Lambda_{2,1}^{L=1}(\ii k)$ varies quadratically near its minimum. } this numerical artifact forces to significantly increase $x_{\rm max}$, fortunately only for four-body angular momentum channels $\ell\leq 5$ as long as $\alpha\leq 13.3$. Then, a physical consequence: a second singularity appears in $I_{3,1}(0^+)$, of the form $(\alpha_c^{2,1}-\alpha)^{1/2}$, again a square root, now centered on the three-body Efimovian threshold. \footnote{To the left of this threshold, we write \`a la Weierstrass 
$\ln\Lambda_{2,1}^{(L=1)}(\ii k)=\ln[(k^2+\sigma_0^2)/(k^2+1)]+\ldots$
where the ellipse is a regular function of $\alpha$ even at the threshold and $\sigma_0>0$ is the scaling exponent of the $2+1$-body problem whose square vanishes by changing sign at the threshold. Very close to the threshold, $\sigma_0\ll 1$ and, assuming as in reference \cite{PRA} that 
$\theta_{L=1}(k,S)\sim b(S) k$
when $k\to 0$, where the scattering radius $b(S)$ is an unknown function of $S$, we find that 
$I_{3,1}^{(\ell,\veps)\neq(1,+)}(0^+)$
contains a singularity 
$\propto\int_0^{+\infty}\dd k[k^2/(k^2+\sigma_0^2)-k^2/(k^2+1)]\propto 1-\sigma_0$ where $\sigma_0\sim \gamma_0 (\alpha_c^{2,1}-\alpha)^{1/2}$ and $\gamma_0\simeq 0.438$.
We have confirmed this prediction by a specific numerical calculation {\rouge of $I_{3,1}(0^+)$} in the $\ell=0$ channel very close to the threshold, going up to $\alpha=13.59$; a nice linear {\yvan law} in $Y$ is observed and a fit gives $I_{3,1}^{(\ell=0)}(0^+)\simeq 0.026+0.040 Y$ where the variable $Y$ is the one in equation (\ref{eq155}).} This three-body singularity occurs at a point very close to the four-body singularity, which casts doubt on the accuracy of the extrapolation made in figure \ref{fig3}a which did not take it into account. We remedy this by fitting functions including the two singularities, polynomial in the quantities 
$X=(\alpha_c^{3,1}-\alpha)^{1/2}$ and $Y=(\alpha_c^{2,1}-\alpha)^{1/2}$:
\be
\label{eq155}
I_{3,1}(0^+)=A_0+A_1 X+A_2 (Y-Y_0) + A_3 X^2 \quad\mbox{and}\quad I_{3,1}(0^+)=A_0+A_1 X+A_2 (Y-Y_0) + A_3 X^2+A_4 X^3+A_5(Y^3-Y_0^3)
\ee
where 
$Y_0=(Y^2-X^2)^{1/2}=(\alpha_c^{2,1}-\alpha_c^{3,1})^{1/2}$
is also the value of $Y$ at $X=0$. On the interval $10\leq\alpha\leq 13.3$, this leads to $A_0=2.495$ and $A_0=2.447$ hence the final result (\ref{eq144}), which supports the more naive one in figure \ref{fig3}a.

\paragraph{\rouge Case $\omb\neq 0$} To conclude this section, let us briefly study the dependence {\yvan on} trap stiffness of the cluster coefficient $B_{3,1}(\omb)$. One {\yvan experimentally useful} way to account for this is to calculate the first deviation from the zero stiffness limit, of even degree in $\omb$ since the integrand of equation (\ref{eq003}) is an even function of $\omb$: 
\be
\label{eq158}
B_{3,1}(\bar{\omega}) \underset{\omb\to 0^+}{=} B_{3,1}(0^+) + \frac{1}{2} B_{3,1}''(0^+) \bar{\omega}^2 + O(\bar{\omega}^4)
\ee
This allows to quantify the error due to the local density approximation, systematically used in the experiment and which {\bleu amounts to keeping} only the first term of Taylor expansion (\ref{eq158}). It is easy to see that the small parameter controlling this approximation can only be $\omb$ in the unitary limit: the approximation only makes sense if the equilibrium correlation length of the homogeneous gas in each {\yvan spin} state $\sigma$, i.e.\ {\yvan the de Broglie thermal length $\lambda_\sigma$} in the nondegenerate case {\yvan due to} scale invariance, is much smaller than the spatial radius of the trapped gas 
$R_\sigma=(k_B T/m_\sigma\omega^2)^{1/2}$; one has indeed $\lambda_\sigma/R_\sigma\propto\hbar\omega/k_B T=\omb\ll 1$.
To obtain an integral expression for the second derivative $B_{3, 1}''(0^+)$ and implement convergence acceleration, we take twice the derivative with respect to $\omb$ of equation (\ref{eq003}) under the integral sign, equation (\ref{eq105}) under the sum sign and expressions (\ref{eq110},\ref{eq125}) and then make $\omb$ tend to zero.

\begin{figure}[t]
\includegraphics[width=7.5cm,clip=]{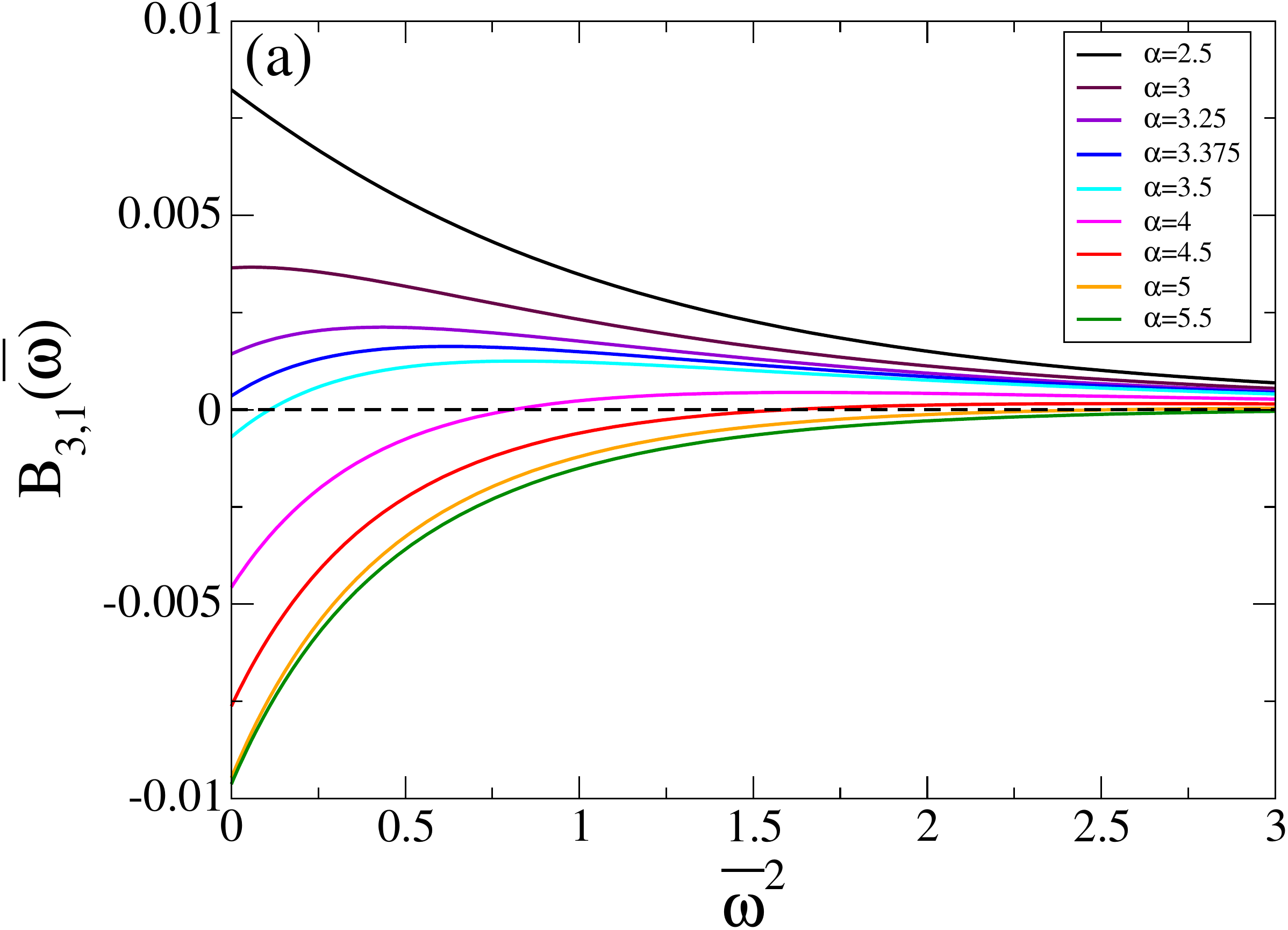}\hspace{5mm}\includegraphics[width=7.5cm,clip=]{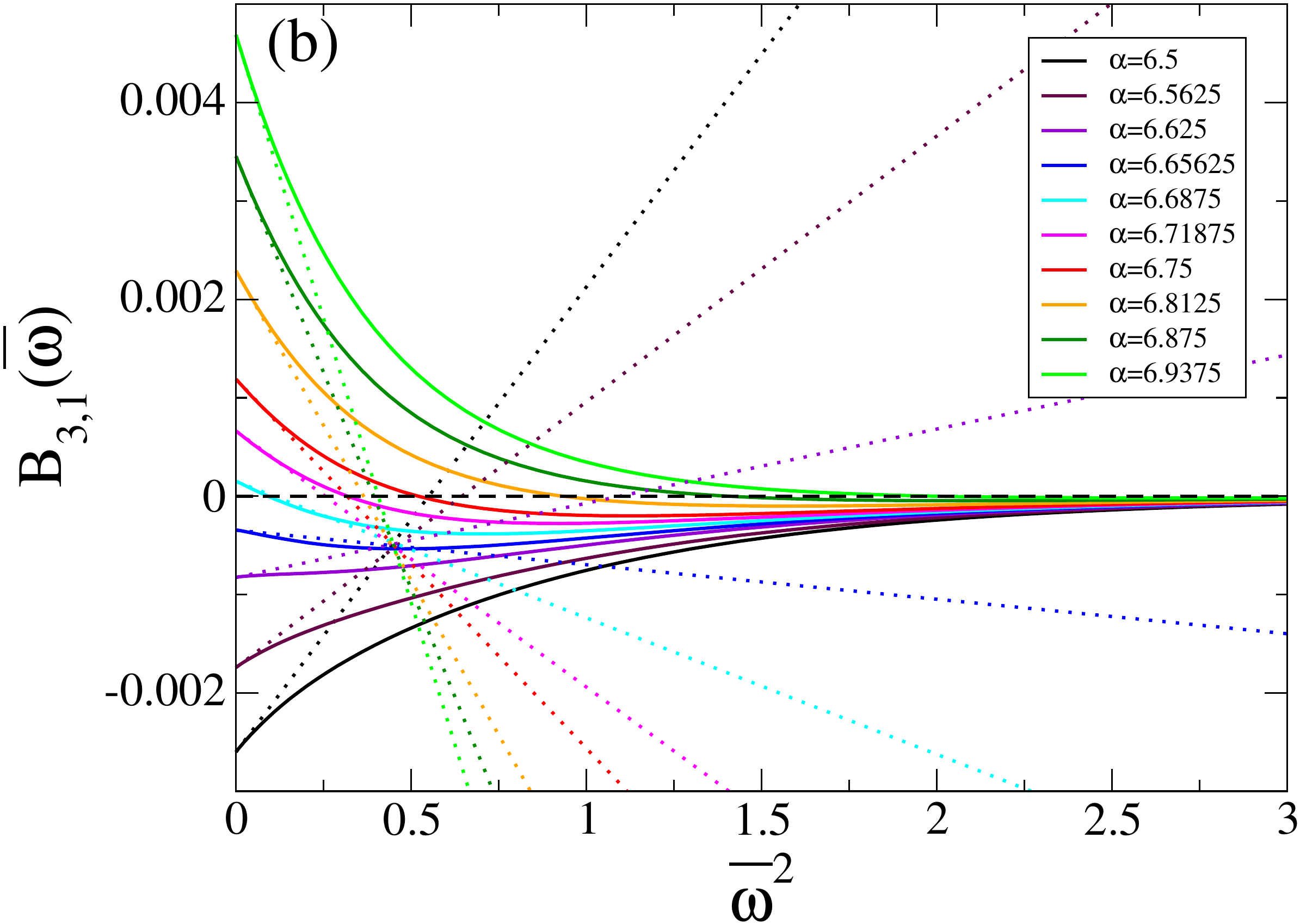}
\caption{Fourth cluster coefficient $B_{3,1}(\omb)$ of the trapped unitary Fermi gas as a function of the reduced stiffness $\omb^2=(\hbar\omega/k_B T)^2$ of the harmonic {\rouge potentials}, for different mass ratios $\alpha=m_\uparrow/m_\downarrow$ close to the nodal points $\alpha_1\simeq 3.412$ (a) and $\alpha_2\simeq 6.678$ (b) of $B_{3,1}(0^+)$, which is useful for understanding the {\yvan enlargements} in figure \ref{fig3}b, i.e., the relationship between the signs of the value and curvature of $B_{3,1}(\omb)$ at the origin. In (b), the slopes of the tangents at the origin (dotted) are taken from figure \ref{fig3}b.}
\label{fig5}
\end{figure}

The result is plotted as a function of the mass ratio in figure \ref{fig3}b. \footnote{For values of $\alpha$ close to $\alpha_2$, we improve the convergence acceleration method by approximating 
$[I^{(\ell,\veps)}_{3,1}]''(0^+)-[J^{(\ell,\veps)}_{3,1}]''(0^+)$ for $\ell>\ell_{\rm max}$ by $A\ell^\gamma\re(\eee^{\ii\psi}z_0^\ell)$
rather than by $0$ as in equation (\ref{eq105}); the {\rouge real} parameters $A$, $\psi$ and $\gamma$ are obtained by fitting on the interval $6\leq\ell\leq\ell_{\rm max}$, and the complex number $z_0$ is that of the asymptotic law (\ref{eq333}). }  Let us try to interpret it in a {\bleu naive} scenario: as in the case of equal masses $m_\uparrow=m_\downarrow$ studied in reference \cite{JPA}, $B_{3,1}(\omb)$ would {\bleu simply} be a monotonic function of $\omb$, of course of zero limit at {\bleu infinite $\omb$}.  The second derivative at the origin would then always have the opposite sign to the value at {\bleu $\omb=0$}. This explains figure \ref{fig3}b if we look on a large scale: $B_{3,1}''(0^+)$ seems indeed to vanish by changing sign at the same mass ratios $\alpha=\alpha_1$ and $\alpha=\alpha_2$ as $B_{3,1}(0^+)$. However, we can see on a first enlargement, in inset in figure \ref{fig3}b, that this scenario {\yvan fails} at $\alpha_1$ (it still seems to hold at $\alpha_2$). Indeed, as shown in figure \ref{fig5}a, when $\alpha$ approaches $\alpha_1\simeq 3.412$ {\bleu from} lower values (top to bottom curves), the function $\omb\mapsto B_{3, 1}(0^+)$ ceases to be monotonic, becomes convex near the origin and reaches an absolute (positive) maximum at a point $\omb_0>0$ {\rouge that departed} from zero before $B_{3,1}(0^+)$ becomes $<0$. This absolute maximum {\yvan persists} when $\alpha$ continues to grow beyond $\alpha_1$, but its position moves toward $+\infty$ and out of figure \ref{fig5}a. A second enlargement, in the inset of figure \ref{fig3}b, shows that the simple scenario also fails (but just barely) at the second nodal point $\alpha_2\simeq 6.678$: when $\alpha$ approaches $\alpha_2$ by higher values (top-down curves in figure \ref{fig5}b \footnote{By varying all truncation and discretization parameters in the numerical calculation, we estimate the uncertainty on the curves in figure \ref{fig5}b closest to the horizontal axis to be $2\times 10^{-5}$ (e.g., for $\alpha=6.6875$); the {\yvan plotted} dependencies {\yvan on} $\omb$ are thus significant.}), an absolute (negative) minimum coming from the $\omb=+\infty$ side approaches the origin and becomes more pronounced, before {\rouge $B_{3,1}(0^+)$} in turn becomes $<0$. {\bleu For convenience, we give the values of $B_{3,1}(\omb)$ and its half-second derivative {\yvan at} $\omb=0^+$ in numerical form in table \ref{tab:bbs}, for the mass ratios of figure \ref{fig5}. To conclude,} let us finally point out that the simple scenario {\bleu assuming monotonicity of $B_{3,1}(\omb)$} was actually highly improbable because it implied that the $3+1$-body cluster coefficient in the trap was zero for any stiffness, $B_{3,1}(\omb)\equiv 0$, at mass ratios where $B_{3,1}(0^+)=0$.

\begin{table}[t]
{\footnotesize
\begin{center}
\begin{tabular}{|c|c|c|c|c|c|c|c|c|c|c|}
\hline
$\alpha$ & 2.5 & 3 & 3.25 & 3.375 & 3.5 & 4 & 4.5 & 5 & 5.5 & \\
\hline
$B_{3,1}$  & 0.00823 & 0.00365 & 0.00143 & 0.00035 & $-$0.00070 & $-$0.00457 & $-$0.00763 & $-$0.00947 & $-$0.00963 & \\
$\frac{1}{2}B''_{3,1}$  & $-$0.00660 & 0.00055 & 0.00420 & 0.00596   & 0.00766 & 0.0140 & 0.0190 & 0.0217 & 0.0212 & \\
\hline
$\alpha$ & 6.5 & 6.5625 & 6.625 & 6.65625 & 6.6875 & 6.71875 & 6.75 & 6.8125 & 6.875 & 6.9375 \\
\hline
$B_{3,1}$  & $-$0.00260 & $-$0.00174 & $-$0.00083 & $-$0.00035 & 0.00015 & 0.00066 & 0.00119 & 0.00229 & 0.00346 & 0.00469 \\
$\frac{1}{2}B''_{3,1}$  & 0.0047 & 0.0027 & 0.0007 & $-$0.0003 & $-$0.0013 & $-$0.0026 & $-$0.0037 & $-$0.0063 & $-$0.0088 & $-$0.0115 \\
\hline
\end{tabular}
\end{center}
}
\caption{\bleu Numerical values of the fourth cluster coefficient $B_{3,1}(0^+)$ and its half-second derivative $(1/2)B_{3,1}''(0^+)$ with respect to $\omb$, given with their significant digits, for the mass ratios of figure \ref{fig5}. The half-derivatives are plotted (some with an error bar) in figure \ref{fig3}b.}
\label{tab:bbs}
\end{table}
\section{Cluster coefficient for $2+2$ fermions in a trap}
\label{sec3}

We explain in this section how to perform an efficient numerical calculation of the quantity $I_{2,2}(\omb)$ defined by equation (\ref{eq003}), limiting ourselves for simplicity to the case $\omb=0^+$, i.e. to a trap of infinitesimal stiffness. The {\rouge unitary gas} cluster coefficient $B_{2,2}(0^+)$ can be deduced by conjecture (\ref{eq004}). As the two {\yvan spin} states $\uparrow$ and $\downarrow$ play perfectly symmetric roles here, $B_{2,2}(0^+)$ is invariant by changing the mass ratio $\alpha$ into its inverse $1/\alpha$ and one can limit the numerical calculations to the case $\alpha\geq 1$. In contrast to the previous section \ref{sec2}, one can go up to the three-body Efimovian threshold, $\alpha<\alpha_c^{2,1}$, since there is no $2+2$-body Efimov effect \cite{PRA}.

\paragraph{\rouge Formulation of the problem} Recall the expression of the Hermitian operator $M_{2,2}^{(\ell,\veps)}(\ii S)$ given in reference \cite{PRA} for angular momentum $\ell$, parity $\veps=\pm$ (restricted to $+$ if $\ell=0$) and the pure imaginary scale exponent $\ii S$, with the same Schr\"odinger-Dirac notation as in equation (\ref{eq100}) but with the difference that the variable $x$ varies on the whole real {\yvan axis}: 
\be
\label{eq200}
\boxed{
\langle x,u|\langle \ell,m_z| M_{2,2}^{(\ell,\veps)}(\ii S)|f\rangle = \mD_{2,2}(x,u) f_{m_z}(x,u) + \int_{-\infty}^{+\infty}\dd x' \int_{-1}^{1}\dd u' \sum_{m_z' \ |\ (-1)^{m_z'}=\veps}  K^{(\ell)}_{2,2} (x,u,m_z;x',u',m_z') f_{m_z'}(x',u')
}
\ee
with a diagonal part independent of $\ii S$, angular momentum and magnetic quantum number $m_z$: 
\be
\label{eq201}
\mD_{2,2}(x,u)=\left[\frac{\alpha}{(1+\alpha)^2}\left(1+\frac{u}{\ch x}\right)+\frac{\eee^{-x}+\alpha\eee^{x}}{2(\alpha+1)\ch x}\right]^{1/2} 
\ee
and a matrix integral kernel decomposed into three contributions $K_{2,2}=K_1+K_2+K_3$ written line by line in this order: 
\begin{multline}
\label{eq202}
K_{2,2}^{(\ell)}(x,u,m_z;x',u',m_z')=\left(\frac{\eee^x\ch x'}{\eee^{x'}\ch x}\right)^{\ii S/2} \left(\frac{\eee^{x+x'}}{4\ch x\ch x'}\right)^{1/4}
\int_0^{2\pi} \frac{\dd\phi}{(2\pi)^2} \frac{\eee^{-\ii m_z\theta/2}\langle \ell,m_z|\eee^{\ii\phi L_x/\hbar}|\ell,m_z'\rangle \eee^{im_z'\theta'/2}}{\ch(x-x')+\frac{1}{1+\alpha} [{(u+\eee^{-x})(u'+\eee^{-x'})} + v v'\cos\phi]} \\
+\left(\frac{\eee^{-x}\ch x'}{\eee^{-x'}\ch x}\right)^{\ii S/2} \left(\frac{\eee^{-x-x'}}{4\ch x\ch x'}\right)^{1/4} \int_0^{2\pi} \frac{\dd\phi}{(2\pi)^2} \frac{\eee^{\ii m_z\theta/2}\langle \ell,m_z|\eee^{\ii\phi L_x/\hbar}|\ell,m_z'\rangle \eee^{-\ii m_z'\theta'/2}}{\ch(x-x') +\frac{\alpha}{1+\alpha} [ {(u+\eee^{x})(u'+\eee^{x'})} + v v'\cos\phi]}  \\
+\frac{(-1)^{\ell+1}}{4\pi[(u+\ch x) (u'+\ch x')\ch x\ch x']^{1/4}} \left(\frac{(u'+\ch x')\ch x'}{(u+\ch x)\ch x}\right)^{\ii S/2} \frac{\eee^{\ii m_z\gamma(x,u)}\langle \ell,m_z|\ell,m_x=0\rangle\langle \ell,m_x=0|\ell,m_z'\rangle \eee^{-\ii m_z'\gamma(x',u')}} {\left(\frac{\eee^{-x'}+\alpha \eee^{x'}}{1+\alpha}\right) (u+\ch x) +\left(\frac{\eee^{-x}+\alpha \eee^x}{1+\alpha}\right)(u'+\ch x')}
\end{multline}
with the angle 
$\gamma(x,u)=\atan\{\thf(x/2)[(1-u)/(1+u)]^{1/2}\}$,
the angle $\theta\in[0,\pi]$ such that $u=\cos\theta$ and the notation $v=\sin\theta$. The third contribution $K_3$ is nonzero only in the parity sector $\veps=(-1)^\ell$ (in the other sector, we have $\langle\ell,m_x=0|\ell,m_z\rangle\equiv 0$). Contrary to the first two contributions, it is a singular function of the coordinates $(x,u)$ and $(x',u')$, diverging at the point $(x=0,u=-1)$, or {\yvan equivalently} $(x=0,\theta=\pi)$. In the numerical calculation, limited to the rectangle $(x,\theta)\in[-x_{\rm max},x_{\rm max}]\times[0,\pi]$, we isolate the singularity by a half disk of radius $R$ and center $(0,\pi)$, in which we use a logarithmic-polar grid as in reference \cite{PRA} \footnote{\label{note8} We set $(x,\pi-\theta)=R\,\eee^t(\cos\psi,\sin\psi)$, where $t_{\rm min}<t<0$ and $0<\psi<\pi$. We have chosen $R=2/5$. The integration on $\psi$ is done by the Gauss-Legendre method with 15 points (series $0\leq\ell\leq 4$) or 25 points (series $0\leq\ell\leq 6$), the integration on $t$ with the midpoint rule and a step $\dd t=0.1$ or $\dd t=0.2$. We extrapolate quadratically to $t_{\rm min}=-\infty$ from the three choices $t_{\rm min}=-10$, $t_{\rm min}=-20$ and $t_{\rm min}=-40$.}, and outside of which we use the same type of grid as in section \ref{sec2}.

\begin{figure}[t]
\begin{center}
\includegraphics[width=5cm,clip=]{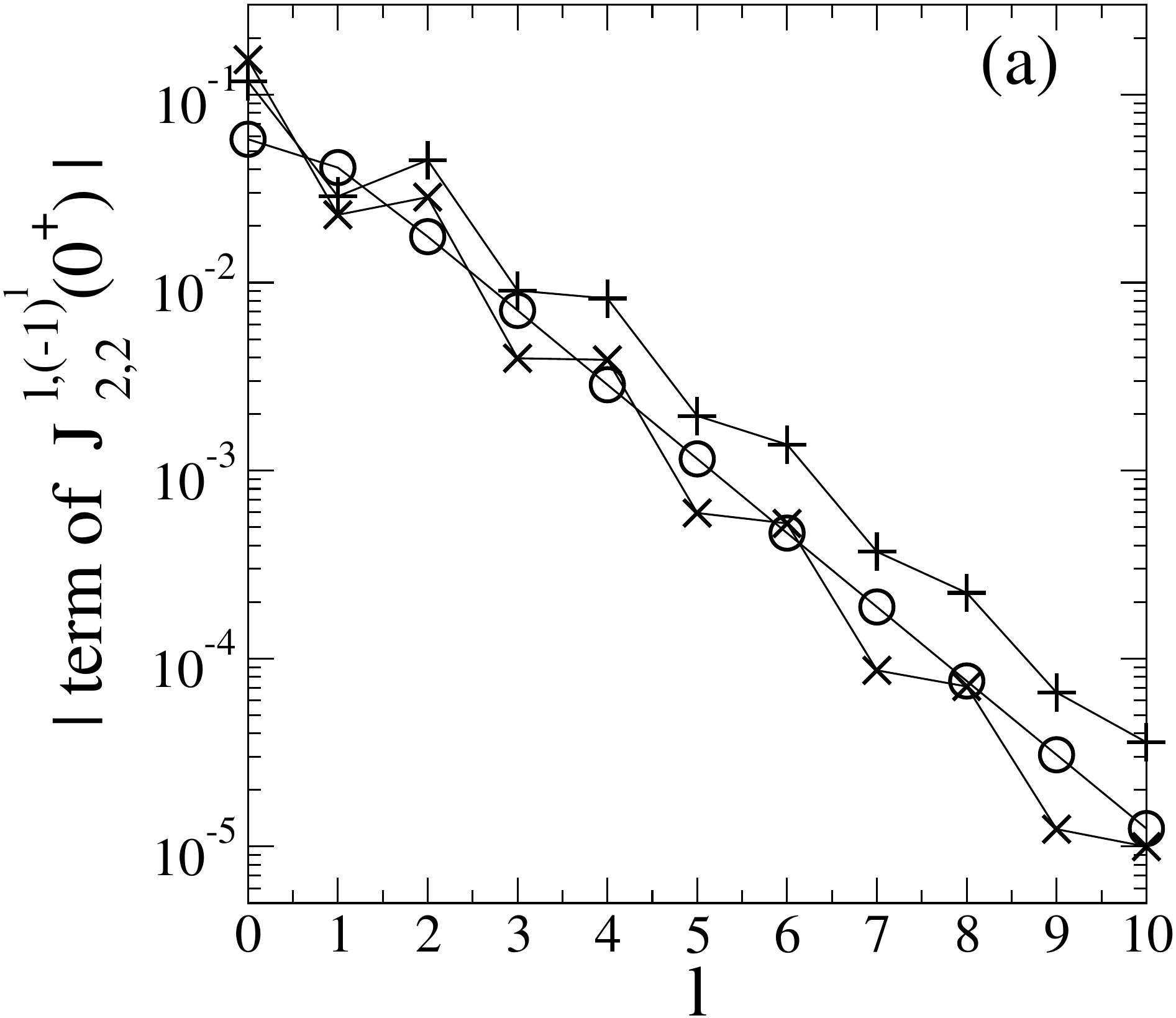}\hspace{5mm}\includegraphics[height=4.4cm,clip=]{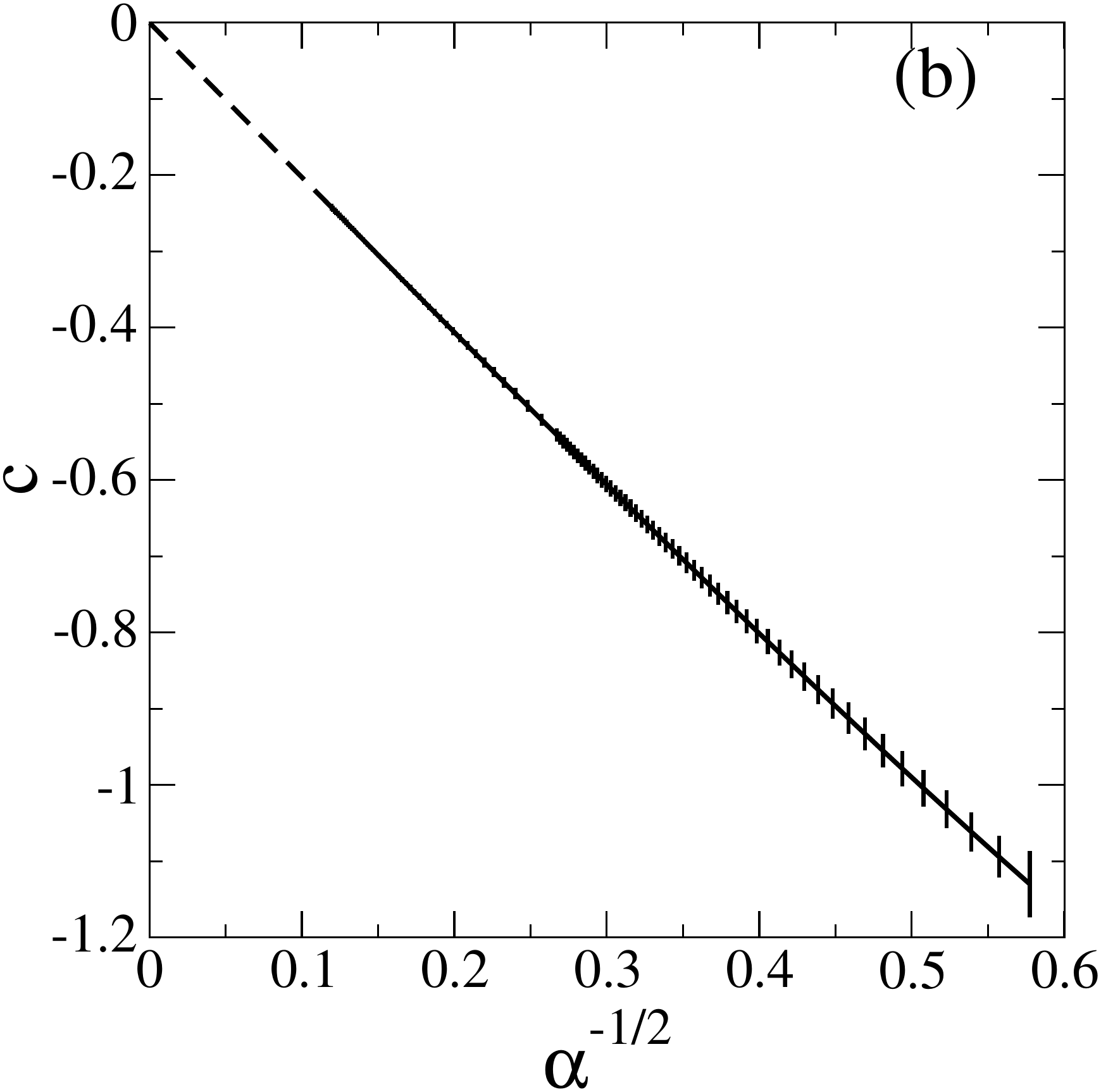}\hspace{5mm}\includegraphics[width=4.9cm,clip=]{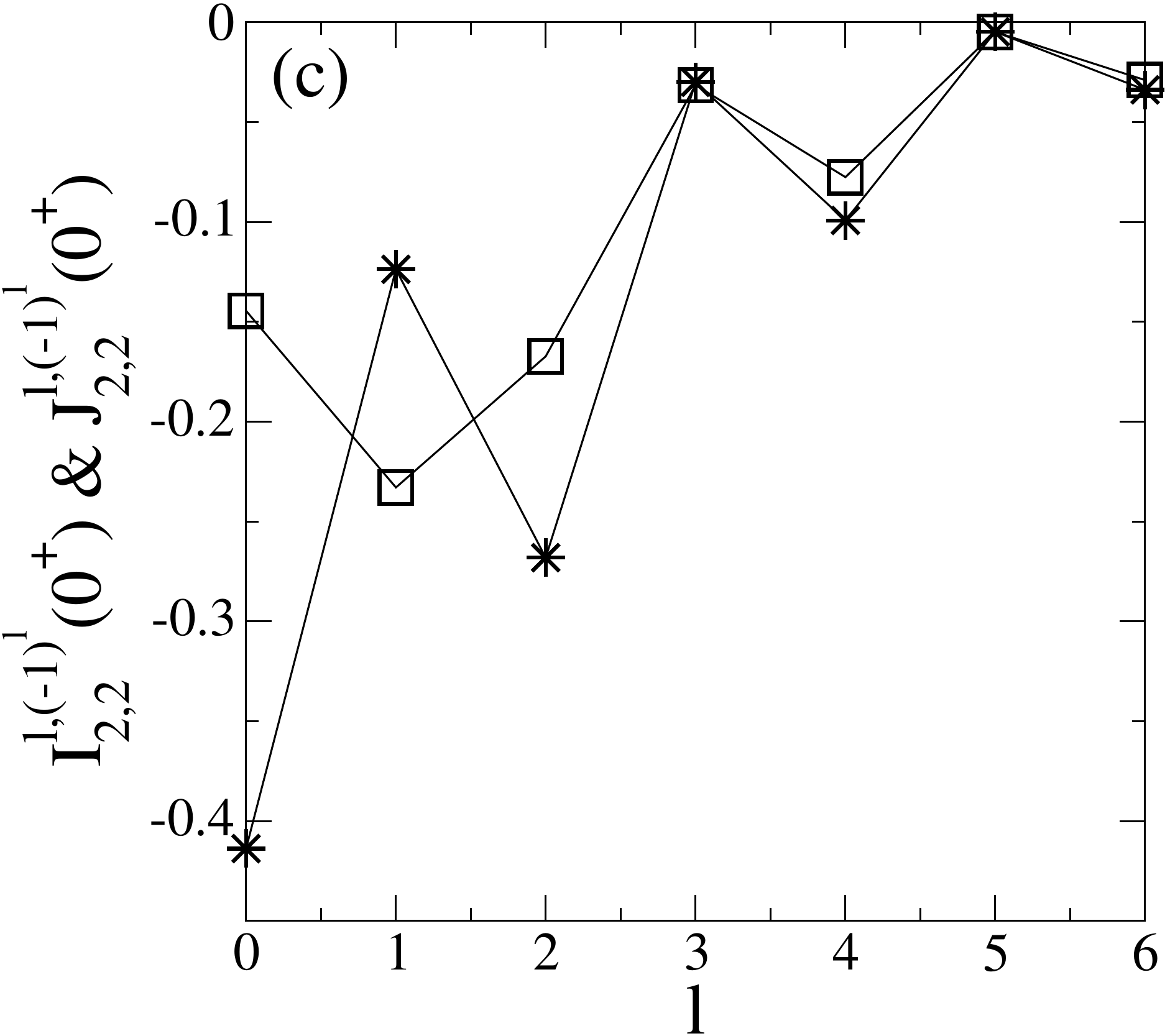}
\end{center}
\caption{Some numerical results on the asymptotic approximant $J_{2,2}^{(\ell,\veps)}(0^+)$ of the contribution $I_{2,2}^{(\ell,\veps)}(0^+)$ of angular momentum $\ell$ and parity $\veps$ to the quantity $I_{2,2}(0^+)$ in equation (\ref{eq003}). (a) For a mass ratio $\alpha=5$ and the {\yvan challenging natural-parity} sector $\veps=(-1)^\ell$, absolute value of the three terms of $J_{2,2}^{(\ell,\veps)}(0^+)$ in expression (\ref{eq400}) (symbols $\circ$, $+$, and $\times$ in that order) as a function of angular momentum $\ell$. (b) Rate $c$ of exponential decay with $\ell$ of $J_{2,2}^{(\ell,\veps)}(0^+)$ as in equation (\ref{eq210}), as a function of the square root of the inverse mass ratio. The error bars give the deviation between values of $c$ from various fitting functions (for $\veps=(-1)^{\ell-1}$, $\ln|J_{2,2}^{(\ell,\veps)}(0^+)|$ is taken as an affine function of $\ell$ plus one term $\propto 1/\ell$, or plus one term $\propto\ln\ell$, or plus both). The dashed line is an affine extrapolation to $\alpha^{-1/2}=0$. (c) For $\alpha=13.3$ and $\veps=(-1)^\ell$, exact value $I_{2,2}^{(\ell,\veps)}(0^+)$ obtained numerically (squares) and asymptotic approximant $J_{2,2}^{(\ell,\veps)}(0^+)$ (stars) as functions of $\ell$.}
\label{fig6}
\end{figure}

\paragraph{\rouge Asymptotic approximant} It remains to implement the same convergence acceleration technique as in equation (\ref{eq105}) in the particular case $\omb=0^+$, 
\be
\label{eq214}
I_{2,2}(0^+) \simeq \sum_{\ell=0}^{\ell_{\rm max}}\sum_\veps \left[I_{2,2}^{(\ell,\veps)}(0^+)-J_{2,2}^{(\ell,\veps)}(0^+)\right] + \sum_{\ell=0}^{+\infty}\sum_\veps J_{2,2}^{(\ell,\veps)}(0^+)
\ee
where, as in section \ref{sec2}, $I^{(\ell,\veps)}_{2,2}(0^+)$ is the contribution of the angular momentum $\ell$ and parity $\veps$ channel to the desired quantity $I_{2, 2}(0^+)$, and $J_{2,2}^{(\ell,\veps)}(0^+)$ is an asymptotic approximant, valid at large angular momentum. Reference \cite{PRA} gives such an approximant only in the {\yvan tractable unnatural-parity} sector $\veps=(-1)^{l-1}$ where $K_3\equiv 0$, by perturbatively treating $K_1$ and $K_2$ to second order in the logarithm of the determinant of $M_{2,2}^{(\ell,\veps)}(\ii S)$. The calculation is very similar to equation (\ref{eq106}), with the simplification that the linear terms and the square terms are independent of $S$ and can be absorbed into the constant, so that only the crossed terms remain \cite{PRA}: 
\be
\label{eq204}
\ln\det[M_{2,2}^{(\ell,\veps)}(\ii S)]\stackrel{\veps=(-1)^{\ell-1}}{=}\mathrm{const}-\mbox{Tr}^{\veps}\,\left[\mD_{2,2}^{-1}K_1\mD_{2,2}^{-1}K_2\right]+\ldots
\ee
In the {\yvan more challenging natural-parity} sector $\veps=(-1)^\ell$, a difficulty arises: because of its divergent character, we cannot treat $K_3$ perturbatively. We first perform a gauge transform on $K_j$ eliminating the $S$-dependence of $K_3$ without changing the determinant, which is indicated by a tilde, 
\be
\label{eq205}
\tilde{K}_j(x,u,m_z;x',u',m_z') \equiv \frac{\eee^{-\ii m_z \gamma(x,u)}}{[(u+\ch x)\ch x]^{-\ii S/2}} K_j(x,u,m_z;x',u',m_z') \frac{\eee^{\ii m_z'\gamma(x',u')}}{[(u'+\ch x')\ch x']^{\ii S/2}}
\ee
then we isolate a purely \g{external} part of $\tilde{K}_3$, acting on the \g{orbital} space of $(x,u)$ but not on that of {\yvan the $|\ell,m_z\rangle$'s}, i.e.\ such that 
$\tilde{K}_3=\tilde{K}_3^{\rm ext}\otimes |\ell,m_x=0\rangle\langle\ell,m_x=0|$.
Finally, we carry out an expansion of the logarithm of the determinant in powers of $K_1$ and $K_2$ without any hypothesis on $\tilde{K}_3^{\rm ext}$: 
\begin{multline}
\label{eq206}
\ln\det[M_{2,2}^{(\ell,\veps)}(\ii S)] \stackrel{\veps=(-1)^\ell}{=}\mbox{const}+\ln\det\left[\,\identit+\frac{1}{\mD_{2,2}+\tilde{K}_3^{\rm ext}\otimes|\ell,m_x=0\rangle\langle\ell,m_x=0|}\left(\tilde{K}_1+\tilde{K}_2\right)\right]= \\
\mbox{const}-\mbox{Tr}_{x,u}\left[\mD_{2,2}^{-1}\tilde{\KKK}_3^{\rm ext}\mD_{2,2}^{-1}\langle\ell,m_x=0|(\tilde{K}_1+\tilde{K}_2)|\ell,m_x=0\rangle\right]-\mbox{Tr}^{\veps}\left[\mD_{2,2}^{-1}\tilde{K}_1\mD_{2,2}^{-1}\tilde{K}_2\right] \\
 +\mbox{Tr}_{x,u}\left[\mD_{2,2}^{-1}\tilde{\KKK}_3^{\rm ext}\mD_{2,2}^{-1}\langle\ell,m_x=0|(\tilde{K}_1+\tilde{K}_2)\mD_{2,2}^{-1}(\tilde{K}_1+\tilde{K}_2)|\ell,m_x=0\rangle\right] \\
-\frac{1}{2}\mbox{Tr}_{\rouge x,u}\left[\mD_{2,2}^{-1}\tilde{\KKK}_3^{\rm ext}\mD_{2,2}^{-1}\langle\ell,m_x=0|(\tilde{K}_1+\tilde{K}_2)|\ell,m_x=0\rangle\mD_{2,2}^{-1}\tilde{\KKK}_3^{\rm ext}\mD_{2,2}^{-1}\langle\ell,m_x=0|(\tilde{K}_1+\tilde{K}_2)|\ell,m_x=0\rangle\right]+\ldots
\end{multline}
In this expression, a resummed form of the external part {\yvan appears}: 
\be
\label{eq208}
\tilde{\KKK}_3^{\rm ext}\equiv\mD_{22}[\mD_{22}^{-1}-(\mD_{22}+\tilde{K}_3^{\rm ext})^{-1}]\mD_{22}=\tilde{K}_3^{\rm ext}-\tilde{K}_3^{\rm ext}(\mD_{22}+\tilde{K}_3^{\rm ext})^{-1}\tilde{K}_3^{\rm ext}
\ee
An optimal writing of $\tilde{\KKK}_3^{\rm ext}$ is obtained by reparameterization of the Faddeev ansatz of the $2+2$-body problem: we consider that the functions on which {\rouge operator $M_{2,2}$ acts} now depend on the relative wave vectors and the center of mass of the particles $2$ and $4$ as in reference \cite{Ludo} instead of the {\rouge single-particle} wave vectors $\kk_2$ and $\kk_4$ as in references \cite{PRL, PRA}; {\yvan this} avoids the half-disk around the singularity and leads us to numerically invert an operator acting on a single real variable instead of the two variables $(x,u)$. This leads to a considerable saving of {\rouge computation} time and simplification. The reader is referred to \ref{apC} for more details. We find numerically that the second term in the third side of equation (\ref{eq206}), formally of the first order, is actually of the same order of magnitude as the third term (their contributions to $J_{2,2}^{(\ell,\veps)}(0^+)$ tend to zero exponentially with $\ell$ with the same rate, see figure \ref{fig6}a). The resummed kernel $\tilde{\KKK}_3^{\rm ext}$ is thus, like $K_1$ and $K_2$, a first-order infinitesimal; consequently, we neglect the fourth and fifth terms of equation (\ref{eq206}) to keep 
\be
\label{eq209}
\boxed{
\ln\det[M^{(\ell,\veps)}(\ii S)] \stackrel{\veps=(-1)^\ell}{=} \mbox{const}-\mbox{Tr}_{x,u}\left[\mD_{2,2}^{-1}\tilde{\KKK}_3^{\rm ext}\mD_{2,2}^{-1}\langle\ell,m_x=0|(\tilde{K}_1+\tilde{K}_2)|\ell,m_x=0\rangle\right]-\mbox{Tr}^{\veps}\left[\mD_{2,2}^{-1}\tilde{K}_1\mD_{2,2}^{-1}\tilde{K}_2\right]+\ldots
}
\ee
{\rouge If one wants, one can undo the gauge transform (\ref{eq205}) in the third term without changing the trace, to recover contribution (\ref{eq204}).} In \ref{apB}, we give a more explicit expression of the asymptotic approximant $J_{2,2}^{(\ell,\veps)}(0^+)$ deduced from expansions (\ref{eq204}) and (\ref{eq209}), see equations (\ref{eq400},\ref{eq401},\ref{eq407},\ref{eq409}), as well as its sum over $\ell$ and over $\veps$, see equations (\ref{eq413},\ref{eq425}).

\begin{SCfigure}
\includegraphics[width=8cm,clip=]{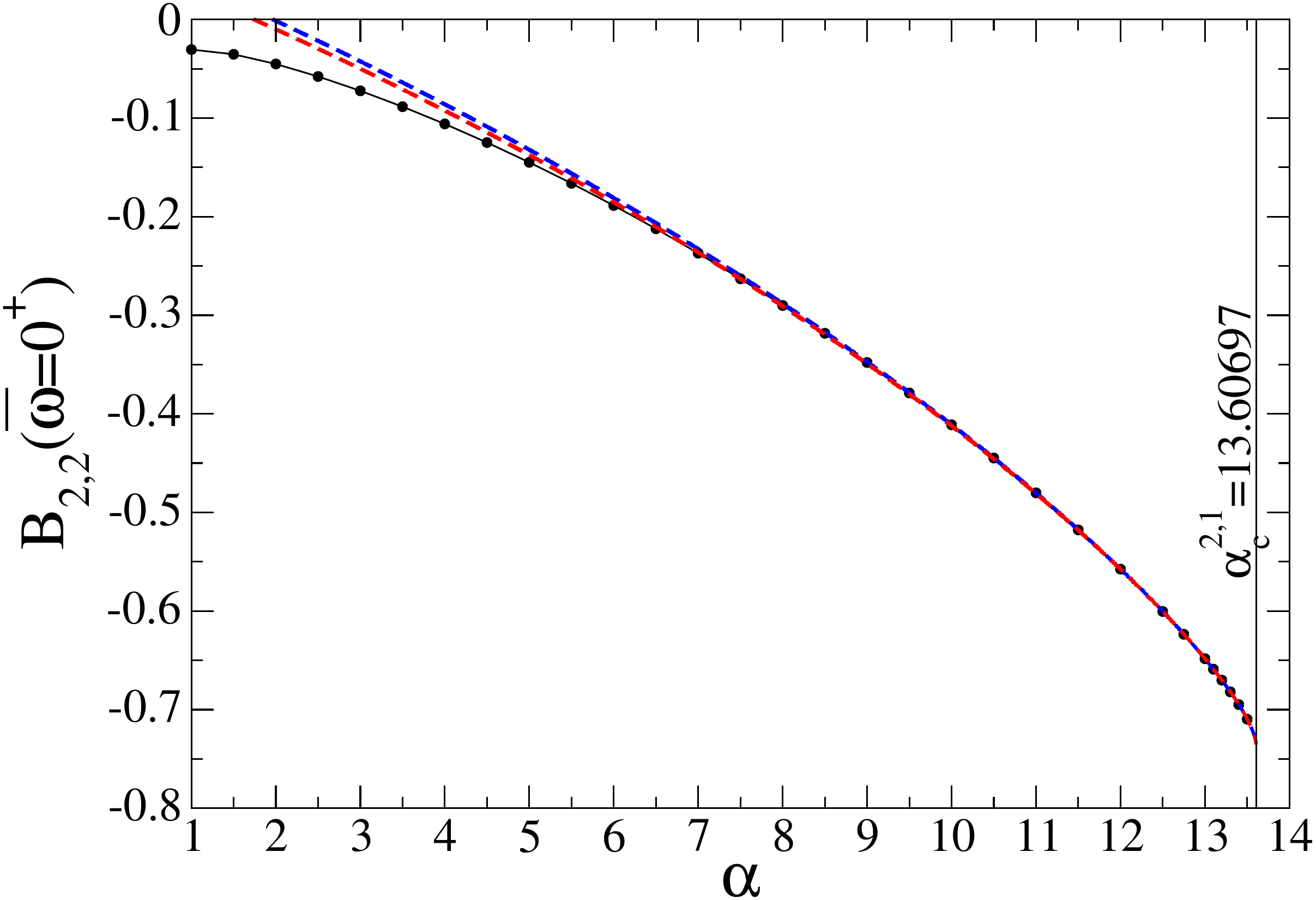}
\caption{Fourth cluster coefficient $B_{2,2}(0^+)$ of the unitary Fermi gas in a harmonic trap of infinitesimal stiffness, as a function of the mass ratio $\alpha=m_\uparrow/m_\downarrow$ restricted to $\alpha\geq 1$ by $\alpha\leftrightarrow 1/\alpha$ symmetry. Disks (connected by a thin line): Our numerical calculation of expression (\ref{eq003}) for $(n_\uparrow,n_\downarrow)=(2,2)$ and application of conjecture (\ref{eq004}), with parameters $x_{\rm max}=-x_{\rm min}=20$ and $n_\theta=15$ in the series $0\leq\ell\leq 4$, $x_{\rm max}=-x_{\rm min}=5$ and $n_\theta=25$ in the series $0\leq\ell\leq 6$ (the overlap of the two series serves as a check), and in all cases $\dd x=1/15$, $S_{\rm max}=12$, $\dd S=12/100$ (see footnote \ref{note8} for the treatment of the singularity of the integral kernel at $(x,\theta)=(0,\pi)$). We use the convergence acceleration method (\ref{eq214}) and the same extrapolation to $S_{\rm max}=+\infty$ as in section \ref{sec2}. Thin vertical line: three-body Efimovian threshold $\alpha=\alpha_c^{2,1}$. Blue dashed line: cubic fit in the variable $Y=(\alpha_c^{2,1}-\alpha)^{1/2}$ on the interval $10\leq\alpha\leq 13.5$, i.e. \protect
$B_{2,2}(0^+)=-0.73742+0.061673Y+0.074061Y^2-0.0084386Y^3$.
Red dashed line: same on the interval $12.5\leq\alpha\leq 13.5$, i.e. \protect
$B_{2,2}(0^+)=-0.73761+0.062206Y+0.073901Y^2-0.0086514Y^3$.
}
\label{fig7}
\end{SCfigure}

\paragraph{\rouge Application and results} The sum over $\ell$ in $I_{2,2}(0^+)$ shows the same phenomenon of slow convergence as in $I_{3,1}(0^+)$ at large values of the mass ratio $\alpha$. We again find that the asymptotic approximant, here $J_{2,2}^{(\ell,\veps)}(0^+)$, tends to zero exponentially in $\ell$, with a rate $c$ that we compute numerically for convenience in the {\yvan tractable} parity sector $\veps=(-1)^{\ell-1}$ and for a mass ratio $\alpha\gtrsim 3$: \footnote{The dominant behaviors in the two parity sectors differ only in their power laws in $\ell$. When the mass ratio is too close to $1$, $J_{2,2}^{(\ell,\veps=(-1)^{\ell-1})}(0^+)$ tends to zero while oscillating which makes numerical rate extraction more difficult.} 
\be
\label{eq210}
J_{2,2}^{(\ell,\veps)}(0^+) \stackrel{\veps=(-1)^{\ell-1}}{\underset{\ell\to+\infty}{=}} \veps \exp[-c \ell + O(\ln\ell)]
\ee
The rate $c$ is plotted as a function of the mass ratio in figure \ref{fig6}b. Here, as in section \ref{sec2}, a $c\propto\alpha^{-1/2}$ law is observed at large mass ratios. In this regime, the convergence acceleration method is an indispensable {\bleu aid to} numerical calculation, as shown in figure \ref{fig6}c. It allows us to obtain the fourth cluster coefficient $B_{2,2}(0^+)$ of the trapped system, {\rouge given in numerical form in table \ref{tab:2p2} and} plotted as a function of the mass ratio in figure \ref{fig7}, with an uncertainty of less than one percent in a reasonable computation time. Because of the $2+1$-body Efimov effect, this coefficient has a $(\alpha_c^{2,1}-\alpha)^{1/2}$ singularity near the threshold, for the same reason as discussed around equation (\ref{eq149}) in section \ref{sec2}. We take this into account in the dashed fits in figure \ref{fig7}, to obtain the extrapolation 
\be
\label{eq211}
\boxed{
B_{2,2}(0^+) \underset{\alpha\to(\alpha_c^{\rm 3\,body})^-}{{\rouge\longrightarrow}} -0.737 \pm 0.007
}
\ee

\begin{table}[t]
{\footnotesize
\begin{tabular}{|c|c|c|c|c|c|c|c|c|c|c|c|}
\hline
$\alpha$ & 1 & 1.5 & 2 & 2.5 & 3 & 3.5 & 4 & 4.5 & 5 & 5.5 &  \\
\hline
$B_{2,2}$ & $-$0.03056 & $-$0.03529 & $-$0.04518 & $-$0.05777 & $-$0.07233 & $-$0.08845 & $-$0.1060 & $-$0.1248 & $-$0.1449 & $-$0.1662 & \\
\hline
$\alpha$ & 6 & 6.5 & 7 & 7.5 & 8 & 8.5 & 9 & 9.5 & 10 & 10.5 &\\
\hline
$B_{2,2}$ & $-$0.1886 & $-$0.2122 & $-$0.2370 & $-$0.2629 & $-$0.2900 & $-$0.3183 & $-$0.3478 & $-$0.3788 & $-$0.4110 & $-$0.4448 & \\
\hline
$\alpha$ & 11 & 11.5 & 12 & 12.5 & 12.75 & 13 & 13.1 & 13.2 & 13.3 & 13.4 & 13.5\\
\hline
$B_{2,2}$ & $-$0.4801 & $-$0.5177 & $-$0.5575 & $-$0.6004 & $-$0.6235 & $-$0.6484 & $-$0.6590 & $-$0.6701 & $-$0.6819 & $-$0.6948 & $-$0.7097\\
\hline
\end{tabular}
}
\caption{\rouge Numerical values of the fourth cluster coefficient $B_{2,2}(0^+)$ of the unitary Fermi gas in an infinitesimal-stiffness trap, tabulated as functions of the mass ratio $\alpha=m_\uparrow/m_\downarrow$. Since the coefficient is invariant by change of $\alpha$ to $1/\alpha$, we restricted ourselves to $\alpha\geq 1$. The uncertainties, not specified, are less than one percent. {\bleu These values are shown graphically in figure \ref{fig7}.}}
\label{tab:2p2}
\end{table}

\noindent{\bf Acknowledgements:} This work was supported by Japanese \g{Grants-in-Aid for Scientific Research} KAKENHI Grant {\yvan Numbers} 21H00116 and {\yvan 22K03492}.

\appendix
\section{Asymptotic approximant of {\rouge $I_{3,1}^{(\ell,\veps)}(\omb)$}, its sum over $\ell$ and $\veps$, its dominant behavior}
\label{apA}

\paragraph{The asymptotic approximant} To obtain an approximation $J_{3,1}^{(\ell,\veps)}(\omb)$ at large $\ell$ of the contribution $I_{3,1}^{(\ell, \veps)}(\omb)$ of angular momentum $\ell$ and parity $\veps$ to the quantity $I_{3,1}(\omb)$ of equation (\ref{eq003}), we use the notations of section \ref{sec2} and start from expansion (\ref{eq106}). A clever calculation of the trace on $x>0$ allows us to collect contributions making $x$ and $-x$ appear and to reduce to integrals on the whole real {\yvan axis}, in which we only have to take the traces on the variables $u$ and $m_z$: 
\begin{multline}
\label{eq107}
\ln\det M_{3,1}^{(\ell,\veps)}=\mbox{const}-\int_\mathbb{R}\dd x\, \eee^{\ii S x}\mathrm{Tr}_{u,m_z}^{\veps}[\langle x|\bar{K}_{3,1}^{(\ell)}|\!-\!x\rangle \eee^{\ii\pi L_x/\hbar}]{\rouge +}\int_{\mathbb{R}^2}\dd x\,\dd x' \eee^{\ii S x} \mathrm{Tr}_{u,m_z}^{\veps}[\langle x|\bar{K}_{3,1}^{(\ell)}|x'\rangle\langle x'|\bar{K}_{3,1}^{(\ell)}|\!-\!x\rangle \eee^{\ii\pi L_x/\hbar}] \\
-\frac{1}{2} \int_{\mathbb{R}^2}\dd x\,\dd x' \eee^{\ii S (x-x')} \mathrm{Tr}_{u,m_z}^{\veps}[\langle x|\bar{K}_{3,1}^{(\ell)}|x'\rangle \eee^{\ii\pi L_x/\hbar} \langle\!-x'|\bar{K}_{3,1}^{(\ell)}|\!-\!x\rangle \eee^{\ii\pi L_x/\hbar}]+\ldots
\end{multline}
where the rotation operator of axis $Ox$ of angle $\pi$ comes from footnote \ref{note4} and we have introduced the primitive kernel (\ref{eq103}) divided on the left by the diagonal part (\ref{eq101}) and taken with zero scale exponent, i.e. 
$\bar{K}_{3,1}^{(\ell)}\equiv\mD_{3,1}^{-1}K_{3,1}^{(\ell)}(\ii S=0)$.
The contributions kept to the second side of equation (\ref{eq107}) constitute an even, regular, real-valued and rapidly {\yvan decreasing} function of $S$, that is $-\phi^{(\ell,\veps)}(S)$. So we can integrate by parts over $S$ in equation (\ref{eq003}) and we recognize the Fourier component of $-\phi^{(\ell,\veps)}(S)$ at frequency $\omb$: 
\be
\label{eq109}
J_{3,1}^{(\ell,\veps)}(\omb)=\frac{(\ell+1/2)\omb}{\sh\omb} \int_{-\infty}^{+\infty}\frac{\dd S}{2\pi} \eee^{-\ii\omb S}\phi^{(\ell,\veps)}(S)
\ee
{\yvan which is} easy to obtain from (\ref{eq107}) by means of the identity in the sense of distributions 
$\int_\mathbb{R}\exp(\ii k S)\,\dd S=2\pi\delta(k)$.
A long but not difficult calculation, treating the contributions to $\phi^{(\ell,\veps)}(S)$ in the order they appear, finally gives: 
\begin{multline}
\label{eq110}
J_{3,1}^{(\ell,\veps)}(\omb)= \frac{(\ell+1/2)\omb}{\pi\sh\omb\sqrt{2\ch\omb}} \int_{-1}^{1}\frac{\dd u}{\mD_{3,1}(\omb,u)} \int_{0}^{2\pi}\frac{\dd\phi}{2\pi} \frac{\mT^{(\ell,\veps)}(\theta;\phi+\pi)}{1+2\ch(2\omb)+\frac{2\alpha}{1+\alpha}(2u\ch\omb+u^2+v^2\cos\phi)} \\
-\frac{(\ell+1/2)\omb}{\pi^2\sh\omb\sqrt{2\ch\omb}} \int_{\mathbb{R}} \dd x' \int_{-1}^{1}\!\!\dd u \dd u'\!\!\int_{0}^{2\pi}\frac{\dd\phi\dd\phi'}{(2\pi)^2}\frac{\left(\lambda'^{2+1/2}/\sqrt{2\ch x'}\right)[\mD_{3,1}(\omb,u)\mD_{3,1}(x',u')]^{-1}\mT^{(\ell,\veps)}(\theta;\phi+\phi'+\pi)}{[1\!+\!\lambda_{\omb}^2\!+\!\lambda'^2\!+\!\frac{2\alpha}{1+\alpha}(u\lambda_{\omb}\!+\!u'\lambda'\!+\!\lambda_{\omb}\lambda'(uu'\!+\!vv'\cos\phi))][(\lambda_{\omb},\phi)\to(\lambda_{\omb}^{-1},\phi')]} \\
+\frac{(\ell+1/2)\omb}{2\pi^2\sh\omb}\int_{\mathbb{R}} \dd X \int_{-1}^{1}\dd u \dd u'\int_{0}^{2\pi}\frac{\dd\phi\dd\phi'}{(2\pi)^2} 
\frac{\left[2\sqrt{\ch X_+\ch X_-}\mD_{3,1}(X_+,u)\mD_{3,1}(X_-,u')\right]^{-1} \mT^{(\ell,\veps)}(\theta,\theta';\phi,\phi')}{[1\!+\!\lambda_+^2\!+\!\lambda_-^2\!+\!\frac{2\alpha}{1+\alpha}(\lambda_+u\!+\!\lambda_-u'\!+\!\lambda_+\lambda_-(uu'\!+\!vv'\cos\phi))][(\lambda_\pm,\phi)\to(\lambda_{\pm}^{-1},\phi')]}
\end{multline}
with the notations 
$X_\pm=X\pm\omb/2$, $\lambda_{\omb}=\exp(\omb)$, $\lambda_\pm=\exp(X_\pm)$
(these last two modeled on $\lambda=\exp(x)$) completing those of equation (\ref{eq104}) and the angular functions\footnote{{\rouge\label{note12} {\yvan One may} object that the definition of $\mT^{(\ell,\veps)}(\theta,\theta';\phi,\phi')$ should in principle include an orthogonal projector on the subspace of parity $(-1)^{L_z/\hbar}=\veps$ next to the operator $\eee^{\ii\theta'L_z/\hbar}$. The parity in $\phi$ and $\phi'$ of the denominator in the third contribution to (\ref{eq110}), however, allows us to do without it. For example, only the even part of $\eee^{\ii\phi L_x/\hbar}$ contributes after integration over $\phi$; it indeed preserves the parity of $L_z/\hbar$.}} 
\be
\label{eq112}
\mT^{(\ell,\veps)}(\theta;\phi)\equiv \mathrm{Tr}_{m_z}^\veps\left(\eee^{-\ii\theta L_z/\hbar}\eee^{\ii\phi L_x/\hbar}\right)\quad\mbox{and}\quad \mT^{(\ell,\veps)}(\theta,\theta';\phi,\phi')\equiv \mathrm{Tr}_{m_z}^\veps\left(\eee^{-\ii\theta L_z/\hbar}\eee^{\ii\phi L_x/\hbar}\eee^{\ii\theta'L_z/\hbar}\eee^{\ii\phi'L_x/\hbar}\right)
\ee
The sums over parity $\veps$ of the quantities defined in (\ref{eq112}) have simple expressions in terms of angles $\xi\in[0,\pi]$, see reference \cite{PRA}, \footnote{\label{note5} Turning to the half-angles, we note that we have more simply {\rouge $\cos(\xi/2)=\cos(\theta/2)|\cos(\phi/2)|$} in equation (\ref{eq113a}). } 
\be
\label{eq113a}
\mT^{(\ell)}(\theta;\phi)=\frac{\sin[(\ell+1/2)\xi]}{\sin(\xi/2)} \quad\mbox{with}\quad 1+2\cos\xi=u(1+\cos\phi)+\cos\phi
\ee
\be
\label{eq113b}
\mT^{(\ell)}(\theta;\phi)=\frac{\sin[(\ell+1/2)\xi]}{\sin(\xi/2)} \quad\mbox{with}\quad 1+2\cos\xi = uu'(1+\cos\phi \cos\phi')-(u+u')\sin\phi \sin\phi'+vv'(\cos\phi + \cos\phi')+\cos\phi\cos\phi'
\ee
which also shows how to return to fixed parity, {\rouge for example} 
\be
\label{eq114}
{\rouge \mT^{(\ell,\veps)}(\theta;\phi)=\frac{1}{2}\sum_{n=0}^{1}\veps^{n}\mT^{(\ell)}(\theta+n\pi;\phi)}
\ee
However, we prefer to reserve equations (\ref{eq113a},\ref{eq113b},\ref{eq114}) to analytical studies; for numerical computation, we evaluate the traces of (\ref{eq112}) in the eigenbasis of $L_z$ (under the constraint $(-1)^{m_z}=\veps$) after insertion of closure relations in the eigenbasis of $L_x$ (obtained by numerical diagonalization of its tridiagonal matrix in the $|\ell, m_z\rangle$ basis) at the location of the $\phi$- or $\phi'$-angle rotation operators, then we use the value of the integral\footnote{\label{note7} For optimization, we can (i) replace $\exp[\ii(-m_z\theta+m_z'\theta')]$ by its real part $\cos(m_z\theta)\cos(m_z'\theta')+\sin(m_z\theta)\sin(m_z'\theta')$ and tabulate the corresponding sines and cosines, (ii) reduce to a single index loop $m_x=m_x'$ and use the symmetry $m_x\leftrightarrow-m_x$, (iii) use the symmetry $(m_z,m_z')\leftrightarrow(-m_z,-m_z')$ {\rouge (we restrict ourselves to $(-1)^{m_z'}=\veps$, see footnote \ref{note12}),} (iv) tabulate the powers of degree $|m_x|$ appearing in (\ref{eq115}).} 
\be
\label{eq115}
\int_0^{2\pi} \frac{\dd\phi}{2\pi} \frac{\eee^{\ii n\phi}}{b_0+b_1 \cos\phi} = \frac{1}{\sqrt{b_0^2-b_1^2}} \left(\frac{-b_1}{b_0+\sqrt{b_0^2-b_1^2}}\right)^{|n|} \quad \forall n\in\mathbb{Z},\ \forall b_0>0,\ \forall b_1\in]-b_0,b_0[
\ee

\paragraph{Its sum on $\ell$ and $\veps$} The sum of $J_{3,1}^{(\ell,\veps)}(\omb)$ on the parity $\veps$ is straightforwardly done, see equations (\ref{eq113a},\ref{eq113b}). The sum on $\ell\in\mathbb{N}$ gives rise, in the sense of distributions, to the Fourier series of a Dirac comb \footnote{\label{note6} One simply writes 
$\sin[(\ell+1/2)\xi]=\{\cos(\ell\xi)-\cos[(\ell+1)\xi]\}/[2\sin(\xi/2)]$
to introduce the partially telescopic sum 
$\sum_{\ell\in\mathbb{N}} (\ell+1/2)\{\cos(\ell\xi)-\cos[(\ell+1)\xi]\}=(-1/2)+\sum_{\ell\in\mathbb{N}} \cos(\ell\xi)$.
} and is calculated thanks to the identity: 
\be
\label{eq116}
\sum_{\ell=0}^{+\infty} (2\ell+1) \frac{\sin[(\ell+1/2)\xi]}{\sin(\xi/2)} = \frac{\pi\delta(\xi)}{\sin^2(\xi/2)}
\ee
It thus remains to find the cases of cancellation of the angle $\xi$, knowing that the polar angles $\theta$, $\theta'$ (from now on taken as integration variables in preference to $u$,$u'$) span $[0,\pi]$ and that the azimuthal angles $\phi$, $\phi'$ can be conveniently taken in $[-\pi,\pi]$. In equation (\ref{eq113a}), we find as the only nodal point $(\theta,\phi)=(0,0)$; in the neighborhood of this point, $\xi^2\simeq\theta^2+\phi^2$ so that the action of $\delta(\xi)$ is easily evaluated in polar coordinates $(\rho,\gamma)$ in the plane $(\theta,\phi)$, for any regular function $f$: 
\be
\label{eq119}
\int_0^\pi\dd\theta \,\sin\theta \int_{-\pi}^{\pi} \frac{\dd\phi}{2\pi} f(\theta,\phi) \frac{\pi\delta(\xi)}{\sin^2(\xi/2)} = f(0,0) \int_0^{+\infty} \dd\rho \,\rho\, \int_{-\pi/2}^{\pi/2}\dd\gamma\, \rho\cos\gamma \frac{\delta(\rho)}{2(\rho/2)^2}=2 f(0,0)
\ee
In equation (\ref{eq113b}), we find three nodal lines\footnote{If $\xi=0$, $\mT^{(1)}(\theta,\theta';\phi,\phi')=3$; however, this is the trace of a rotation matrix in $\mathbb{R}^3$ which must therefore reduce to the identity: we have 
$\mR_z(\theta)\mR_x(-\phi)\mR_z(-\theta')\mR_x(-\phi')=\identit$
where $\mR_x(\theta)$ is the rotation of angle $\theta$ of axis $Ox$, etc. By taking the $zx$ matrix element of this relation, we get the condition $\sin\theta'\sin\phi=0$. Similarly, by changing the order of the operators by circular permutation under the trace, we end up with $\sin\theta\sin\phi'=0$. The nodal lines $l_1$ and $l_2$ in (\ref{eq120}) correspond to $\sin\theta=\sin\theta'=0$ that is $\theta=\theta'=0[\pi]$, {\rouge the cases $\theta=0,\theta'=\pi$ and $\theta=\pi,\theta'=0$ being trivially not suitable}. The line $l_3$ corresponds to $\sin\phi=\sin\phi'=0$ i.e. $\phi=\phi'=0[\pi]$, which in practice reduces to $\phi=\phi'=0$; indeed, {\rouge the cases $\phi=0,\phi'=\pi$ and $\phi=\pi,\phi'=0$ are obviously not suitable, and} the case $\phi=\phi'=\pi$ imposes $\cos(\theta+\theta')=1$ so $\theta=\theta'=0$ or $\pi$, which corresponds to nodal points, of zero contribution to the integral on $\theta,\theta',\phi,\phi'$. The remaining cases $\sin\theta=\sin\phi=0$ or $\sin\theta'=\sin\phi'=0$ lead only to nodal points and do not contribute either.} 
\be
\label{eq120}
l_1: \theta=\theta'=0,\phi+\phi'=0\quad ; \quad l_2: \theta=\theta'=\pi, \phi-\phi'=0 ; \quad l_3: \theta=\theta', \phi=\phi'=0
\ee
In the integral we are dealing with, which involves a regular function $g$ of the four angles, 
\be
\label{eq122}
\mathcal{J} = \int_0^\pi \dd\theta\, \sin\theta\, \int_0^\pi\dd\theta'\, \sin\theta'\, \int_{-\pi}^{\pi} \frac{\dd\phi}{2\pi} \int_{-\pi}^{\pi} \frac{\dd\phi'}{2\pi} g(\theta,\theta',\phi,\phi') \frac{\pi\delta(\xi)}{\sin^2(\xi/2)}
\ee
the lines $l_1$ and $l_2$ have a zero contribution because of the Jacobians $\sin\theta$ and $\sin\theta'$. In the vicinity of line $l_3$, we quadratize the quantity $\xi$ into the variables $\phi,\phi'$ and $\theta'-\theta$; the integral at fixed $\theta$ 
\be
\label{eq123}
\int_{-\infty}^{+\infty}\dd\phi \int_{-\infty}^{+\infty}\dd\phi' \int_{-\infty}^{+\infty}\dd(\theta'-\theta)\, \frac{\delta(\sqrt{\phi^2+\phi'^2+2\phi\phi'\cos\theta+(\theta'-\theta)^2})}{\phi^2+\phi'^2+2\phi\phi'\cos\theta+(\theta'-\theta)^2} = \frac{2\pi}{\sin\theta}
\ee
is well computed in the eigenbasis of the quadratic form appearing in the denominator and under the square root (it has eigenvalues $1$ and $1\pm\cos\theta$). It remains simply 
\be
\label{eq124}
\mathcal{J}=2\int_0^{\pi}\dd\theta\sin\theta\,g(\theta,\theta,0,0)
\ee
from which the expression of the desired sum (respecting the order of the contributions in equation (\ref{eq110})):\footnote{Where $\phi+\pi$ appears in the argument of $\mT^{(\ell)}$, we perform the change of variable $\phi\to\phi-\pi$.} 
\begin{multline}
\label{eq125}
\sum_{\ell\in\mathbb{N}}\sum_\veps J_{3,1}^{(\ell,\veps)}(\omb)=\frac{\omb}{\pi\sh\omb\sqrt{2\ch\omb}}\frac{[\mD_{3,1}(\omb,u=1)]^{-1}}{1+2\ch(2\omb)+\frac{2\alpha}{1+\alpha}(1+2\ch\omb)} \\
-\frac{\omb}{2\pi^2\sh\omb\sqrt{\ch\omb}} \int_{\mathbb{R}} \dd x'\int_{-1}^{1}\dd u' \frac{\lambda'^{2+1/2}}{\sqrt{\ch x'}}\frac{[\mD_{3,1}(\omb,u=1)\mD_{3,1}(x',u')]^{-1}}{[1+\lambda_{\omb}^2+\lambda'^2+\frac{2\alpha}{1+\alpha}(\lambda_{\omb}+u'\lambda'+\lambda_{\omb}\lambda'u')] [\lambda_{\omb}\to \lambda_{\omb}^{-1}]}\\
+\frac{\omb}{4\pi^2\sh\omb}\int_{\mathbb{R}} \dd X \int_{-1}^{1}\dd u \frac{[\sqrt{\ch X_+ \ch X_-}\mD_{3,1}(X_+,u)\mD_{3,1}(X_-,u)]^{-1}} {[1+\lambda_+^2+\lambda_-^2+\frac{2\alpha}{1+\alpha}((\lambda_++\lambda_-)u+\lambda_+\lambda_-)][\lambda_\pm\to\lambda_\pm^{-1}]}
\end{multline}
In practice, the most important case corresponds to $\omb=0^+$, because it gives access to the cluster coefficient $b_{3,1}$ of the homogeneous gas. We give the corresponding explicit expression of (\ref{eq125}), always respecting the order of the contributions, in terms of trigonometric functions and the dilogarithm or Bose function $g_2(z)=\sum_{n>0} z^n/n^2$: 
\begin{multline}
\label{eq126}
\sum_{\ell\in\mathbb{N}}\sum_\veps J_{3,1}^{(\ell,\veps)}(\omb=0^+)=\frac{1}{3\pi\sqrt{2}}\frac{(1+\alpha)^2}{(1+3\alpha)^{3/2}} -\frac{(1+\alpha)^5}{\sqrt{2}\pi^2\alpha(1+2\alpha)(1+3\alpha)^{3/2}}\left[\atan\frac{\alpha}{\sqrt{1+2\alpha}} -\frac{\alpha\argth\frac{\sqrt{1+3\alpha}}{\sqrt{2(1+2\alpha)}}}{(\alpha+1)\sqrt{2(1+3\alpha)}}\right] \\
+ \frac{(1+\alpha)^4}{2\pi^2\alpha(1+3\alpha)^2}\left\{\pi\asin\frac{\alpha}{1+2\alpha}+2\re\left[g_2\left(\frac{-1}{\zeta}\right)-g_2\left(\frac{1}{\zeta}\right)\right]-\re\left[g_2\left(\frac{2\zeta}{1+\zeta}\right)-g_2\left(\frac{-2\zeta}{1-\zeta}\right)+g_2\left(\frac{2}{1-\zeta}\right)-g_2\left(\frac{2}{1+\zeta}\right)\right]\right\} 
\end{multline}
where $\zeta=(2\alpha+\ii\sqrt{2(1+3\alpha)})/(1+\alpha)$.

\begin{SCfigure}
\includegraphics[width=5cm,clip=]{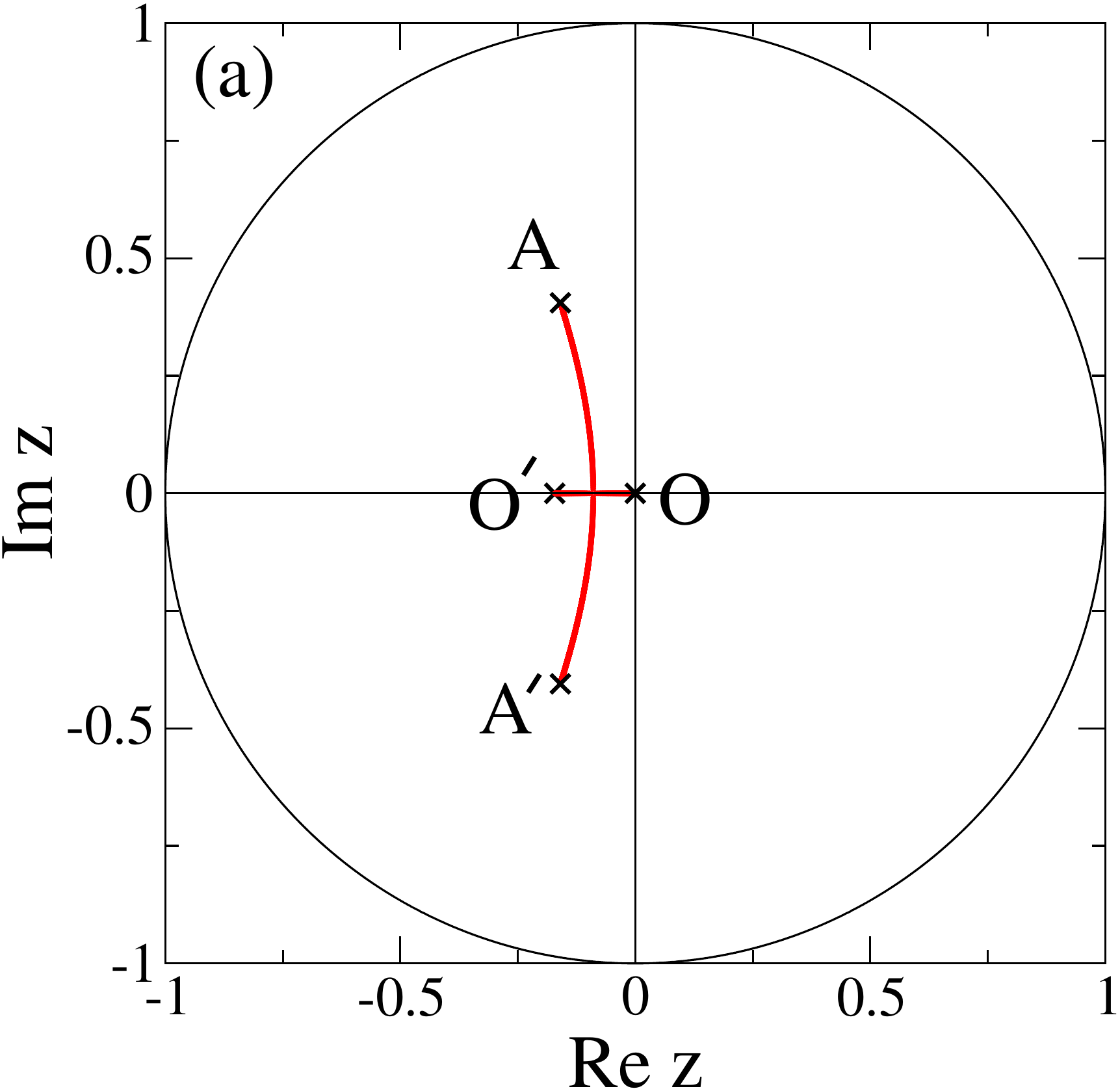}\hspace{1cm}\includegraphics[width=5cm,clip=]{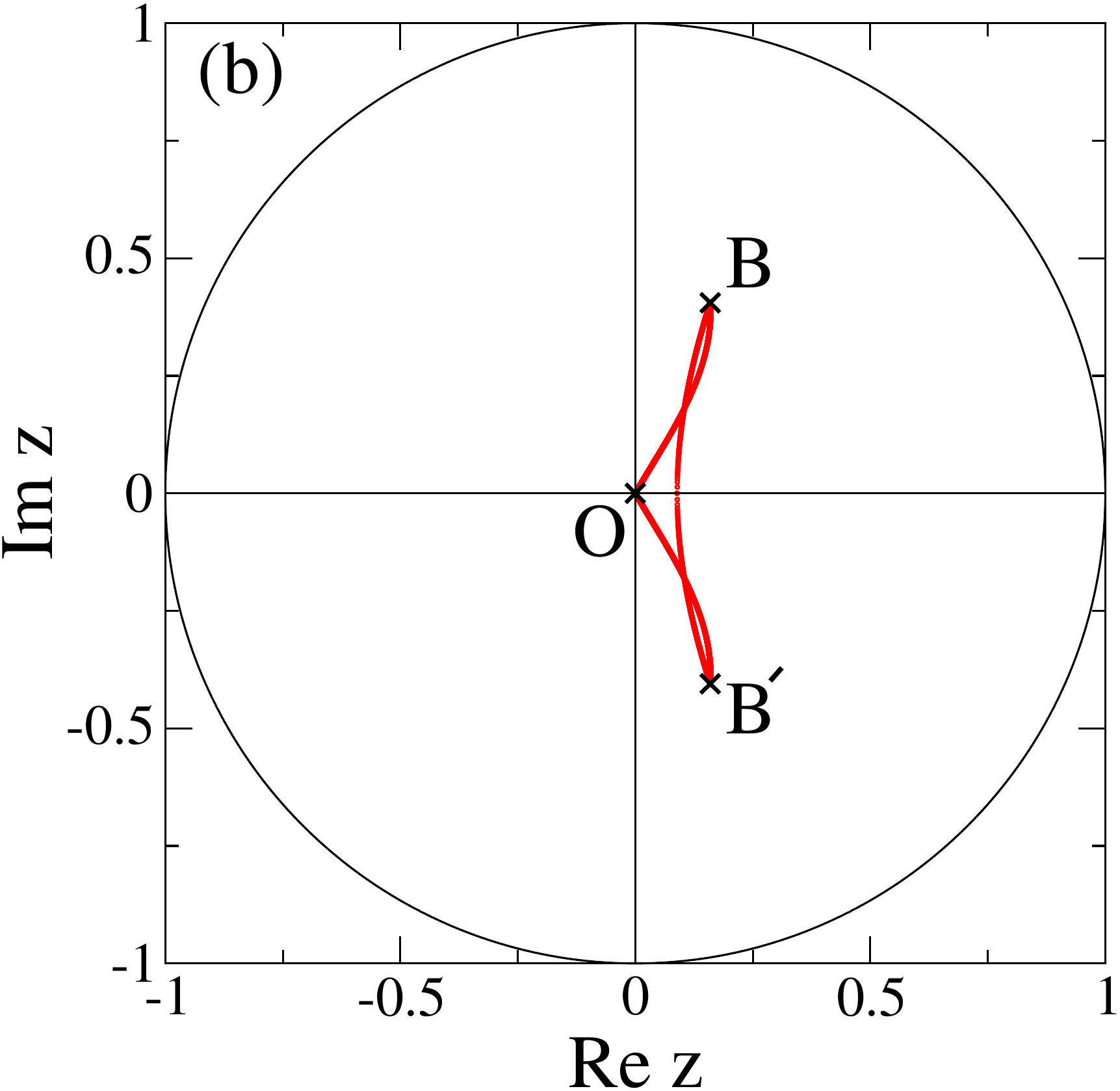}
\caption{Branch cuts of the integrand of equation (\ref{eq131}) in the unit disk (thick red line) in the case (a) $n=0$ and (b) $n=1$, obtained numerically for a mass ratio $\alpha=1$ taken as example.}
\label{fig2}
\end{SCfigure}

\paragraph{Its dominant behavior} To obtain an {\rouge asymptotic} equivalent of $J_{3,1}^{(\ell,\veps)}(\omb)$ {\rouge or, what amounts to the same thing, of $I_{3,1}^{(\ell,\veps)}(\omb)$} for large angular momentum, it suffices to restrict ourselves to the first contribution to the second side of equation (\ref{eq110}), which comes from the term {\yvan of order one} in the kernel {\rouge $\mK_{3,1}^{(\ell)}$} in expansion (\ref{eq106}). We then write it as an integral over the angle $\xi$ by inserting a Dirac delta $\delta(\xi-\xi(u,\phi))$ linking $\xi$ to $u$ and $\phi$ as in equation (\ref{eq113a}), then explicitly calculating the integral over $\phi$ and then over $u$. \footnote{After having reduced to an integral on $\phi\in[0,\pi]$ by periodicity and parity of the integrand, we use the relation 
$\delta(2\cos\xi+1-u(1+\cos\phi)-\cos\phi)=\delta(\phi-\phi_0)/[(1+u)\sin\phi_0]$ with $\sin\phi_0=2(u-\cos\xi)^{1/2}(1+\cos\xi)^{1/2}/(1+u)$
where the root $\phi_0$ is in $[0,\pi]$ if and only if $0\leq\theta\leq\xi$ given footnote \ref{note5}. {\rouge We also use relation (\ref{eq114}).}} To simplify, we restrict ourselves to $\omb=0^+$ and find: 
\be
\label{eq127}
I_{3,1}^{(\ell,\veps)}(0^+)\underset{\ell\to+\infty}{\sim} \int_0^\pi \dd\xi \, \frac{\sin[(\ell+1/2)\xi]}{2\sin(\xi/2)} [\rho_0(\xi)+\veps\rho_1(\xi)] \quad\mbox{where}\quad \rho_n(\xi)=(2\ell+1)A_n\frac{\sin^2(\xi/2)}{w_n(\cos\xi)}F_n(f_n(\cos\xi))
\ee
with the notation $\beta=1/\alpha$, the prefactors and the auxiliary functions\footnote{\label{note10} When we move to the complex plane, we mean that 
$\atan Z=\ln[(1+\ii Z)/(1-\ii Z)]/(2\ii)$.
As a result, the branch cut of the function $F_0(Z)$ is $]-\infty,-1]$ and that of $F_1(Z)$ is $\ii]-\infty,-1]\cup\ii[1,+\infty[$. Note that, despite appearances, $F_0(Z)$ is analytic in the unit disk, as shown by its series expansion at $Z=0$. {\rouge Note also that with this definition of the $\atan$ function, we have 
$\pi/2-\atan(\ii x)=-\ii \argth(1/x)$ $\forall x\in\mathbb{R}\setminus[-1,1]$,
which is already used in the integral (\ref{eq127}) on the real axis because the quantity under the last square root in the denominator of (\ref{eq129c}) can become negative; we have made here the choices 
$\ln(-1)=\ii\pi$ and $\sqrt{-1}=\ii$.
}} 
\bea
\label{eq129a}
\!\!\!\!\!\!\!\! A_0&\!\!\!=\!\!\!& \frac{\sqrt{2}(1+\alpha)^2}{\pi^2\alpha\sqrt{1+3\alpha}} \quad;\quad w_0(Z)=3\beta+(1+2Z)^2 \quad ; \quad F_0(Z)=\frac{\atan\sqrt{Z}}{\sqrt{Z}} \quad;\quad f_0(Z)=\frac{(1-Z)(5+4Z)}{w_0(Z)} \\
\label{eq129b}
\!\!\!\!\!\!\!\! A_1&\!\!\!=\!\!\!&\frac{\sqrt{2}(1+\alpha)^{5/2}}{\pi^2\alpha^2} \quad;\quad w_1(Z)=8(1+\beta)Z^2-(5\beta+7)Z+\beta(3\beta+1) \quad ;\quad F_1(Z)=Z\,\left(\frac{\pi}{2}-\atan Z\right) \\ 
\label{eq129c}
\!\!\!\!\!\!\!\! f_1(Z)&\!\!\!=\!\!\!&\frac{w_1(Z)}{4\sqrt{1+\beta}\sqrt{1-Z}\{[\frac{3\beta+1}{4}+(1+\beta)Z][3\beta+(1-2Z)^2]\}^{1/2}}
\eea
To reduce to contour integrals on the unit circle $C$, of the form 
\be
\label{eq131}
\mathcal{I}_{n}=\int_C\frac{\dd z}{\ii z} (z-1)z^\ell \frac{F_n\left(f_n\left(\frac{z+1/z}{2}\right)\right)}{w_n\left(\frac{z+1/z}{2}\right)}
\ee
we proceed in three steps: (i) we eliminate the factors $\sin(\xi/2)$ in the integrand of (\ref{eq127}), first by simplifying the sine in the denominator in front of the brackets with a sine in the numerator of $\rho_n(\xi)$, then by {\yvan using the remaining sine to make} $\cos[(\ell+1)\xi]-\cos(\ell\xi)$ appear as in footnote \ref{note6}; (ii) we extend the integral on $\xi$ to $[-\pi,\pi]$ by parity of the integrand; (iii) we take as new integration variable $z=\exp(\ii\xi)$, which {\yvan spans} the unit circle, so that $\cos\xi=(z+1/z)/2$, and we can write $\cos[(\ell+1)\xi]-\cos(\ell\xi)={\rouge\re(z^{\ell+1}-z^\ell)}$ and take out the function $\re$ of the integral since the remainder of the integrand is real-valued on $C$. For the usual determinations of the logarithm and the square root in the complex plane (branch cut on $\mathbb{R}^-$), we find that the integrand of $\mI_0$ has as singularity in the unit disk a trident-shaped branch cut, union of a segment $OO'$ and an arc $AA'$, see figure \ref{fig2}a, and that of $\mI_1$ a {\yvan flower-shaped} branch cut, union of three arcs $OB$, $OB'$ and $BB'$, see figure \ref{fig2}b. \footnote{In this second case, we expect to have three branch cuts, the one $L_1$ coming from the function $F_1(Z)$, the ones $L_3$ and $L_2$ coming from $\sqrt{1-Z}$ and the other square root in the denominator of the function $f_1(Z)$. In reality, $L_3$ is included in $L_1$. Moreover, the intersections $L_1\cap L_2$ and $L_1\cap L_3$ are not branch cuts of the integrand, for the reason that $z\mapsto z^{1/2}(\pi/2-\atan z^{\rouge 1/2})$ has only $[-1,0]$ as a branch cut (for $z\in]-\infty,-1]$, the change of sign of $\pi/2-\atan$ compensates for that of the square root). The arcs $OB$ and $OB'$ are made up of the points lying only on $L_2$, {\yvan and} the arc $BB'$ collects the points belonging only to $L_1$.} The point $O$ is here the origin of the coordinates, $A$ and $A'$ are the points of affixes $z_0$ and $z_0^*$, $B$ and $B'$ are the points of affixes $z_1=-z_0^*$ and $z_1^*$, and $z_0$ is the unique solution of the equation 
$z+z^{-1}+1+\ii\sqrt{3\beta}=0$
in the unit disk {\rouge (so that $w_0(\frac{z_0+1/z_0}{2})=0$)}: 
\be
\label{eq133}
z_0=\frac{1}{2}\left[\ii\sqrt{1-\ii\sqrt{3\beta}}\sqrt{3+\ii\sqrt{3\beta}}-(1+\ii\sqrt{3\beta})\right]
\ee
Using Cauchy's integral theorem, we shrink the integration contour until it fits the branch cuts, without changing the value of $\mI_n$. In the limit $\ell\to+\infty$, {\rouge because of the factor $z^\ell$,} the integral is then dominated by the neighborhood of the singularity points farthest from $O$, namely $A$ and $A'$ for $\mI_0$, $B$ and $B'$ for $\mI_1$, this being true for any mass ratio $\alpha$.  {\yvan $\hookrightarrow$} For the computation of $\mI_0$ in the neighborhood of $z_0$, let $z=z_0(1+\eta u)$, where $u$ is a complex number and $\eta>0$ is an infinitesimal, and approximate the horn of the trident by its right semitangent at the vertex $A$; we then have the equivalents $f_0((z+1/z)/2)\sim C_0/(\eta u)$ and $z^{\rouge\ell}\sim z_0^\ell \exp(\ell\eta u)$, and the {\rouge semitangent} becomes the half line of origin $O$ of direction $-C_0$ in the $u$ space {\rouge (considering footnote \ref{note10})}, that is the half line $\mathbb{R}^-$ in the space of $v=\eta u/C_0$, so that 
\be
\label{eq134}
\mI_0^{\mbox{\scriptsize\textcircled{$z_0$}}} \underset{\ell\to+\infty}{\sim}\frac{(z_0-1)z_0^\ell}{2\sqrt{3\beta}(z_0-1/z_0)} \int_{\mC_u} \frac{\dd u}{u} \eee^{\ell\eta u} F_0\left(\frac{C_0}{\eta u}\right)=\frac{z_0^{\ell+1}}{2\sqrt{3\beta}(1+z_0)} \int_{\mC_v}\frac{\dd v}{v} \eee^{\ell C_0 v} F_0(1/v) 
\ee
where the paths $\mC_u$ and $\mC_v$ surround the half-tangents counterclockwise in $u$ or $v$ space. The expression for $C_0$ is given in equation (\ref{eq334}). The integral {\yvan in} the last side of (\ref{eq134}) is easily computed: 
\be
\label{eq135}
\int_{\mC_v}\frac{\dd v}{v} \eee^{\ell C_0 v} F_0(1/v) = \int_{-\infty}^{0} \frac{\dd x}{x} \eee^{\ell x C_0} \left[F_0\left(\frac{1}{x-\ii0^+}\right)-F_0\left(\frac{1}{x+\ii 0^+}\right)\right]=  \int_{{\rouge -1}}^{0} \frac{\dd x}{x} \eee^{\ell x C_0} (-\ii\pi|x|^{1/2}){\rouge\underset{\ell\to+\infty}{\sim}}\frac{\ii\pi^{3/2}}{\sqrt{\ell C_0}}
\ee
The horn $A'$ of affix $z_0^*$ gives a complex conjugate contribution of the horn $A$. We get the first part of equation (\ref{eq333}). $\hookrightarrow$ For the calculation of $\mI_1$ in the neighborhood of $z_1$, we proceed in the same way, by setting $z=z_1(1+\eta u)$, $\eta\to 0^+$. Now the local behavior $f_1((z+1/z)/2)\sim \mA/(\mB\eta u)^{1/2}$ has itself a branch cut because of the square root and is characterized by two complex amplitudes $\mA$ and $\mB$, hence the structure with two branch cuts, the one coming from $f_1$ and the one coming from $\atan f_1$ in $F_1(f_1)$.\footnote{In the previous case $n=0$, this does not occur: In the function $F_0(z)$, on either side of the branch cut $\mathbb{R}^-$ of the square root, $\sqrt{z}$ can of course take two opposite values but $\atan$ is an odd function. As a consequence, the branch cut of $F_0(z)$ is imposed by the function $\atan$ and is $]-\infty,-1]$, in agreement with footnote \ref{note10}.} {\yvan They are} approximated by two semitangents corresponding in the space of $v=\eta\mB u/\mA^2$ simply to {\rouge $\mA^{-2}\mathbb{R}^-$ and, in light of footnote \ref{note10}, to} $\mathbb{R}^-$. One has 
\be
\label{eq136}
\mI_1^{\mbox{\scriptsize\textcircled{$z_1$}}} \underset{\ell\to+\infty}{\sim} \frac{(z_1-1)z_1^\ell\mA^2}{\ii\mB {\rouge w_1(\frac{z_1+1/z_1}{2})}}\int_{\mC_v} \dd v \,\eee^{\ell \mA^2 v/\mB} \frac{\mA}{\sqrt{v\mA^2}} \left(\frac{\pi}{2}-\atan \frac{\mA}{\sqrt{v\mA^2}}\right)
\ee
where the path $\mC_v$ surrounds (in the third quadrant) the half line $\mA^{-2}\mathbb{R}^-$ counterclockwise and then vertically joins the half line $\mathbb{R}^-$ from below to surround it also counterclockwise. The calculation leads to\footnote{One indeed has $\sqrt{\mA^2/\mB}=-\mA/\sqrt{\mB}$ for any mass ratio $\alpha$.} 
\begin{multline}
\label{eq137}
\int_{\mC_v}\dd v \,\eee^{\ell {\mA}^2 v/\mB}F_1\left(\frac{\mA}{\sqrt{v{\mA}^2}}\right) = \int_{-\infty}^{0}\frac{\dd x}{{\mA}^2} \eee^{\ell x/\mB}\left[F_1\left(\frac{\mA}{\sqrt{x-\ii 0^+}}\right)-F_1\left(\frac{\mA}{\sqrt{x+\ii 0^+}}\right)\right] 
+ \int_{-\infty}^{0} \dd x\, \eee^{\ell{\mA^2}x/\mB} \left[F_1\left(\frac{\mA}{\sqrt{(x-\ii 0^+){\mA^2}}}\right)\right. \\ \left.-F_1\left(\frac{\mA}{\sqrt{(x+\ii 0^+){\mA^2}}}\right)\right] {\rouge\underset{\ell\to+\infty}{\sim}} \int_{-\infty}^{0} \frac{\dd x}{\mA} \eee^{\ell x/\mB} \frac{\ii\pi}{\sqrt{-x}} +\int_{-\infty}^{0} \dd x\, \eee^{\ell{\mA^2}x/\mB} \frac{(-\ii\pi)}{\sqrt{-x}}=\frac{\ii\pi^{3/2}}{\sqrt{\ell}}\left(\frac{\sqrt{\mB}}{\mA}-\frac{1}{\sqrt{{\mA}^2/\mB}}\right)=\frac{2\ii\pi^{3/2}\mB}{\mA\sqrt{\ell\mB}}
\end{multline}
which gives the second part of equation (\ref{eq333}), {\yvan where} the coefficient $\mB$ {\yvan is} called $C_1$ and given by equation (\ref{eq334}).

\section{Asymptotic approximant of $I_{2,2}^{(\ell,\veps)}(0^+)$, its sum over $\ell$ and $\veps$}
\label{apB}

\paragraph{The asymptotic approximant} The aim here is to write as explicitly as possible an approximation $J_{2,2}^{(\ell,\veps)}(0^+)$ at large $\ell$ of the contribution $I_{2, 2}^{(\ell,\veps)}(0^+)$ of angular momentum $\ell$ and parity $\veps$ to the quantity $I_{2,2}(0^+)$ of equation (\ref{eq003}). We take as a starting point expansions (\ref{eq204},\ref{eq209}) according to the value of $\veps$. Their insertion into the integrand of (\ref{eq003}) leads, after integration on $S$, to the structure 
\be
\label{eq400}
J_{2,2}^{(\ell,\veps)}(0^+) = \mathcal{J}_{K_1,K_2}^{(\ell,\veps)} + \delta_{\veps,(-1)^\ell} \left[\mathcal{J}_{K_1,K_3}^{(\ell)}+\mathcal{J}_{K_2,K_3}^{(\ell)}\right]
\ee
where the subscript indicates that it is a cross contribution of the integral kernels $K_i$ and $K_j$. The case $(i,j)=(1,2)$ was treated in reference \cite{PRA} for $\veps=(-1)^{\ell-1}$, using forms (\ref{eq202}) of the kernels; here we simply extend it to the case $\veps=(-1)^\ell$, which is just a rewriting, and {\yvan we} copy the result: 
\begin{multline}
\label{eq401}
\mathcal{J}_{K_1,K_2}^{(\ell,\veps)}= \int_{\mathbb{R}} \dd x \int_{-1}^{1} \dd u\dd u' \int_0^{2\pi} \frac{\dd\phi\dd\phi'}{(2\pi)^2} \frac{[(2\ell+1)/(4\pi)^2][\ch x{\rouge\,}\mD_{2,2}(x,u) \mD_{2,2}(x,u')]^{-1}\mT^{(\ell,\veps)}(\theta,\theta';\phi,\phi')}{\left\{1\!+\!\frac{1}{1\!+\!\alpha}\left[(u\!+\!e^{-x})(u'\!+\!e^{-x})\!+\!v v'\cos\phi\right]\right\}\left\{1\!+\!\frac{\alpha}{1\!+\!\alpha}\left[(u\!+\!e^{x})(u'\!+\!e^{x})\!+\!v v' \cos\phi'\right]\right\}}
\end{multline}
with notations (\ref{eq104}) and (\ref{eq112}). The case $(i,j)=(1,3)$ is much simpler to handle (especially numerically) by using forms (\ref{eq318}) of the kernels, marked by a Czech accent. Only one factor depends on the scaling exponent $\ii S$, on which we have to integrate in (\ref{eq003}). In the notations of (\ref{eq318}) it comes after an integration by parts: 
\be
\label{eq402}
\int_{\mathbb{R}}\frac{\dd S}{2\pi}  (|z_1'|/|z_1|)^{\ii S} = 2\delta(\ln|z_1'|^2-\ln|z_1|^2)=2|z_1|\,|z_1'|\, \delta({\rouge |z_1|^2-|z_1'|^2})=2|z_1|\,|z_1'|\, \delta(\eee^{2x}-\eee^{2x'}-\sqrt{2\alpha}\eee^x u+\sqrt{2\alpha}\eee^{x'}u')
\ee
We decide to integrate the Dirac distribution on $u'$ at {\yvan fixed} $u$, $x$, $x'$, which links the value of $u'$ to these other variables: 
\be
\label{eq403}
u'=\eee^{x-x'}u+\frac{\eee^{2 x'}-\eee^{2 x}}{\sqrt{2\alpha}\eee^{x'}}
\ee
As $u'$ must be between $-1$ and $1$, the integration interval on $u$ is constrained and reduces to $[u_{\rm min},u_{\rm max}]$ with 
\be
\label{eq404}
u_{\rm min}=\max\left(-1,\frac{\eee^{2 x}-\eee^{2 x'}}{\sqrt{2\alpha}\eee^{x}}-\eee^{x'-x}\right) \quad ; \quad u_{\rm max}=\min\left(1,\frac{\eee^{2 x}-\eee^{2 x'}}{\sqrt{2\alpha}\eee^{x}}+\eee^{x'-x}\right)\quad;\quad u_{\rm min}<u_{\rm max}\Leftrightarrow |\eee^{x'}-\eee^{x}|<\sqrt{2\alpha} 
\ee
To compute the resummed kernel (\ref{eq208}), we recognize in the external part of $\check{K}_3$ (last contribution of (\ref{eq318})) a form factorized into an operator on $x$-space and an orthogonal projector of rank one on $u$-space, which reduces the problem to the inversion of an operator on $x$-space only: \footnote{There is no gauge transform {\rouge of type (\ref{eq205})} to be made here because $\check{K}_3$ is directly independent of the scale exponent.{\rouge We also notice {\yvan in} equation (\ref{eq316}) that $\check{\mD}_{22}\equiv 1$.}} 
\be
\label{eq405}
\check{K}_3^{\rm ext}={\rouge\hat{k}}\otimes |\chi\rangle\langle\chi| \quad\mbox{and}\quad \langle \chi|\chi\rangle=1 \Longrightarrow \check{\KKK}_3^{\rm ext}=[\,\identit-(\,\identit+{\rouge\hat{k}})^{-1}]\otimes |\chi\rangle\langle\chi|
\ee
with 
\be
\label{eq406}
\langle u|\chi\rangle=\frac{1}{\sqrt{2}} \quad \mbox{and}\quad \langle x|\hat{k}|x'\rangle=\frac{(-1)^{l+1}}{\pi} \frac{2\eee^{x+x'}}{\eee^{2x}+\eee^{2x'}+1} \frac{1}{(1+\eee^{-2x})^{1/4} (1+\eee^{-2x'})^{1/4}} 
\ee
We finally obtain 
\be
\label{eq407}
\boxed{
\mJ_{K_1,K_3}^{(\ell)}= \frac{(1+\alpha)^3}{8\pi\alpha^2} \int_{-\infty}^{+\infty}\dd x \int_{-\infty}^{+\infty} \dd x' \frac{\mathcal{R}(x,x')\langle x'|\,\identit-(\,\identit+\hat{k})^{-1}|x\rangle}{(1+\eee^{-2x})^{1/4} (1+\eee^{-2x'})^{1/4}}
}
\ee
where we have introduced the symmetric function of $x$ and $x'$ taking into account (\ref{eq404}) by a Heaviside function $Y$: 
\be
\label{eq408}
\boxed{
\mathcal{R}(x,x')\equiv Y(\sqrt{2\alpha}-|\eee^{x'}-\eee^x|) \int_{u_{\rm min}}^{u_{\rm max}}\dd u \frac{\eee^x}{|z_1|} \int_0^{2\pi}\frac{\dd\phi}{2\pi} \frac{(2\ell+1)\langle l,m_x=0|\eee^{\ii\tau_1 L_z/\hbar} \eee^{\ii\phi L_x/\hbar} \eee^{-\ii\tau_1' L_z/\hbar}|l,m_x=0\rangle}{|z_2|^2+|z_2'|^2+(1+\alpha^{-1})(\cos\tau_1\cos\tau_1'+\sin\tau_1\sin\tau_1'\cos\phi)}
}
\ee
The complex numbers $z_i$ and $z_i'$ and their arguments $\tau_i$ and $\tau_i'$ are defined below equation (\ref{eq314}) as functions of the angles $\theta$ and $\theta'$ in the interval $[0, \pi]$ such that $u=\cos\theta$ and $u'=\cos\theta'$, {\yvan where} the angle $\theta'$ {\yvan is} related to $\theta$ by equation (\ref{eq403}); the writing of the denominator of the integrand in (\ref{eq408}) takes into account the equality of the moduli $|z'_1|=|z_1|$. \footnote{\label{note9} In practice, the angular integral in (\ref{eq408}) is done numerically on the angle $\theta$ (rather than on $u$) with the 41-point Gauss-Legendre method; if $x$ is close enough to but different from the singularity point $x_{\rm sing}=\ln{\rouge\sqrt{\alpha/2}}$ where the first denominator $|z_1|$ can vanish, i.e. $0<|x-x_{\rm sing}|<\theta_c$ with $\theta_c=0.15$ for example, a narrow structure in $\theta$ appears and the integration on $[\theta_{\rm min},\theta_c]$ is performed with the change of variable $\theta=|x-x_{\rm sing}| \sh t$, with the midpoint rule on the variable $t$ discretized in 100 equidistant points. On the other hand, the second denominator in (\ref{eq408}) always remains greater than $1+\min(\alpha,\alpha^{-1})$ on the integration domain and cannot vanish. The integration on $x$ and $x'$ is done with the midpoint method on a truncation interval $[x_{\rm min},x_{\rm max}]$, by arranging that $x_{\rm sing}$ is the edge of one of the subintervals of width $\dd x$; we make the fixed choice 
$x_{\rm min}=-5$ since $\mR(x,x')=O(\min(\exp(x),\exp(x'))$ when $x,x'\to-\infty$,
but we extrapolate to $x_{\rm max}=+\infty$ cubically in $1/x_{\rm max}$ from the four values 
$x^{(0)}_{\rm max}=7$, $x^{(1)}_{\rm max}=30$, $x^{(2)}_{\rm max}=52$ and $x^{(3)}_{\rm max}=75$.
To take advantage of the fact that $\mR(x, x')=O(\exp(-2x))$ on an increasingly narrow support $x-x'=O(\exp(-x))$ when $x\to+\infty$, we separate the matrix $A$, discretized version of the operator $\,\identit+\hat{k}$, in blocks $A_{ij}$ between the intervals number one $[x_{\rm min}, x_{\rm max}^{(0)}]$ and number two $[x_{\rm max}^{(0)},x_{\rm max}^{(n)}]$, neglecting the blocks $12$, $21$ and $22$ of its inverse $A^{-1}$ and {\yvan we then} compute its block $11$ by the expression 
$(A^{-1})_{11}=[A_{11}-A_{12}(A_{22})^{-1}A_{21}]^{-1}$
which includes the effect of the non-diagonal coupling in the form of a second-order effective Hamiltonian (it would be incorrect to neglect it completely because $\langle x|\hat{k}|x'\rangle$ does not tend to zero when $x,x'\to+\infty$ at $x-x'$ fixed). As the function $\mR(x,x')$ shows near $x=x'=x_{\rm sing}$ on the left more and more rapid oscillations when $\ell$ increases, we cannot take a fixed value of the step $\dd x$ but we have to use an iterative method of Romberg type: From the initial choice $\dd x=1/20$, we reduce the step $\dd x$ by a factor of $2$ and extrapolate at zero step linearly in $(\dd x)^2$ until the extrapolated value is stable at the percent level (or $\dd x$ falls below the very small value $2\times 10^{-3}$). The matrix element in $|\ell,m_x=0\rangle$ is computed as above equation (\ref{eq115}) with the same symmetry tricks as in footnote \ref{note7} and the integral over $\phi$ is deduced from equation (\ref{eq115}).} Finally, the last contribution in (\ref{eq400}) is deduced from the second contribution by changing everywhere $\alpha$ into its inverse $1/\alpha$: 
\be
\label{eq409}
\mJ_{K_2,K_3}^{(\ell)}(\alpha)=\mJ_{K_1,K_3}^{(\ell)}(1/\alpha)
\ee

\paragraph{Its sum on $\ell$ and $\veps$} It remains to compute the sum on all channels $(\ell,\veps)$ of the asymptotic approximant (\ref{eq400}). The contribution of type $K_1 K_2$ is treated exactly as in \ref{apA}: 
\begin{multline}
\label{eq413}
\sum_{\ell\in\mathbb{N}}\sum_\veps \mJ_{K_1,K_2}^{(\ell,\veps)}={\rouge \int_{\mathbb{R}} \dd x \int_{-1}^{1}\dd u \frac{[8\pi^2\ch x\, \mD_{2,2}^{2}(x,u)]^{-1}}{[1+\frac{1}{1+\alpha}(1+\eee^{-2x}+2u\eee^{-x})][1+\frac{\alpha}{1+\alpha}(1+\eee^{2x}+2u\eee^{x})]}}\\
{\rouge =}\frac{(1+\alpha)^2}{8\pi^2\alpha}\left\{2\pi\asin\left(\frac{\alpha}{(1+2\alpha)(\alpha+2)}\right)^{1/2}-\re\left[g_2\left(\frac{2z}{1+z}\right)-g_2\left(\frac{-2z}{1-z}\right)+g_2\left(\frac{2}{1-z}\right)-g_2\left(\frac{2}{1+z}\right)\right.\right. \\
\left.\left.+2g_2\left(\frac{1}{z}\right)-2g_2\left(\frac{-1}{z}\right)\right] -\re\left[g_2\left(\frac{2z'}{1+z'}\right)-g_2\left(\frac{-2z'}{1-z'}\right)+g_2\left(\frac{2}{1-z'}\right)-g_2\left(\frac{2}{1+z'}\right)+2g_2\left(\frac{1}{z'}\right)-2g_2\left(\frac{-1}{z'}\right)\right] \right\}
\end{multline}
with $z=\sqrt{\alpha}+\ii\sqrt{1+\alpha}$, $z'=\sqrt{\alpha^{-1}}+\ii\sqrt{1+\alpha^{-1}}$
and $g_2$ the dilogarithm function. In the $K_1 K_3$ type contribution, let us first sum over the angular momentum $\ell$ of fixed parity  $(-1)^\ell=\eta$, so that the operator $\hat{k}$ in (\ref{eq407}) takes the fixed value $\hat{k}_\eta$. Let us transform the quantum average in state $|\ell,m_x=0\rangle$ in the numerator of the integrand of (\ref{eq408}) by inserting a closure relation {\yvan in the eigenbasis} of $L_x$ and using the expression of the corresponding matrix elements deduced from equations (7.2--9) on page 101, (7.3--15) on page 105 and (7.4--7) on page 109 of reference \cite{WuKi}: 
\be
\label{eq415}
\langle\ell,m_x=0|\eee^{\ii\tau L_z/\hbar}|\ell,m_x\rangle \eee^{\ii\psi m_x} = \left(\frac{4\pi}{2\ell+1}\right)^{1/2} Y_\ell^{m_x}(\tau,\psi) \quad\forall \tau\in[0,\pi],\forall\psi\in\mathbb{R}
\ee
where $Y_\ell^m$ are the usual spherical harmonics. {\yvan It then leads} to \footnote{The relation (\ref{eq415}) is used twice, in its direct form with $(\tau,\psi)=(\tau_1,\phi)$ and in its conjugated form with $(\tau,\psi)=(\tau_1',0)$. If $\tau_1$ is in $[-\pi,0]$, the relation does not apply, but it is then sufficient to change $\tau_1$ into $-\tau_1$, which amounts to changing the integration variable $\phi$ into $\phi+\pi$ {\rouge in (\ref{eq408})} (indeed, 
$\exp(-\ii\tau_1 L_z/\hbar)=\exp(\ii\pi L_x/\hbar) \exp(\ii\tau_1 L_z/\hbar) \exp(\ii\pi L_x/\hbar)$
) and does not modify the value of the integral. We proceed in the same way if $\tau_1'\in[-\pi,0]$. In the following, we can therefore assume that $\tau_1$ and $\tau_1'$ are in the interval $[0,\pi]$.}\footnote{Using the addition theorem for spherical harmonics, see equation (8.6--3) on page 145 of reference \cite{WuKi}, we show that the second side of (\ref{eq417}) is also written $(2\ell+1)P_\ell(\cos\delta)$ where $P_\ell$ is a Legendre polynomial and $\delta$ is the angle between the unit vectors of polar coordinates $(\tau_1,\phi)$ and $(\tau_1',0)$ ; its cosine 
$\cos\delta=\cos\tau_1\cos\tau_1'+\sin\tau_1\sin\tau_1'\cos\phi$
appears in the denominator of the integrand of (\ref{eq408}). } 
\be
\label{eq417}
(2\ell+1)\langle l,m_x=0|\eee^{\ii\tau_1 L_z/\hbar} \eee^{\ii\phi L_x/\hbar} \eee^{-\ii\tau_1' L_z/\hbar}|l,m_x=0\rangle = 4\pi \sum_{m_x=-\ell}^{\ell} Y_{\ell}^{m_x}(\tau_1,\phi)[Y_{\ell}^{m_x}(\tau_1',0)]^*
\ee
It remains to invoke the closure relation (8.6{\rouge--}10) on page 146 of reference \cite{WuKi} on spherical harmonics and the {\rouge spatial} parity property 
$Y_\ell^m(\pi-\theta,\phi+\pi)=(-1)^\ell Y_\ell^m(\theta,\phi)$
to obtain the closure relation with fixed $\ell$ parity: \footnote{{\rouge\label{note13}} In the Dirac distributions $\delta(\phi-\phi_0)$, $\phi$ has a meaning modulo $2\pi$; thus, one can replace the arbitrary integration interval $[0,2\pi]$ of equation (\ref{eq408}) by the interval of length $2\pi$ centered on $\phi_0$. } 
\begin{multline}
\label{eq420}
\sum_{\ell\geq 0\ |\ (-1)^\ell=\eta}\sum_{m_x=-\ell}^{\ell} Y_{\ell}^{m_x}(\tau_1,\phi)[Y_{\ell}^{m_x}(\tau_1',0)]^* = \frac{1}{2}\left[\delta(\cos\tau_1-\cos\tau_1')\delta(\phi)+\eta\,\delta(\cos\tau_1+\cos\tau_1')\delta(\phi-\pi)\right]\\
=\frac{|z_1|}{4}\left[\sqrt{2\alpha}\,\eee^{-2x}\delta(x-x')\,\delta(\phi)+\eta\,\eee^{-x}\delta(u-u_0)\,\delta(\phi-\pi)\right]\quad\mbox{where}\quad u_0\equiv\frac{2\alpha+\eee^{2x}-\eee^{2x'}}{2\sqrt{2\alpha}\eee^x}
\end{multline}
where we have replaced in the third side $\cos\tau_1\pm\cos\tau_1'$ by its value, {\rouge remembering that $|z_1|=|z_1'|$ and using (\ref{eq403})}. The integration on $\phi$ is straightforward {\rouge in view of footnote \ref{note13}}. To integrate over $u$, we need to know if the root $u_0$ is in the interval $[u_{\rm min},u_{\rm max}]$. For this purpose, we divide the support of the Heaviside function in (\ref{eq408}) into four distinct areas: 
(i) $\exp(x)<\exp(x')<\exp(x)+\sqrt{2\alpha}$ and $\exp(x)+\exp(x')>\sqrt{2\alpha}$, (ii) $\exp(x')<\exp(x)<\exp(x')+\sqrt{2\alpha}$ and $\exp(x)+\exp(x')>\sqrt{2\alpha}$, (iii) $\exp(x)<\exp(x')<\sqrt{2\alpha}-\exp(x)$, (iv) $\exp(x')<\exp(x)<\sqrt{2\alpha}-\exp(x')$.
{\yvan If} the expressions depending on $x$ in the definitions (\ref{eq404}) {\yvan are denoted by} $u_{\rm min}^{\rm\rouge exp}$ and $u_{\rm max}^{\rm exp}$, we find that we systematically have 
$u_{\rm min}=-1<u_0<u_{\rm max}=u_{\rm max}^{\rm exp}$ in zone (i), $u_{\rm min}=u_{\rm min}^{\rm exp}<u_0<u_{\rm max}=1$ in zone (ii), $u_{\rm min}=-1<u_{\rm max}=1<u_0$ in zone (iii), $u_{\rm min}=u_{\rm min}^{\rm exp}<u_{\rm max}=u_{\rm max}^{\rm exp}<u_0$ in zone (iv).
In other words, the integral of $\delta(u-u_0)$ over $u$ is always equal to one in the first two areas and to zero in the last two. We deduce the sum of the quantity $\mR(x,x')$ on all $\ell$ of fixed parity: 
\be
\label{eq424}
S_\eta(x,x')=\frac{1}{2}\left[\frac{1}{2}\alpha\,\eee^{-2x} \ln \frac{\eee^{2x}+\sqrt{2/\alpha}\,\eee^x+\alpha^{-1}+1/2}{\eee^{2x}-\sqrt{2/\alpha}\,\eee^x+\alpha^{-1}+1/2} \delta(x-x') + \eta \frac{Y(\sqrt{2\alpha}-|\eee^x-\eee^{x'}|)Y(\eee^x+\eee^{x'}-\sqrt{2\alpha})}{\eee^{2x}+\eee^{2x'}+1}\right]
\ee
It remains to sum on $\eta=\pm 1$ to arrive at the desired result: 
\be
\label{eq425}
\boxed{
\sum_{\ell=0}^{+\infty} \mJ_{K_1,K_3}^{(\ell)} = \frac{(1+\alpha)^3}{8\pi\alpha^2}\sum_{\eta=\pm} \int_{\mathbb{R}^2} \dd x\, \dd x' \frac{S_\eta(x,x')\langle x'|\,\identit-(\,\identit+\hat{k}_\eta)^{-1}|x\rangle}{(1+\eee^{-2x})^{1/4}(1+\eee^{-2x'})^{1/4}}
}
\ee
{\yvan Again,} the operator $\hat{k}_\eta$ is deduced from equation (\ref{eq406}) by replacing in the second side $(-1)^{\ell+1}$ by $(-\eta)$; the numerical inversion of the operators 
$\,\identit+\hat{k}_\eta$
and the integration over $x$ and $x'$ are done with the same techniques and tricks as in footnote \ref{note9} (on the other hand, there is no more integration to do on $u$). Finally, as shown in equation (\ref{eq409}), 
$\sum_{\ell\in\mathbb{N}} \mJ_{K_2,K_3}^{(\ell)}$
is deduced from expression (\ref{eq425}) by changing everywhere $\alpha$ to $1/\alpha$ {\rouge ({\yvan also} in $S_\eta(x,x')$)}.

\section{The operator $M_{2,2}^{\rouge(\ell,\veps)}(\ii S)$ in the formulation of reference \cite{Ludo}}
\label{apC}

To obtain the operator $M_{2,2}^{(\ell,\veps)}(\ii S)$ at the basis of conjecture (\ref{eq003},\ref{eq004}) on the cluster coefficient $B_{2,2}(\omb)$ of the trapped system, we first write {\rouge a Faddeev} ansatz for an eigenstate of the unitary $2+2$-body $\uparrow\uparrow\downarrow\downarrow$ problem of zero energy and zero momentum in free space. This ansatz is expressed in terms of an unknown function $D$ of two wave vectors. Taking into account the Wigner-Bethe-Peierls two-body $\uparrow\downarrow$ contact conditions (as in footnote \ref{note0}) leads to an integral equation {\yvan for} this function. Then, we use the {\yvan rotational invariance} to project the equation on the subspace of angular momentum $\ell$ and parity $\veps$ as reference \cite{CRAS} explains it in detail. Finally, we use the scale invariance of the unitary problem (in the absence of three-body Efimov effect) to choose a function $D$ with a well-defined scale exponent $s$. The integral equation is then reduced to the condition $\det M_{2,2}^{(\ell,\veps)}(s)=0$ where $M_{2,2}^{(\ell,\veps)}(s)$ is a kernel operator, which we have to extend on the pure imaginary axis $s=\ii S$ to evaluate expression (\ref{eq003}).

Expression (\ref{eq200},{\rouge\ref{eq201},\ref{eq202}}) of $M_{2,2}^{(\ell,\veps)}(\ii S)$ corresponds to the choice {\rouge of unknown function} $D(\kk_2,\kk_4)$ where $\kk_2$ and $\kk_4$ are the wave vectors of two opposite-spin fermions, as in reference \cite{PRA} (the wave vectors $\kk_1$ and $\kk_3$ disappear in the limit {\yvan expressing} the contact condition). {\rouge The starting integral equation is given by equation (13) of reference \cite{PRA}.} Another choice is made in reference \cite{Ludo}, corresponding to the {\yvan change of} function 
\be
\label{eq300}
D(\kk_2,\kk_4)=F(\uu\equiv-(\kk_2+\kk_4),\vv\equiv\frac{m_\downarrow\kk_2-m_\uparrow\kk_4}{m_\uparrow+m_\downarrow})
\ee
{\yvan It has} the advantage of providing a much simpler expression of the singular integral kernel $K_3$, i.e.\ of the third contribution in (\ref{eq202}) (but not of the kernels $K_1$ and $K_2$, which justifies in the end keeping (\ref{eq202}) in the numerical calculation {\rouge of (\ref{eq003})}).  Up to a sign, the new variables are simply the wave vector of the center of mass and the relative motion of particles $2$ and $4$.  In the following, we use the notation $\alpha=m_\uparrow/m_\downarrow$.

Let's follow the previously stated steps. The starting integral equation is written {\rouge in parameterization (\ref{eq300}) \cite{Ludo}} 
\be
\label{eq301}
\frac{\kappa}{4\pi} F(\uu,\vv)+ \int \frac{\dd^3 u'}{(2\pi)^3} \left[\frac{F(\uu',\vv_{14})}{\kappa^2+\sigma_{14}^2} + \frac{F(\uu',\vv_{23})}{\kappa^2+\sigma_{23}^2}\right]{\rouge -\int\frac{\dd^3 v'}{(2\pi)^3} \frac{F(-\uu,-\vv')}{\kappa^2+v'^2}}=0
\ee
with the notations taken from reference \cite{Ludo},\footnote{Reference \cite{Ludo} arranges the fermions in the order 
$\uparrow\downarrow\uparrow\downarrow$.
We renumber the particles accordingly.} 
\begin{multline}
\label{eq302}
\kappa = \left(v^2 + \frac{2\alpha u^2}{(1+\alpha)^2}\right)^{1/2} \quad ; \quad \sigg_{14} = \vv +\frac{1-\alpha}{1+\alpha}\uu + \uu' \quad ; \quad \vv_{14}=\vv-\frac{\alpha}{1+\alpha}\uu + \frac{\alpha}{1+\alpha} \uu' \quad ; \\
\quad \sigg_{23}=\vv + \frac{1-\alpha}{1+\alpha}\uu - \uu' \quad ; \quad \vv_{23} = \vv+\frac{\uu}{1+\alpha} -\frac{\uu'}{1+\alpha}
\end{multline}
by correcting what seems to us to be a sign error in the coefficient of the vector $\kk$ (here called $\uu'$) in the definition of $\vv_{23}$ (third row and first column of table III of this reference) and in the expression of $E_{\rm coll}$ (notation not introduced here) just above equation (141) of this reference. As in reference \cite{PRA}, to make future transformations simpler, we adopt a variational formulation of equation (\ref{eq301}), 
$\delta\mathcal{E}/\delta F^*(\uu,\vv)=0$
with the functional of $F$ and $F^*$ that follows: 
\begin{multline}
\label{eq304}
\mathcal{E}{\rouge\equiv}\int\dd^3u\dd^3v\frac{\kappa}{4\pi} F^*(\uu,\vv)F(\uu,\vv)+\int\frac{\dd^3u\dd^3v\dd^3u'\dd^3v'}{(2\pi)^3} F^*(\uu,\vv) F(\uu',\vv')\left[\frac{\delta(\vv'-\vv_{14})}{\kappa^2+\sigma_{14}^2}+\frac{\delta(\vv'-\vv_{23})}{\kappa^2+\sigma_{23}^2}-{\rouge\veps}\frac{\delta(\uu-\uu')}{\kappa^2+v'^2}\right]
\end{multline}
where the parity $\veps=\pm 1$ of the solution has been introduced to make the sign $-$ disappear in front of $\uu$ {\rouge and $\vv'$} in {\rouge $F(-\uu,-\vv')$}. Let us now consider rotational invariance, restricting ourselves to the subspace of total angular momentum $\ell\in\mathbb{N}$, with a zero angular momentum component along the quantization axis $Oz$. The solution $F(\uu,\vv)$ is then expressed in terms of $2\ell+1$ functions $f_{m_z}^{(\ell)}(u,v,w)$ ($-\ell\leq m_z\leq\ell$) of only three real variables, the moduli $u$ and $v$ of the two vectors and the non-oriented angle $\theta=\widehat{(\uu,\vv)}\in[0,\pi]$ between them or, what amounts to the same thing, its cosine $w=\cos\theta$; to fix the parity at $\veps$ is to impose $(-1)^{m_z}=\veps$ thus to decouple the problem into $\ell$ and $\ell+1$ unknown functions, which we indicate by an exponent $\veps$ on the sum sign below. Our ansatz for $F$ is that of equation (14) of reference \cite{PRA}. We insert it into functional (\ref{eq304}) and integrate over the variables other than the arguments of the functions $f_{m_z}^{(\ell)}$ in the same way as in that reference. For example, let us apply equation (45) of \cite{PRA} to the last contribution of functional $\mathcal{E}$, the one {\rouge with $\veps$ in factor}, that gives rise to the singular kernel $K_3$. {\rouge First, a direct reference trihedron $\mT$ with polar axis $\uu$ and another $\mT'$ with polar axis $\uu'$ are chosen. The integration on $\vv$ (or on $\vv'$) is performed in the spherical coordinates associated to $\mT$ (or to $\mT'$), the cosine of the corresponding polar angle being $w$ (or $w'$).} Integrating over the other variables then amounts to {\rouge taking the average over the orientations of $\mT$ and $\mT'$ which can be done by} fixing the direction of the vector $\uu$ along the convenient direction $Ox$ and replacing the integration over the direction of $\uu'$ by an integration over {\rouge rotation $\mR$ mapping $\mT$ to $\mT'$ in} the SO(3) group with an invariant measure, explicit 
$\dd\mR=\dd a(\sin b)\dd b\dd c/8\pi^2$
in the Euler parameterization 
$\mR=\mR_Z(a)\mR_Y(b)\mR_Z(c)$,
where angles $a$ and $c$ {\yvan span} an interval of length $2\pi$, angle $b$ spans $[0,\pi]$ and the direct Cartesian reference frame $OXYZ$ is of any orientation with respect to the reference frame $Oxyz$ (see section 8.2 of reference \cite{WuKi}); it follows that 
\begin{multline}
\label{eq305}
\int\frac{\dd^3u\dd^3v\dd^3u'\dd^3v'}{(2\pi)^3} F^*(\uu,\vv) F(\uu',\vv')\frac{\delta(\uu-\uu')}{\kappa^2+v'^2}= {\sum_{m_z,m_z'}}^{\!\!\veps} 2\int_0^{+\infty}\!\!\!\!\!\!\dd u u^2 \dd v v^2 \dd u' u'^2 \dd v' v'^2 \int_{-1}^{1} \!\!\dd w \dd w' f_{m_z}^{(l)*}(u,v,w) f_{m_z'}^{(l)}(u',v',w') \\
\times\int_{\rm SO(3)}\dd\mR (\langle\ell,m_z|R|\ell,m_z'\rangle)^* \frac{\delta(u\mathbf{e}_x-u'\mR \mathbf{e}_x)}{v^2+v'^2+\frac{2\alpha u^2}{(1+\alpha)^2}}
\end{multline}
where operator $R$ represents rotation $\mR$ in the Hilbert space of a quantum particle. The choice of axes $OZ=Oz$ and $OY=Ox$ leads to 
${\rouge(\sin b)}\delta(u\mathbf{e}_x-u'\mR\mathbf{e}_x)={\rouge(\sin b)}\delta(u\mR_z(-a)\mathbf{e}_x-u'\mR_x(b)\mR_z(c)\mathbf{e}_x)=\delta(u\cos a-u'\cos c)\delta(-u\sin a-u'\sin c\cos b)\delta(-u'\sin c{\rouge)}=[\delta(u-u')/(u u'{\rouge)}] [\delta(a)\delta(c)+\delta(a-\pi)\delta(c-\pi)]$
where we successively used the rotational invariance of the three dimensional Dirac distribution and decomposed its action into Dirac distributions along $Ox$, $Oy$ and $Oz$. Integration {\rouge in} SO(3) simply reduces to {\rouge the line} $\mR=\mR_x(b)$ if $a=c=0$ and to {\rouge the line} $\mR=\mR_x(-b)$ if $a=c=\pi$, which is taken into account by an integration on $b$ extended to $[-\pi,\pi]$, {\yvan giving} rise to a projector on the state of zero angular momentum along $Ox$, 
$\int_{-\pi}^{\pi}\frac{\dd b}{2\pi}\langle\ell,m_z|\eee^{-\ii b L_x/\hbar}|\ell,m_z'\rangle=\langle\ell,m_z|\ell,m_x=0\rangle\langle\ell,m_x=0|\ell,m_z'\rangle$.
We get 
\begin{multline}
\label{eq310}
\mE={\sum_{m_z}}^{\!\veps} \int_0^{+\infty}\!\!\!\dd u u^2 \dd v v^2 \int_{-1}^1\!\!\dd w f_{m_z}^{(l)*}(u,v,w) f_{m_z}^{(l)}(u,v,w) \frac{1}{2}\left(v^2+\frac{2\alpha u^2}{(1+\alpha)^2}\right)^{1/2}+{\sum_{m_z,m_z'}}^{\!\!\veps}\int_0^{+\infty}\!\!\!\dd u u^2 \dd v v^2 \dd u' u'^2 \dd v' v'^2 \\
\times \int_{-1}^{1}\!\!\dd w \dd w'\, f_{m_z}^{(l)*}(u,v,w) f_{m_z'}^{(l)}(u',v',w') \int_0^{2\pi}\frac{\dd\phi}{(2\pi)^2} \left[\frac{\delta(|Z_1|-|Z_1'|)(|Z_1|\,|Z_1'|)^{-1} \eee^{\ii T_1 m_z} \langle l,m_z|\eee^{\ii\phi L_x/\hbar}|l,m_z'\rangle \eee^{-\ii T_1' m_z'}}{|Z_2|^2+|Z_2'|^2+\frac{2uu'}{1+\alpha}(\cos T_1\cos T_1'+\sin T_1\sin T_1'\cos\phi)}
\right.\\ \left.+ \frac{\delta(|Z_2|-|Z_2'|)(|Z_2|\,|Z_2'|)^{-1} \eee^{\ii T_2 m_z} \langle l,m_z|\eee^{\ii\phi L_x/\hbar}|l,m_z'\rangle \eee^{-\ii T_2' m_z'}}{|Z_1|^2+|Z_1'|^2+\frac{2\alpha uu'}{1+\alpha}(\cos T_2\cos T_2'+\sin T_2\sin T_2'\cos\phi)}+\frac{(-\veps)\delta(u-u')}{u u'}\frac{\langle\ell,m_z|\ell,m_x=0\rangle\langle\ell,m_x=0|\ell,m_z'\rangle}{v^2+v'^2+\frac{2\alpha u^2}{(1+\alpha)^2}}\right] 
\end{multline}
where we have introduced the complex numbers 
$Z_1\equiv \alpha u/(1+\alpha)-v\exp(\ii\theta)=|Z_1|\exp(\ii T_1)$, $Z_2=u/(1+\alpha)+v\exp(\ii\theta)=|Z_2|\exp(\ii T_2)$
and their counterparts for the primed variables {\rouge
$Z_1'\equiv \alpha u'/(1+\alpha)-v'\exp(\ii\theta')=|Z_1'|\exp(\ii T_1')$, $Z_2'=u'/(1+\alpha)+v'\exp(\ii\theta')=|Z_2'|\exp(\ii T_2')$.
} Finally, let us take advantage of scale invariance by means of the ansatz 
\be
\label{eq312}
f_{m_z}^{(l)}(u,v,w) = \frac{(\ch x)^{s+3/2}(1+\eee^{-2x})^{-s/2} \Phi_{m_z}^{(l)}(x,w)}{(u^2+v^2/{\rouge\varsigma}^2)^{(s+7/2)/2}} \quad \mbox{where}\quad v={\rouge\varsigma}\eee^x u\quad\mbox{and}\quad {\rouge\varsigma}=\frac{\sqrt{2\alpha}}{1+\alpha}
\ee
chosen cleverly so that the diagonal part of the functional (first contribution in (\ref{eq310})) becomes scalar (independent of any variable) and so that the singular part (last contribution) does not depend on the mass ratio or even on the scaling exponent $s$. The integration on $u$ brings out as in reference \cite{PRA} an infinite constant factor, here 
$({\rouge\varsigma}^4/16)\int_0^{+\infty}\dd u/u$,
to give the regularized functional 
\begin{multline}
\label{eq314}
\bar{\mE}={\sum_{m_z}}^{\!\veps} \int_{-\infty}^{+\infty} \dd x \int_{-1}^{1} \dd w\, \Phi_{m_z}^{(l)*}(x,w) \Phi_{m_z}^{(l)}(x,w)+\frac{2}{\rouge\varsigma^3} {\sum_{m_z,m_z'}}^{\!\!\veps}\int_{-\infty}^{+\infty} \dd x \dd x' \int_{-1}^{1} \dd w\dd w'\frac{\eee^{x+x'}\Phi_{m_z}^{(l)*}(x,w) \Phi_{m_z'}^{(l)}(x',w')}{(1+\eee^{-2x})^{1/4}(1+\eee^{-2x'})^{1/4}}\int_0^{2\pi}\frac{\dd\phi}{(2\pi)^2}\\
\left[\frac{(|z_1'|/|z_1|)^s(|z_1|\,|z_1'|)^{-1/2}\eee^{\ii\tau_1 m_z} \langle l,m_z|\eee^{\ii\phi L_x/\hbar}|l,m_z'\rangle \eee^{-\ii\tau_1' m_z'}}{|z_1'|^2|z_2|^2\!+\!|z_1|^2|z_2'|^2\!+\!(1\!+\!\beta)(\re z_1\re z_1'\!+\!\im z_1\im z_1'\cos\phi)}+
\frac{(|z_2'|/|z_2|)^s(|z_2|\,|z_2'|)^{-1/2}\eee^{\ii\tau_2 m_z} \langle l,m_z|\eee^{\ii\phi L_x/\hbar}|l,m_z'\rangle \eee^{-\ii\tau_2' m_z'}}{|z_2'|^2|z_1|^2\!+\!|z_2|^2|z_1'|^2\!+\!(1\!+\!\alpha)(\re z_2\re z_2'\!+\!\im z_2\im z_2'\cos\phi)} \right.  \\
\left.+(-\veps){\rouge\varsigma}^3\frac{\langle\ell,m_z|\ell,m_x=0\rangle\langle\ell,m_x=0|\ell,m_z'\rangle}{\eee^{2x}+\eee^{2x'}+1}\right]
\end{multline}
where we have set 
$\beta=1/\alpha$, $z_1\equiv(\alpha/2)^{1/2}-\exp(x\!+\!\ii\theta)=|z_1|\exp(\ii\tau_1)$, $z_2\equiv(\beta/2)^{1/2}+\exp(x\!+\!\ii\theta)=|z_2|\exp(\ii\tau_2)$
and the equivalent relations for the primed variables, 
$z_1'\equiv(\alpha/2)^{1/2}\!-\!\exp(x'\!+\!\ii\theta')=|z_1'|\exp(\ii\tau_1')$, $z_2'\equiv(\beta/2)^{1/2}\!+\!\exp(x'\!+\!\ii\theta')=|z_2'|\exp(\ii\tau_2')$
and where we recall that 
$\theta=\acos w$ and $\theta'=\acos w'$.
A simple functional derivation of (\ref{eq314}) with respect to $\Phi_{m_z}^{(\ell)*}(x,w)$, taking a pure imaginary scale exponent and returning to the notation $u=\cos\theta$ of the main article gives the new form of the operator $M^{(\ell,\veps)}_{2,2}(\ii S)$, marked with a Czech accent to avoid confusion with (\ref{eq200},{\rouge\ref{eq201},\ref{eq202}}): 
\be
\label{eq316}
\langle x,u|\langle \ell,m_z| \check{M}_{2,2}^{(\ell,\veps)}(\ii S)|f\rangle = f_{m_z}(x,u) + \int_{-\infty}^{+\infty}\dd x' \int_{-1}^{1}\dd u' \sum_{m_z' \ |\ (-1)^{m_z'}=\veps}  \check{K}^{(\ell)}_{2,2} (x,u,m_z;x',u',m_z') f_{m_z'}(x',u')
\ee
with an integral kernel separated into three contributions 
$\check{K}=\check{K}_1+\check{K}_2+\check{K}_3$
written line by line: 
\begin{multline}
\label{eq318}
\!\!\!\!\!\check{K}^{(\ell)}_{2,2} (x,u,m_z;x',u',m_z')=\frac{2{\rouge\varsigma}^{-3}\eee^{x+x'}(|z_1|\,|z_1'|)^{-1/2}}{(1+\eee^{-2x})^{1/4}(1+\eee^{-2x'})^{1/4}}\!\!\int_0^{2\pi}\!\!\!\!\frac{\dd\phi}{(2\pi)^2}\frac{|z_1|^{-\ii S}\eee^{\ii\tau_1 m_z}\langle l,m_z|\eee^{\ii\phi L_x/\hbar}|l,m_z'\rangle\eee^{-\ii\tau_1' m_z'}|z_1'|^{\ii S}}{|z_1'|^2|z_2|^2\!+\!|z_1|^2|z_2'|^2\!+\!(1\!+\!\beta)(\re z_1\re z_1'\!+\!\im z_1\im z_1'\cos\phi)} \\
+\frac{2{\rouge\varsigma}^{-3}\eee^{x+x'}(|z_2|\,|z_2'|)^{-1/2}}{(1+\eee^{-2x})^{1/4}(1+\eee^{-2x'})^{1/4}}\int_0^{2\pi}\frac{\dd\phi}{(2\pi)^2} \frac{|z_2|^{-\ii S}\eee^{\ii\tau_2 m_z}\langle l,m_z|\eee^{\ii\phi L_x/\hbar}|l,m_z'\rangle\eee^{-\ii\tau_2' m_z'}|z_2'|^{\ii S}}{|z_2'|^2|z_1|^2\!+\!|z_2|^2|z_1'|^2\!+\!(1\!+\!\alpha)(\re z_2\re z_2'\!+\!\im z_2\im z_2'\cos\phi)} \\
+\frac{(-1)^{\ell+1}}{\pi}\frac{\eee^{x+x'}}{\eee^{2x}+\eee^{2x'}+1} \frac{\langle\ell,m_z|\ell,m_x=0\rangle\langle\ell,m_x=0|\ell,m_z'\rangle}{(1+\eee^{-2x})^{1/4}(1+\eee^{-2x'})^{1/4}}
\end{multline}
{\rouge where we remembered that 
{\rouge $\langle\ell,m_z|\ell,m_x=0\rangle=0$ if $\veps\neq(-1)^\ell$.}
} To verify that equation (\ref{eq318}) is written in the same order as equation (\ref{eq202}), i.e. that the kernel $\check{K}_i$ is precisely the kernel $K_i$ written in the formulation of reference \cite{Ludo}, we show numerically that 
$\int_\mathbb{R} \dd S\, \mathrm{Tr}\,[\mD_{2,2}^{-1} K_1(\ii S) \mD_{2,2}^{-1} K_3(\ii S)] = \int_\mathbb{R} \dd S\, \mathrm{Tr}\,[\check{K}_1(\ii S)\check{K}_3]$.


\end{document}